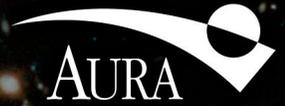

# FROM COSMIC BIRTH TO LIVING EARTHS

## THE FUTURE OF UVOIR SPACE ASTRONOMY

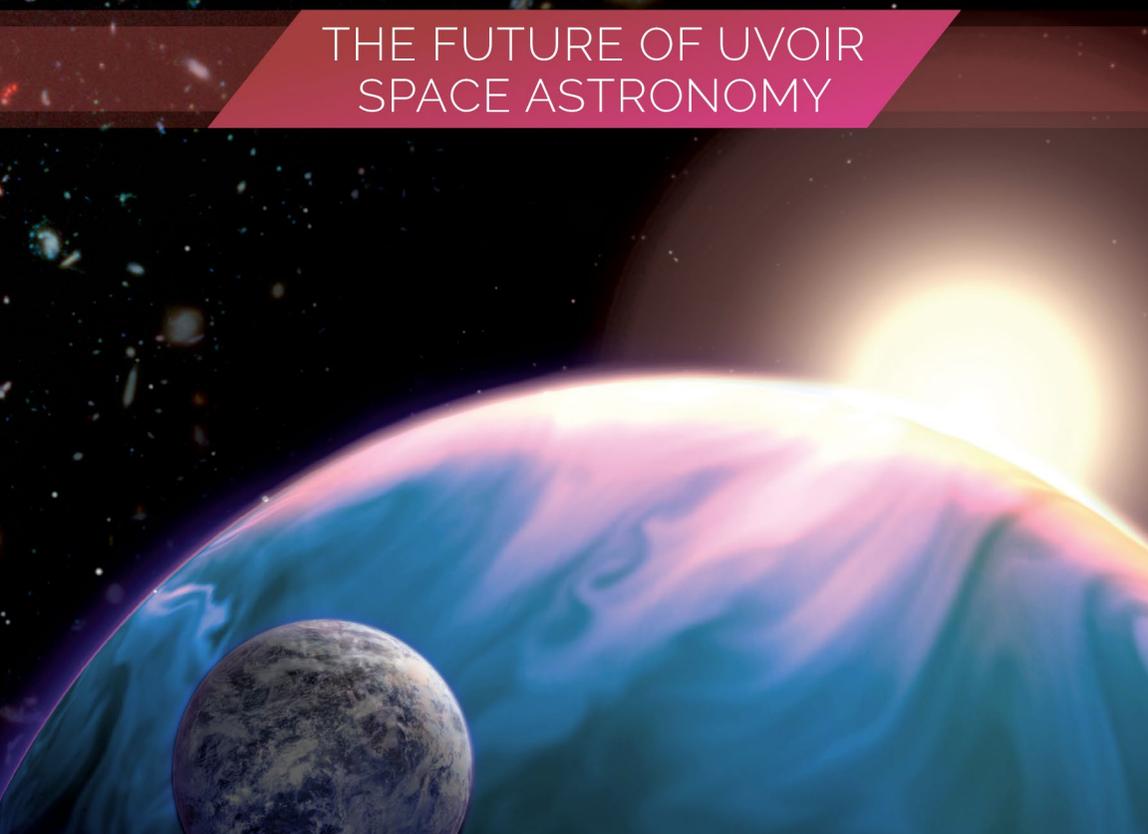

# Contents

*For the first time in history,* humans have reached the point where it is possible to construct a revolutionary space-based observatory that has the capability to find dozens of Earth-like worlds, and possibly some with signs of life. This general purpose, long-lived facility would be the prime tool for generations of astronomers, producing transformational scientific advances in every area of astronomy and astrophysics from black hole physics to galaxy formation, from star and planet formation to the Solar System. The associated inspirational public impact will likely exceed that of all other current and past astronomical endeavors.

## AURA Committee Members

JULIANNE DALCANTON, CO-CHAIR
*University of Washington*

SUZANNE AIGRAIN
*University of Oxford*

STEVE BATTEL
*Battel Engineering, Inc.*

NIEL BRANDT
*Pennsylvania State University*

CHARLIE CONROY
*Harvard University*

LEE FEINBERG
*NASA, Goddard Space Flight Center*

SUVI GEZARI
*University of Maryland, College Park*

OLIVIER GUYON
*University of Arizona/NAOJ*

WALT HARRIS
*University of Arizona/LPL*

SARA SEAGER, CO-CHAIR
*Massachusetts Institute of Technology*

CHRIS HIRATA
*The Ohio State University*

JOHN MATHER
*NASA, Goddard Space Flight Center*

MARC POSTMAN
*Space Telescope Science Institute*

DAVE REDDING
*Jet Propulsion Laboratory/Caltech*

DAVID SCHIMINOVICH
*Columbia University*

H. PHILIP STAHL
*NASA, Marshall Space Flight Center*

JASON TUMLINSON
*Space Telescope Science Institute*

# Chapter 1 The Challenge

Nobel Laureate Dr. Riccardo Giacconi predicted in 1997 that in the next century humanity would have the tools to "study the evolution of the Universe in order to relate causally the physical conditions during the Big Bang to the development of RNA and DNA." The time has arrived to accept this challenge.

Astronomical discoveries emerged at a breathtaking pace over the last two decades. We found that the Universe's expansion is accelerating, powered by as-yet-unknown physics. We saw signs of massive black holes lurking at the heart of nearly every large galaxy. We mapped the earliest ripples in the distribution of matter and traced their development through 14 billion years of star formation and galaxy evolution. We identified thousands of planets outside our Solar System, many of them small and rocky, and used them to show that there are at least as many planets as there are stars in the Milky Way Galaxy.

And yet, with all we know, and with all we have achieved, some of humanity's most compelling questions remain unanswered: Are we alone in the Universe? Are other Earth-like worlds common? Do any have signs of life? How did life emerge from a lifeless cosmic beginning? Curious humans have asked these questions for millennia, but for the first time we can foresee building the astronomical technology required to find dozens of Earth-like planets, to search these planets for signs of life, and to tell the cosmic story of how this life came to be. Over the next decade, progress in instrumentation will rise to the challenge of directly imaging faint planets in orbit around nearby stars and of characterizing their atmospheres. Our understanding of the frequency and size distribution of exoplanets has grown in the past five years to the point where it is statistically very likely that we will find Earth analogs orbiting other stars. These technological and scientific advances, combined with the fundamental quest to understand the evolution of the Universe, now provide an opportunity to meet Dr. Giacconi's challenge to connect cosmic birth to living Earths.

To define a vision for meeting this challenge, the Association of Universities for Research in Astronomy (AURA) commissioned a new study of space-based options for ultraviolet (UV) and optical astronomy in the era following the *James Webb Space Telescope*'s mission. This study follows AURA's charge to promote excellence in astronomical research by providing community access to state-of-the-art facilities. AURA tasked a team of research astronomers and technologists to "assess future space-based options for UV and optical astronomy that can significantly advance our understanding of the origin and evolution of the cosmos and the life within it."

*We conclude, after careful consideration of the science cases and technological approaches, that a 12 meter class space telescope with sufficient stability and the appropriate instrumentation can find and characterize dozens of Earth-like planets and make transformational advances in astrophysics.* Beyond purely scientific considerations, a single observatory covering all areas is also likely to be the optimum programmatic choice. This is true even if, under some assumptions, smaller and more focused observatories could address narrower science aims more efficiently. Exoplanet exploration and astrophysics are stronger together.

The broad outlines of this facility are clear and well-motivated. For example, we know with reasonable accuracy the frequency of planets around other stars, and how many are about the right size and temperature to be similar to the Earth. A 12 m class space-based telescope could detect enough of these planets to perform a robust census of habitable worlds and determine their physical and chemical characteristics: Do their atmospheres contain oxygen, water vapor, and other molecular gases characteristic of life? Is there evidence for continents and oceans on their surfaces? Are there detectable seasonal variations in their surface properties? Telling the full story of life in the cosmos will also require us to show how life's ingredients came together from the diffuse remnants of the Big Bang. The galaxies that form stars, the stars that form the heavy elements, and the planets that host life all have a role in this origins story. We find that a telescope designed to observe from UV to near-infrared wavelengths is not only capable of detecting signs of life on nearby worlds, but can also trace the origins of life's galactic home and raw chemical ingredients back to the earliest epochs of the Universe. *The concept we propose is called the* **"High-Definition Space Telescope"** (*HDST*). *HDST will achieve unprecedented angular spatial resolution and sensitivity in the UV and optical and will reach the extreme contrast required to separate Earth-like planets from the glow of their parent stars and search them for signs of life.* Equipped with a versatile instrument package to optimize its scientific yield, *HDST* would be operated as a general observatory supporting a broad range of investigations, while simultaneously seeking the answer to some of our most profound questions.

The committee did not study all possible concepts for future observatories that might advance exoplanet characterization and astrophysics. We did not consider missions dedicated exclusively to exoplanet detection or to general astrophysics. Such missions have benefits, but as both would be multi-billion dollar flagships with launch dates separated by a decade or more, and as such missions require broad support to be realized, we instead focused our efforts on determining whether the top science requirements for exoplanet characterization and astrophysics, as we envision them now, are compatible with a single great observatory. We find that they are.

This report comprises the outcome of the AURA study. We present the scientific and technical requirements for a space telescope of appropriate size and power to determine whether or not life is common outside the Solar System. We do not propose a specific design for such a telescope, but we describe its required performance, and show that designing and building such a facility is feasible beginning in the next decade—if the necessary strategic investments in technology begin now.

# Chapter 2 From Cosmic Birth to Living Earths

Astronomy exists because it is human nature to observe, to explore, and to understand the Universe. Astronomers gaze to the edges of their vision, and wonder: What else is out there that we cannot yet see? Each new generation pushes forward to answer this question by inventing tools to look beyond, and to explore and understand what was previously unseen. When the unknown comes into view, it is woven into a wondrous tapestry of knowledge about the natural world. Each new addition raises new questions that in turn spur the invention of new tools, repeating and widening the human cycle of discovery. And each new telescope is embraced by humanity because it brings more of the Universe home to us.

> *"Can we find another planet like Earth orbiting a nearby star? To find such a planet would complete the revolution, started by Copernicus nearly 500 years ago, that displaced Earth as the center of the Universe. . . . The observational challenge is great, but armed with new technologies . . . astronomers are poised to rise to it."*
>
> — "New Worlds, New Horizons," Decadal Survey of Astronomy and Astrophysics, 2010

Our telescopes have led us to a remarkable milestone. We now live in a world where children grow up knowing that there are planets around nearby stars, and that some of these planets orbit their stars in the "habitable zone" (HZ), where water would be liquid and the planet could, in principle, support life. Remarkable as this is, it is within our power to leave a far more profound legacy for future generations—the knowledge that not only are there other planets, but that some of these planets are *just like Earth*, with surface oceans and atmospheres with carbon dioxide, and possibly even life. We are the first generation that can reach for this lofty and ambitious goal, because only now are other living Earths orbiting nearby Sun-like stars within our technological grasp.

Almost as fascinating as the possibility of life on other worlds is the certain knowledge that *all* life, here on Earth and elsewhere, is made from atoms forged since the dawn of time in the furnaces of stellar interiors. These heavy elements—carbon, nitrogen, oxygen, iron, and others—were ejected from aging stars or in violent stellar explosions that forced most of these building blocks of life from the galaxies into cold, dark intergalactic space. The slow, gentle nudge of gravity brought them back, gathered them together into new stars and planets, and these planets, after long ages, used these elements and starlight to nurture life into being. These are the astronomical origins of life on Earth and elsewhere in our Galaxy. No matter how different life elsewhere may be from that on Earth, we will have at least this much in common: we are, as Carl Sagan noted, "star stuff," and we share our Galactic home. Thus, the astronomical search for life and its origins in galaxies, stars, and planets are all woven together.

While we now have a small sample of potentially habitable planets around other stars, our current telescopes lack the power to confirm that these alien worlds are truly able to nurture life. This small crop of worlds may have temperate, hospitable surface conditions, like Earth. But they could instead be so aridly cold that all water is frozen, like on Mars, or so hot that all potential life would be suffocated under a massive blanket of clouds, like on Venus. Our current instruments cannot tell the difference for the few rocky planets known today, nor in general, for the larger samples to be collected in the future. Without better tools, we simply cannot see their atmospheres and surfaces, so our knowledge is limited to only the most basic information about the planet's mass and/or size, and an estimate of the energy reaching the top of the planet's atmosphere. But if we could directly observe exoplanet atmospheres, we could search for habitability indicators (such as water vapor from oceans) or for signs of an atmosphere that has been altered by the presence of life (by searching for oxygen, methane, and/or ozone).

Rocky planets are the most promising targets in the search for potentially habitable worlds because their solid surfaces provide a biosphere where the atmosphere, water, minerals, and stellar energy can interact chemically. NASA's *Kepler* mission has identified at least 800 candidate rocky planets that transit their stars. In the next ten years, detections of rocky planets will accelerate with the deployment of two space-based exoplanet survey missions—NASA's *Transiting Exoplanet Survey Satellite* (*TESS*) and ESA's *PLAnetary Transits and Oscillations of*

*Stars* (*PLATO*)—which should identify many habitable-zone rocky planet candidates within 200 light-years of the Sun. Neither will possess the tools to tell a comfortable alien Earth from an uninhabitable Venus, but will provide a next logical step in the search for life by filling out the list of prime targets for future characterization.

If we are lucky, one or a few of the small, rocky planets found by *TESS* or *PLATO* could be studied in some detail in the near future. These planets will be beyond the reach of direct spectroscopy by any planned instruments, but it may be possible to characterize their atmospheres with indirect methods such as transit transmission spectroscopy or, in a few special cases, high-resolution spectroscopic cross-correlation template matching. For example, transit techniques will be applied to rocky planets orbiting small, cool stars (M-dwarfs) at the end of this decade using the *James Webb Space Telescope* (*JWST*) and early in the next decade by the large 20 m–40 m ground-based telescopes (e.g., European Extremely Large Telescope [E-ELT], Giant Magellan Telescope [GMT], Thirty-Meter Telescope [TMT]) currently under construction. These future facilities will excel at examining the atmospheres of giant planets orbiting any kind of star, but at best they will characterize the atmospheres of only a few large rocky planets orbiting stars smaller than the Sun. And unless detectable life exists on virtually every Earth-like planet where it *could* exist, these facilities have only a slim prospect for seeing signs of life on one of the few Earth-like planets within their reach.

The prospect of small samples of rocky exoplanets highlights the core problem in the astronomical search for life: it might be uncommon or rare even on worlds with hospitable conditions. While we now have statistical constraints on the frequency of Earth-sized planets, and on their likelihood for orbiting in their stars' habitable zones, we do not know how many of these worlds will show detectable signs of life. Even a single detection of biosignature gases on an exoEarth would change history, but if life occurs on, for example, only one in ten of such planets, more than twenty will need to be searched to give a good chance of not missing its signatures. Thus the two irreducible requirements for a successful search for life are (1) the ability to detect biosignature gases with spectroscopic remote sensing on (2) a statistically significant sample of relevant planets. As powerful as they are, the new telescopes planned for the 2020s do not meet these requirements. They will show the way, prove technology, verify methods, and promise many discoveries about exoplanetary systems as a population. However, the great task of characterizing the surface and atmospheric features of a large sample of rocky planets around Solar-type stars will be left to more powerful future telescopes that can directly detect and obtain the spectra of those exoplanets.

To meet the challenge posed by direct detection of exoEarth atmospheres, a telescope must meet three key metrics for its sensitivity, resolution, and stability. First, it must be large enough to obtain a low-resolution (R ~70–100) spectrum of an exoEarth with a good signal-to-noise ratio (SNR ~10) at wavelengths covering the habitability and biosignature gas spectral features. This extreme sensitivity implies an optical telescope much larger than *Hubble*, since the broadband reflected light from an exoEarth is not expected to be much more than 10 nJy (equivalent to only 6 photons detected every minute with a 12 m mirror). Second, the telescope must be able to observe planets within the star's habitable zone, which may occupy only 10–35 milliarcseconds (mas) from the central star (the width of a dime viewed from 35–125 km away). Third, and most challenging, the telescope's instruments must be able to suppress the glare from the host star. The hard job of studying faint, distant worlds is made harder by their close proximity to stars that are up to 10 billion times brighter. Suppressing this glare to below the brightness of the nearby planet requires extreme stability in the telescope's image that is unattainable from the ground, where natural fluctuations in the atmosphere occur faster than they can be corrected, even when the technology for compensating for the Earth's atmosphere reaches maturity. Only a large telescope in the quiet of deep space can simultaneously achieve the sensitivity, resolution, and stability required to characterize the atmospheres of a statistically significant number of Earth-like planets around Sun-like stars. A space observatory also enables alternative, novel methods of starlight suppression that are not feasible from the ground—specifically, the use of a specially shaped occulter, or "starshade," that can be flown in front of the telescope to cast a shadow on the mirror, thus blocking the light of the planet's host star.

The foundation for the sensitivity, resolution, and stability needed to survey Earth-like planets is a space telescope with an aperture diameter in excess of 9 m. Although specialized space telescopes as small as 1–2 m might be able to image one to a few exoEarths, their light-gathering power would be insufficient to take spectra and search for biosignature gases in more than a few of those planet atmospheres. Smaller telescopes also cannot survey the hundreds of nearby stars that must be imaged to make the searches for life, or constraints on its non-detection, statistically robust. As we will show later, a 12 m class telescope will search hundreds of stars, find at least 20–50 exoEarths, and, perhaps, signs of life on a few of these. A significantly smaller telescope will search fewer than 100 stars, find perhaps a half-dozen exoEarths, and would need to rely on unusually good fortune to turn up signs of life on even one. If we want to move beyond simple detection of a single *possible* Earth-like planet, a larger sample size—and thus a large telescope—is required to both *confirm* planets' habitability and to place meaningful constraints on the incidence of biosignature gases on such planets. The exact aperture size depends most strongly on the frequency of Earth-sized exoplanets residing in the habitable zones of Sun-like stars and on the levels of exozodiacal dust in the planetary system, as well as on the exact capabilities of the observatory (e.g., inner working angle, bandwidth, throughput, mission lifetime, and the fraction of observing time allocated to planetary searches). These exoEarth yields and their implications for telescope collecting area, resolution, and starlight suppression are detailed in Chapter 3.

Fundamental astrophysics research also faces critical problems that will go unsolved by current instruments. Major advances in our understanding of how galaxies, stars, and planets form and evolve require a telescope with the same overall degree of sensitivity, resolution, and stability needed for discovering and characterizing exoEarth candidates. For fundamental astrophysics, 50–100× the sensitivity of *Hubble* in the UV and optical is needed to observe dwarf galaxies during the epoch of reionization and Solar-type stars in nearby galaxies at 32–34 mag, both fainter than the typical exoEarth. The 10 milliarcsec resolution element of a 12 m telescope (diffraction limited at 0.5 micron) would reach a new threshold in spatial resolution. It would be able to take an optical image or spectrum at about 100 parsec spatial resolution or better, *for any observable object in the entire Universe*. Thus, no matter where a galaxy lies within the cosmic horizon, we would resolve the scale at which the formation and evolution of galaxies becomes the study of their smallest constituent building blocks—their star-forming regions and dwarf satellites. Within the Milky Way, a 12 m telescope would resolve the distance between the Earth and the Sun for any star in the Solar neighborhood, and resolve 100 AU anywhere in the Galaxy. Within our own Solar System, we would resolve structures the size of Manhattan out at the orbit of Jupiter. Finally, the high degree of image stability (and low sky backgrounds) from a space telescope also greatly benefits general astrophysics, by enabling precise photometry at deeper limits and longer time baselines than are possible on the ground. A space telescope also has unique access to powerful ultraviolet diagnostics that are otherwise completely inaccessible.

As detailed in Chapter 4, all these factors of sensitivity, resolution, and stability combine to make a uniquely powerful general astrophysi-cal observatory possessing revolutionary imaging and spectroscopic capabilities. *HDST* will enable detailed studies of the nearest stellar populations, robust measurements of the mass spectrum of new stars, panoramic high-resolution imaging of star formation in the vicinity of supermassive black holes, and detailed dynamical studies of the smallest known galaxies. The unprecedented collecting area of *HDST,* the 100× gain in ultraviolet sensitivity over *Hubble*, and multiplexed instrument modes will detect the nearly invisible material permeating the cosmic web and feeding galaxies, trace the recycling of heavy elements from stars to intergalactic space and back, follow the origins of stellar masses, characterize the composition of planet-forming disks, and monitor geysers on satellites of the outer planets.

The spectral range covered by *HDST* will, at minimum, span 0.1–2 microns and perhaps may extend further into the infrared. Unlike *JWST*, however, this telescope will not need to be operated at cryogenic temperatures as its core science drivers are focused in the wavelength regime shortwards of 2 microns. *HDST* will complement, not duplicate, many of the powerful telescopes that will likely be operating at the end of the next decade and beyond. These synergies are explored further in Chapter 4. Many fields of astronomy will be transformed by these novel capabilities and synergies; none will remain entirely untouched.

A mission at the scale and scientific breadth of *HDST* will be a challenge to design, test and operate. All these challenges are surmountable with an affordable investment in technology development and with strategic international cooperation. NASA is uniquely capable of leading such a mission, given both its large-mission experience and upcoming heavy-lift launch capabilities. Early investments in technology development will be essential in three key areas: starlight-suppression systems for ultra-high-contrast imaging; fabrication of lightweight optics for large telescopes; and active telescope optical and thermo-mechanical control systems to ensure the required stability needed for high-contrast imaging. Much progress has already been made on these fronts, and there is significant technical heritage from which we can draw. These topics are discussed in substantial detail in Chapters 5 and 6, along with plans for developing a versatile, powerful complement of instruments that maximize the scientific output of the observatory.

Telling the story of cosmic birth to living Earths has begun, but it will not be complete without the essential step of performing actual measurements of the habitability of a large sample of rocky exoplanets and determining whether any of those planets may host life. Moreover, while the outline of the story is largely in place, the richness of the tapestry has yet to be realized, and must include revealing the complex physics that brings the ingredients of life from intergalactic space down to protoplanetary systems. A space-based telescope with the capabilities of *HDST* will allow us to make a major step in understanding the astrophysical connections between the origin of cosmic structure and the origins of life. *HDST* will likely be an icon of U.S. technological know-how that will inspire generations of people across the globe, just as *Hubble* has done. For both scientific and inspirational reasons, incremental steps in the next generation of space-based astronomical observatories are insufficient. We require big steps to make the transformational discoveries for which the 21$^{st}$ century will be remembered.

The findings of this report are organized as follows. Chapter 3 details the scientific drivers for exoplanet characterization, including biosignature gas detection. Chapter 4 describes future, transformative astrophysics enabled by *HDST* along with the synergies with other concurrent astronomical observatories. Chapter 5 provides the technology requirements needed for *HDST*, and Chapter 6 gives a plan for strategic investment that allows for the design and construction of this facility on the needed timescale. Chapter 7 discusses the role and costs of ambitious missions in astronomy. Chapter 8 contains our closing thoughts.

## *HDST* at a Glance

*Mirror:* 12 meter class, segmented primary mirror, diffraction limited at 500 nm.

*Observatory operating temperature:* Non-cryogenic, with a sunshade and active thermal control system for stability. Likely range: 250–295 K.

*Wavelength range:* 100 nm to 2 microns (baseline); options to extend bluewards to 90 nm and redwards to 3 to 5 microns (but without resorting to cryogenic telescope structures) to be explored.

*Nominal orbit:* Halo orbit about the Sun-Earth L2 point.

*Stability:* Active wavefront sensing and control system, active thermal control system, internal metrology system, and vibration isolation and disturbance suppression system.

*Serviceability:* Designed with modular sub-systems and science instruments, and potential for later starshade.

## Notional Science Instruments:

- Internal Coronagraph with visible-near-IR IFU (400 nm–2 microns), FOV 10'', $10^{10}$ starlight suppression, 35 milliarcsec inner working angle (3 $\lambda$/D at $\lambda$ = 650 nm, D = 12 m).

- UV Integral Field Spectrometer (90 nm–300 nm), FOV 1'–3', R $\leq$ 100,000.

- Visible Imaging Array (300 nm–1 micron), FOV 6', Nyquist sampled at 500 nm.

- Near-IR imager and spectrograph (1 micron–2 microns), FOV 4', Nyquist sampled at 1.2 microns.

- Multi-object spectrograph (350 nm–1.6 microns), FOV 4', R $\leq$ 2000.

- Mid-IR imager (2.5 microns–5 microns), potential second generation instrument, if it can be implemented without impacting performance of UV or exoplanet instruments.

# Chapter 3 Exoplanets, Planetary Systems, and the Search for Habitable Worlds

## 3.1 Introduction

Looking up at the sky on a clear, dark night, one sees too many stars to count (Figure 3-1). Every star in the sky is a potential sun. Our Sun has eight planets: Mercury, Venus, Earth, Mars, Jupiter, Saturn, Uranus and Neptune (as well as several dwarf planets, asteroids, and many other interesting bodies). If our Sun has a planetary system, it seems logical that other stars should have planetary systems also.

It is now two decades since the discovery of the first exoplanet orbiting a Sun-like star. Over these two decades astronomers have found thousands of exoplanets and exoplanet candidates. Astronomers have also found that most stars host a system of one or more planets. Our Sun is one of hundreds of billions of stars bound together by gravity in the Milky Way, which is only one of the hundreds of billions of galaxies in the observable Universe. Given the vast realm of possibility, one cannot help but wonder: How many planets like Earth might be out there? Are there any life forms on those other Earths, perhaps looking out at the stars in their sky, wondering the same thing?

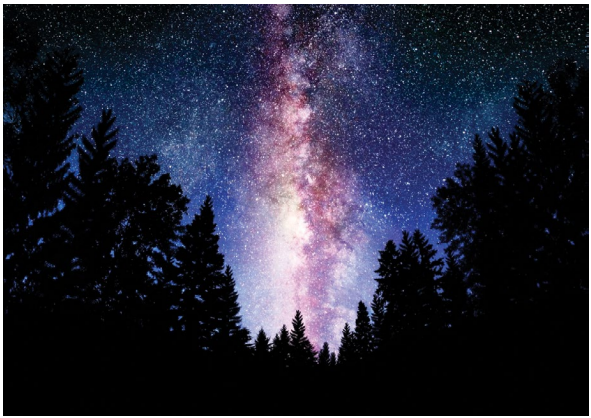

Figure 3-1: Image of a dark, star-filled sky including a view of the stately disk of the Milky Way.

THE FUTURE OF UVOIR SPACE ASTRONOMY

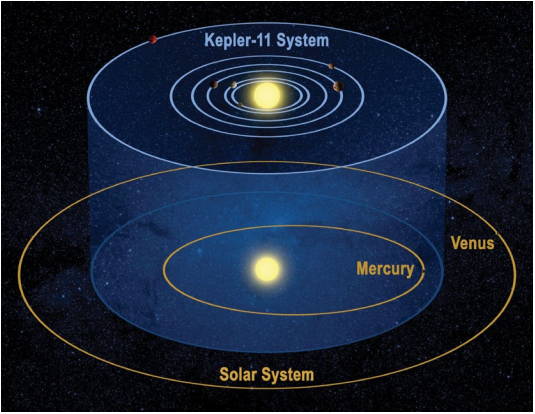

Figure 3-2a: Kepler-11b, the iconic compact multiple planet system, with six planets orbiting interior to where Venus' orbit would be. Credit: NASA.

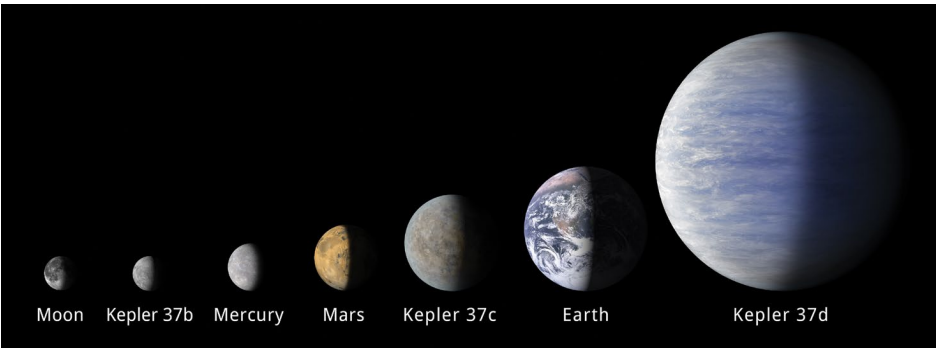

Figure 3-2b: The Moon-sized exoplanet Kepler-37b in comparison to other small planets. Credit: NASA.

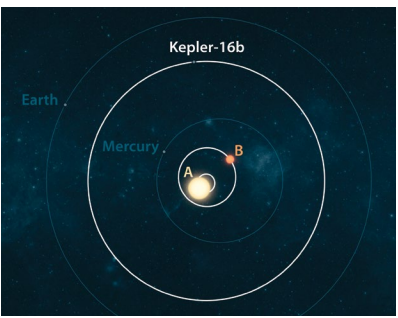

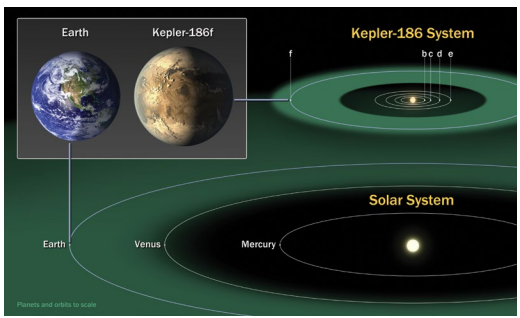

Figure 3-2c: The circumbinary planet Kepler-16b from an overhead view. The eccentric orbits of the two stars Kepler 16A and 16B are also shown. Credit: NASA.

Figure 3-2d: Comparison of the planets in our inner Solar System to those in Kepler 186, a five-planet star system with an M dwarf host star, a star that is half the size and mass of the Sun. Credit: NASA.

A fascinating finding about the known star and exoplanet systems is their huge variety. Some stars have giant planets the size of Jupiter, but orbiting where the Earth would be. Other stars have planets about the size of Earth, but orbiting ten times closer to their star than Mercury is to our Sun. Some stars have planets called "super-Earths," rocky worlds much bigger than Earth, though still smaller than Neptune. Other systems have circumbinary planets that orbit not one, but two stars. Some stars have planetary systems consisting of several planets all orbiting interior to where Venus would be. The most common type of planet found so far is not a giant planet—thought to be the end product of inevitable runaway growth during planet formation—but a planet two to three times the size of Earth, or smaller. There are no planets like this in the Solar System and their formation process is not known. The list of different kinds of planets and their implication for planet formation is a long one. Some interesting, but not unusual exoplanet systems are shown in Figure 3-2. Known exoplanets and planet candidates separated by their mass or size and orbital period are shown in Figure 3-3.

Considering the sheer variety of exoplanets, it seems that almost any kind of planet could exist and is probably out there, somewhere, as long it obeys the laws of physics and chemistry. Out of all the planetary systems found so far, none resemble the Solar System, although Solar System copies remain difficult to find with current planet-finding techniques.

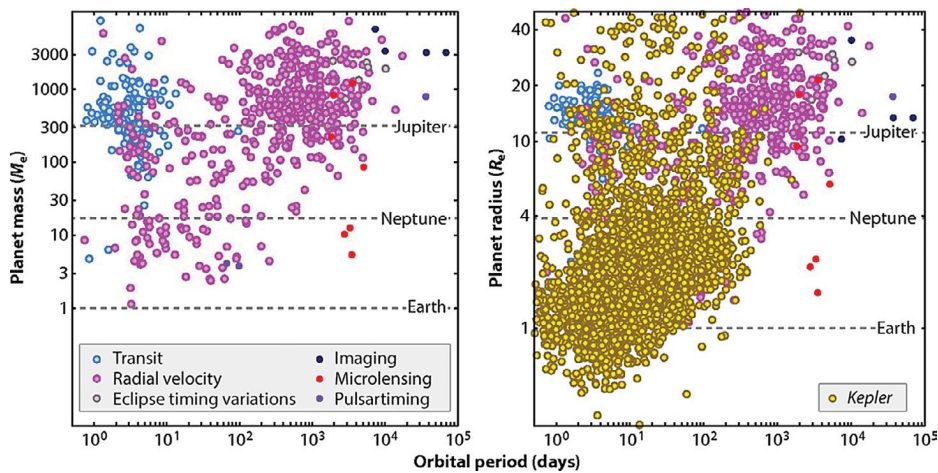

Figure 3-3: Exoplanet discovery space as of 2014. Data points are color-coded according to the planet discovery technique. Plotted as mass vs. orbital period (left) but excluding *Kepler* discoveries. Plotted as radius vs. orbital period (right, using a simplified mass-radius relationship to transform planet mass to radius where needed). A large number of exoplanets and planet candidates are known, but the Earth-size exoplanets in Earth-like orbits still reside in an open part of discovery space. Figure from Batalha (2014).

Thousands of exoplanets are known to exist, including many small planets in their star's habitable zone. The habitable zone is the region around the star where a planet heated by the star is not too hot, not too cold, but just right for life (see Figure 3-4). The planet-detection methods used to date, however, usually measure only the planet's mass and/or size, and cannot yet provide data to infer if a planet is like Earth with oceans, continents, and breathable air—or if instead the planet is like Venus with a thick greenhouse atmosphere making the surface scorching hot and completely inhospitable to life. Indeed, most planet-detection techniques are not sensitive to the presence of an atmosphere at all. While a few planets have had their atmospheres detected with transit spectroscopy, the vast majority have not, and any basic information about their atmospheres is unknown. To distinguish a potential Earth from a potential Venus, the atmospheric conditions must be observed, to con-strain the strength of the atmospheric greenhouse effect, and to search for habitability indicators (water vapor, indicative of liquid-water oceans on a small rocky exoplanet) and even signs of life (such as molecular oxygen). A way to observe exoplanet atmospheres of rocky worlds in the habitable zones of their host stars is needed.

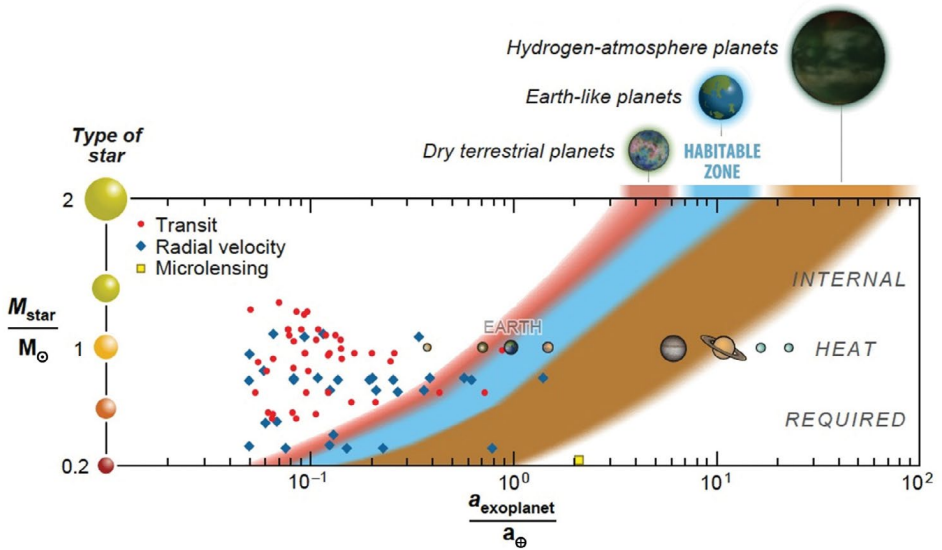

Figure 3-4: Planets in the habitable zone. The Solar System planets are shown with images. Known super-Earths (here planets with a mass or minimum mass less than 10 Earth masses; taken from Rein 2012) are shown as color-coded data points. The light blue region depicts the "conventional" habitable zone for $N_2$-$CO_2$-$H_2O$ atmospheres. The habitable zone could indeed be much wider, depending on the planet's atmospheric properties. The red region shows the habitable zone as extended inward for dry planets, with minimal surface water, a low water-vapor atmospheric abundance, and low atmospheric relative humidity and hence a smaller greenhouse effect. The brown region shows the outer extension of the habitable zone for planets that are massive and cold enough to hold onto molecular hydrogen—a potent greenhouse gas. The habitable zone might even extend out to free-floating planets with no host star. Figure from Seager (2013); see references therein.

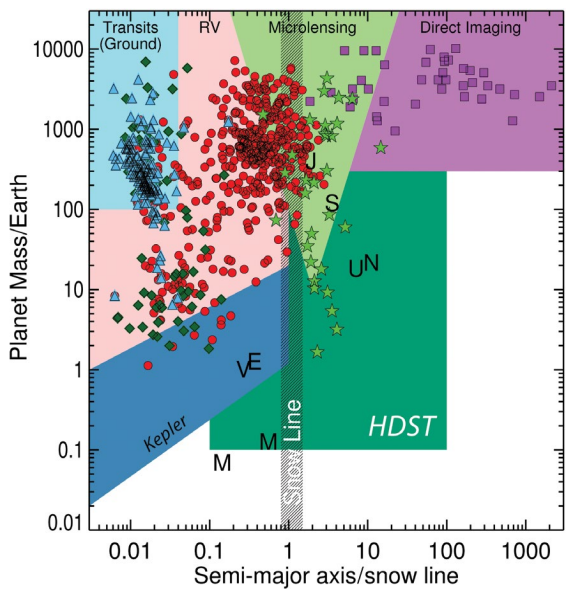

Figure 3-5: The variety of exoplanets as illustrated by their masses and orbital semi-major axes. Different exoplanet-finding techniques' discovery spaces and discovered planets are indicated by colors. Many more exoplanets are known that do not have measured masses (see Figure 3-3). The anticipated parameter space accessible with *HDST* is shown in dark green. Figure adapted from Gaudi and Henderson (private communication) and Wright and Gaudi (2013).

## 3.2 Exoplanet Discovery Space

The last two decades have seen a revolution in exoplanet discovery and characterization: from the first discoveries of exoplanets orbiting Sun-like stars to the present day's census of thousands of known exoplanets that encompass a wide range of physical parameters. Several different planet-finding techniques have unleashed a wave of discoveries. However, each different technique favors finding planets in a particular area of parameter space, biasing samples to specific types of planet masses and/or sizes and planet-star separations (Figure 3-5).

*Radial velocity* surveys are based on a spectroscopic technique for measuring the speed at which stars move toward or away from the observer. Using high-resolution spectrographs, tiny changes in the spectra of stars can be measured through the Doppler effect, itself induced by the planets that orbit those stars. The radial velocity technique inaugurated modern surveys for exoplanets, and is uniquely able to measure planet minimum masses (with an uncertainty linked to inclination angle for non-transiting systems). For transiting planets around nearby bright stars with known orbital inclinations, the radial velocity technique is the natural method for measuring masses.

Radial velocity planet programs continue to search for new planets, pushing to lower and lower-mass planets thanks to ongoing improvements in measurement sensitivity. The future of radial velocity surveys lies in a number of directions: ultra-stable spectrographs on existing telescopes; extensions to the near-IR to search for low-mass planets in the habitable zones of M-dwarf stars; and instrumentation on the next generation of giant ground-based telescopes currently under construction. While these future developments will push measurements to lower masses and longer orbital periods, astrophysical noise sources may provide limits.

*Transit* discoveries first helped confirm the nature of exoplanets already known from radial velocity surveys, but have since become one of the major exoplanet discovery methods. Provided the orbit is aligned with the line of sight, the planet causes a small dimming as it passes in front of its host star, and uniquely enables the size of the planet to be measured if the size of the star is known. Transit observations are possible for the small fraction of exoplanets whose orbits are edge-on to the line of sight to the planet. Even though transits are rare (e.g., occurring for only ~1/10 of close-orbiting exoplanets and ~1/200 for planets in Earth-like orbits), by continuously observing many thousands of stars simultaneously, large numbers of transiting exoplanets have been discovered. The pioneering *Kepler Space Telescope* triggered an avalanche of exoplanet discoveries, finding unexpected new classes of exoplanets and planetary systems and dramatically increasing our statistical understanding of the population as a whole (Figure 3-3). One prominent example is the *Kepler* finding that small planets of two to three times Earth's size are nearly ten times more common than Jupiter-size planets (at eleven times Earth's size) for orbital periods less than around 200 days (Howard 2013; Howard private comm. 2015). Transit-planet surveys have a bright future for planet discovery from both ground-based telescope surveys and space missions (*TESS*, *CHEOPS*, *PLATO*).

*Microlensing* is the only present observation technique that is sensitive to low-mass planets at large orbital separations (~1 to 5 AU). The microlensing discovery technique requires the near-perfect alignment of a telescope, a very distant lens star, and an even more distant source star. The gravity of the lens star magnifies the light of the more distant source star as the two stars move into alignment. The technique works by detecting the gravitational effects of planets orbiting the lens star, which induce subtle perturbations in the brightness of the source star during a microlensing event. Current microlensing discoveries indicate that Neptune-size and smaller planets are more abundant than the giant planets. A large microlensing survey will be undertaken by NASA's *Wide-*

Field Infrared Survey Telescope/Astrophysics Focused Telescope Assets (*WFIRST/AFTA*) observatory, planned for launch early in the next decade. *WFIRST/AFTA* will provide a statistical census of planets in outer orbits (mostly beyond the outer edge of the habitable zone), with sensitivity down to sub-Earth-mass planets. Like the radial velocity and transit discovery techniques, microlensing cannot detect exoplanet atmospheres.

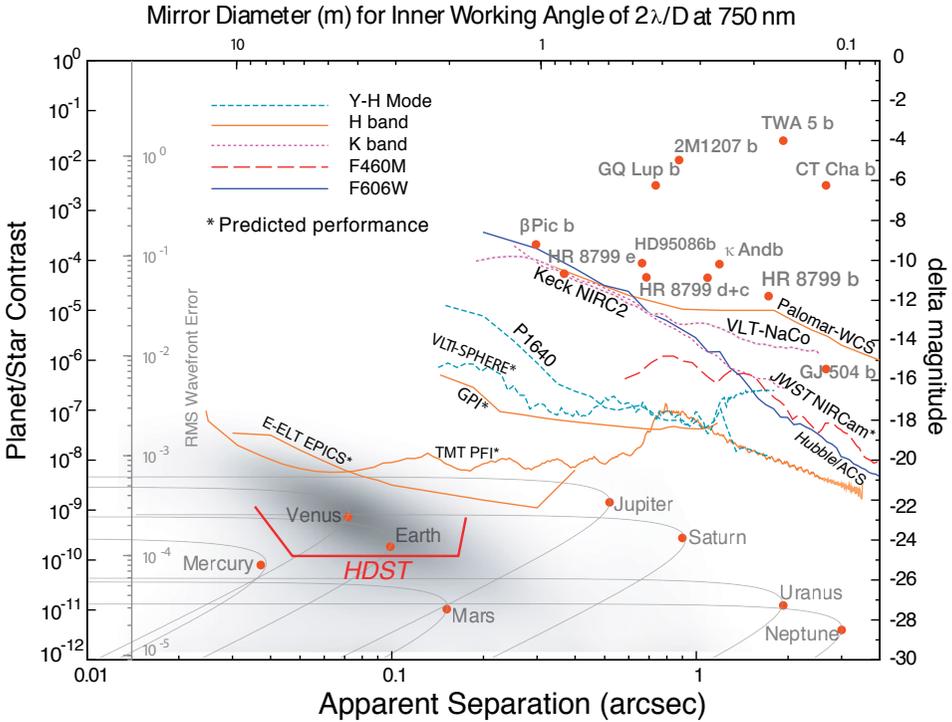

**Figure 3-6:** Direct imaging contrast capabilities of current and future instrumentation. Shown are the 5-σ contrast limits after post-processing one hour's worth of data for various coronagraph instruments. As can be seen in the plot, there are roughly three groupings of curves: 1) state-of-the art instruments in the early 2010s, as represented by Keck near-IRC2, the Palomar Well-Corrected Subaperture, and VLT-NaCo; 2) newly operational state-of-the-art instruments, represented by P1640, GPI, and VLTI-SPHERE; and 3) future extremely large telescopes, represented by TMT PFI and E-ELT EPICS. The contrast curves for *JWST* NIRCam and *Hubble*/ACS are shown for reference. On the top right of the figure are plotted the K-band contrasts of some of the giant exoplanets imaged to date. In the lower part of the figure are plotted our Solar System planets as they would appear in reflected light around a Sun-like star at a distance of 10 pc. The left side of the plot shows the corresponding RMS wavefront error for a coronagraph using a 64 × 64 element deformable mirror. The region above the solid red line would be probed by *HDST*. The gray region in the lower left of the figure shows the predicted locus of terrestrial habitable zone planets for F-G-K (Solar-like) stars. Figure and caption adapted from Lawson et al. (2012), Mawet et al. (2012), and Stapelfeldt (private comm. 2015).

*Direct imaging* of exoplanets—directly observing the planets with a ground- or space-based telescope—differs from radial velocity, transit, and microlensing techniques in a fundamental way. The previous three techniques infer the existence of the planet, and its properties, from its influence on a star, but do not detect the planet's own light emission. (A rare exception is secondary eclipse reflected or thermally emitted light for favorable cases). Direct imaging, by contrast, sees the planet directly, either by reflected light from its host star or by thermally emitted light from the planet. Direct imaging allows both detection of the planets, and spectroscopic characterization of planetary atmospheres. In principle, this technique enables a full determination of the planet's orbital elements through astrometry. Direct imaging is unique in that no other planet-finding technique can accomplish the combination of detection, spectroscopy, and determination of orbital elements.

Direct-imaging instruments must separate the exoplanet light from that of its star, even though the exoplanet is much dimmer than the star. Direct imaging is especially difficult for Earth-like exoplanets (exoEarths) around Sun-like stars for two reasons: the small planet-to-star flux contrast ratios of $10^{-10}$ at visible wavelengths, and the relatively small planet-star separations. Direct-imaging methods must *suppress the starlight* to achieve this extremely high-contrast ratio at the small angular separation between exoplanets and their host stars.

There are two main methods under development for suppressing bright starlight and allowing direct imaging of exoplanets.

*Coronagraphs* (or "internal occulters") are optical instruments that focus the light from a star onto a mask, which acts to cancel or suppress the starlight at the center of the field. The light beam—minus the starlight—is then passed through other optics to a detector, where the planet light is imaged. Coronagraphs can be used with both ground- and space-based telescopes.

*A starshade* (or "external occulter") blocks the light from a star by casting a shadow on a space telescope. A starshade is a large, specially shaped screen (tens of meters in diameter) hosted by its own spacecraft flying in formation at a distance of many tens of thousands of km from the space telescope. The starshade creates a conical shadow, blocking the starlight and enabling only the planet light to enter the telescope. A starshade is very difficult to use with a ground-based telescope due to challenging alignment constraints and the effects of the Earth's atmosphere. For a summary estimate of direct imaging telescope capabilities and discovered planets, see Figure 3-6. To date, planets detected by direct imaging are brighter than Jupiter (because they are hotter, i.e., younger and/or more massive than Jupiter), and are at large separations from their host stars.

*Direct imaging from the ground.* Thanks to recent progress in adaptive optics and starlight-suppression techniques, ground-based instruments (such as the Gemini Planet Imager [GPI], ESO's SPHERE, Subaru Coronagraphic Extreme-AO, Palomar's P1640, the Large Binocular Telescope) can capture the thermal emission from young giant planets (> 1 Jupiter mass) in near-IR wavelength light. Large surveys (most notably GPI and SPHERE) will provide a near-complete census of young massive planets around nearby young stars at large (> 5 AU) separations, and spectrally characterize the atmospheres of many giant planets. The next generation of ~30 m diameter Extremely Large Telescopes (TMT; GMT; E-ELT, hereafter ELTs) will considerably increase both the number of observable giant planets and the ability to characterize their atmospheres. These larger telescopes will also provide the angular resolution to image and characterize giant planets in the habitable zones of nearby young stars.

Direct imaging of habitable planets from the ground is very much more difficult than imaging of giant planets, however, for two reasons. First, habitable planets are small and relatively cool, by virtue of being habitable, and therefore lack the strong thermal emission in the near-IR that makes giant planets (which have some of their own internal heat) detectable. Second, habitable planets have a much smaller angular separation from their host stars than current ground-based telescopes can resolve, even with state-of-the-art adaptive optics and starlight-suppression techniques.

Direct imaging of habitable planets from the ground is out of reach for current large ground-based telescopes. It will be possible with future ELTs, but only for a very particular subregion of parameter space: exoEarths orbiting nearby low-mass stars observed at near-IR wavelengths. M-dwarf stars are relatively bright at near-IR wavelengths as compared to visible wavelengths, and hence their planets are also relatively bright in reflected near-IR light. So while direct imaging of Earth analogs around Sun-like stars requires a detection contrast performance (~$10^{-10}$) that cannot be obtained from the ground, habitable planets around late M-dwarf stars are detectable at ~$10^{-7}$ contrast in reflected light at near-IR wavelengths (note that habitable planets around M5 dwarf stars would have a contrast around $10^{-8}$). The habitable zones for nearby late M-dwarf stars subtend very small angles (< 40 mas) and, in the near-IR, will require a ~30 m-diameter telescope. A dozen or more late M-dwarf stars are accessible, and if higher contrast can be reached, dozens of earlier-type M stars may also be searched. Thus detections of planets orbiting nearby M-dwarf stars in their habitable zone are likely to advance in the era of the ELTs starting in the early 2020s.

Regarding exoEarths in habitable-zone orbits about Sun-like stars, ELTs are not expected to be able to reach the performance required for reflected light detection for systems at ~10 pc. ELTs, however, could potentially image a rocky planet around one of the closest G and K stars such as α Centauri, ε Eridani, τ Ceti (Kasper et al. 2010), although this will remain very challenging.

*Direct imaging from space* can potentially reach very high-contrast levels, far exceeding what can be done from the ground, owing to the lack of atmospheric turbulence. For example, while *Hubble* (with a 2.4 m mirror) was not designed for high-contrast imaging, its ability to image exoplanets is comparable to that achieved with much larger (8 m class) ground-based telescopes. NASA is currently planning to include a coronagraphic instrument on the *WFIRST/AFTA* 2.4 m telescope. The instrument is expected to image at least 16 giant planets in reflected light, planets already known from radial velocity surveys, and to be able to spectrally characterize up to about half of these. The *WFIRST/AFTA* coronagraph is also predicted to be capable of discovering a dozen new planets, including a very small number down to Earth size. The full program will be accomplished in one year of coronagraph observing time (*WFIRST/AFTA* STD2015 Report, Table F-5). NASA has also studied "Probe-class" small-scale (< 2 m telescope) missions dedicated to exoplanet studies (using either a coronagraph or a starshade) with improved science yield. The Probe-class missions aim to survey the nearest, brightest Sun-like stars for exoEarths. As such, the Probe-class missions are valuable scientific and technological pathfinders (as discussed in Chapter 5), however, the limited sensitivity of these small missions confine them to a small exoEarth yield. (See the *Exo-C* and *Exo-S* STDT 2015 Final Reports). Searching hundreds of stars for the aim of detecting and characterizing dozens of potentially habitable planets requires a much larger aperture, optical space telescope.

## 3.3 Characterizing Exoplanet Atmospheres: Current Status and Future Potential

Characterizing exoplanet atmospheres is an essential step in understanding the demographics of exoplanets, in obtaining constraints on their formation, and in detecting potential signs of life. In a technical sense, atmospheric characterization consists of identifying the molecular gas species and any possible clouds in the atmosphere, determining the gas species abundance, and constraining the planetary surface temperature and pressure. From this information, one can infer the atmospheric composition.

### 3.3.1 Motivation to Study Exoplanet Atmospheres

A straightforward example points to the importance of atmospheric characterization: to all other methods of exoplanet study, our comfortable Earth and fiery Venus look identical, but in a spectrum obtained at visible wavelengths, Venus is nearly featureless and completely covered in a photochemically induced haze while Earth has a blue color from Rayleigh scattering, and strong absorption features from water vapor and oxygen. Without the ability to spectroscopically characterize a sample of exoplanets, astronomers will never know how many planets are Earth-like and habitable, or Venus-like and unlikely to be habitable. Without spectroscopic characterization, astronomers would be unable to search for correlations (or lack of) among atmospheric characteristics for planets of different types ("comparative planetology") in the hope to discover clues about planetary formation and evolution.

A major motivation in studying terrestrial planet atmospheres is the search for biosignature gases. Life as we know it produces gases as byproducts from metabolism, and some of these gases will accumulate in a habitable planet's atmosphere and can, in principle, be detected by spectroscopy. Only spectroscopic characterization can provide compelling evidence of life on other worlds via the detection of biosignature gas molecules.

### 3.3.2 Exoplanet Atmosphere Techniques Overview and Potential

Exoplanet atmosphere studies have gone from birth to maturity in the last two decades. The first observations were breakthroughs, finding novel ways to use telescopes and instrumentation not originally designed to measure the tiny signals from exoplanet atmospheres. As removal of systematic noise sources improved and as the community honed observational techniques, many more bright planets suitable for atmosphere observations were discovered. Very basic measurements of dozens of exoplanet atmospheres have been made, yielding a triumphant collection of exoplanet spectra for a handful of hot giant or sub-Neptune-sized planets.

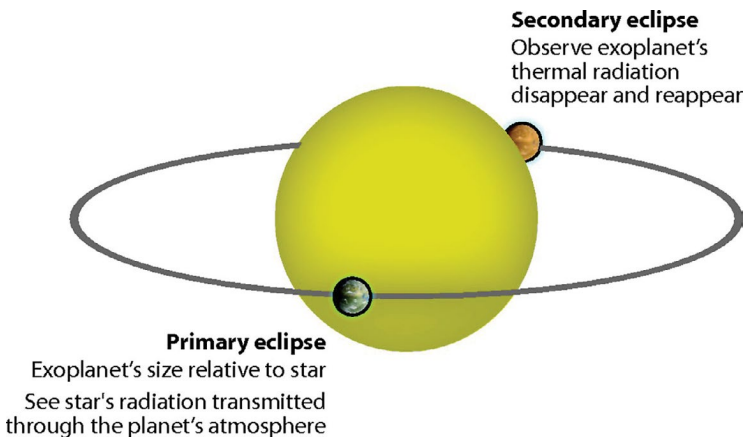

Figure 3-7: Schematic of a transiting exoplanet. Figure credit: D. Beckner.

To place the merits of direct imaging and spectroscopy into context, a brief review of the three primary techniques used today to characterize exoplanet atmospheres is presented.

*Atmospheric characterization techniques for transiting exoplanets.*

The currently most productive set of indirect techniques for observing exoplanet atmospheres are those for systems where the planet transits and is eclipsed by its host star. These techniques do not require high spatial resolution, as the planet and star do not need to be spatially separated on the sky. Observations are made in the combined light of the planet and star (Figure 3-7) with the exoplanet spectrum separated out by comparison with observations of the star alone. As seen from a telescope, when the planet goes in front of the star, the starlight passes through the planet atmosphere and planet atmosphere spectral features are imprinted on the stellar spectrum. This is called transmission photometry or spectroscopy. When the planet goes behind the star, the planet disappears and reappears, adding either reflected light or thermal emission to the combined planet-star radiation. This is referred to as secondary eclipse photometry or spectroscopy.

A half-dozen exoplanets have had their transit transmission spectra observed in detail. Transmission spectra have suggested the presence of high-temperature clouds on hot exoplanets, and they have constrained the water vapor abundance on hot giant and Neptune-sized exoplanets. Dozens of hot exoplanets with high thermal emission have been observed with near-IR and IR photometry in different colors from secondary eclipse spectra. Measurements have constrained the planet day-side temperatures. Spectral measurements to produce reflected light and thermal phase curves provide information about exoplanet albedo and thermal features without requiring transiting geometry, but for now this technique is limited to planets close to their stars that are either bright in reflected light or are thermally hot.

Transit-based atmospheric characterization holds enormous promise, as astronomers are working hard to uncover a sample of rocky planets transiting small stars that are optimized for follow-up with this technique. For example, *JWST* is well-suited for performing transit observations of exoplanets orbiting a subset of small nearby (i.e., bright) non-variable stars. This sample will include a handful of small planets in the habitable zones of small (i.e., M-dwarf) stars. Yet even in ideal circumstances, transit spectroscopy can characterize only the subset of planets that happen to lie in orbits that provide the proper viewing geometry, leaving a large fraction of planetary systems out of reach. It is worth mentioning that atmospheres of exoEarths transiting Sun-sized stars are just not thick enough to give a measureable transmission spectrum signal.

*Doppler-shifted atmospheres for high orbital velocity exoplanets.*

A second indirect technique—very high spectral resolution (R > 100,000) cross-correlation template matching—has identified CO and $H_2O$ on exoplanets orbiting fast enough (~100 km/s) that their atmospheres are Doppler-shifted (Snellen et al. 2010). Detection of molecules in this way is considered robust because of the unique pattern of the many rotational lines within the molecules' vibrational bands. Even winds can be observed if the atmospheric Doppler shift differs from the Doppler shift due to the planet's orbital motion. Studies have postulated the ELTs of the future might be able to use this technique on a small handful of suitable targets to search for signs of habitability and biosignature gases in atmospheres of radial-velocity discovered rocky planets orbiting M stars.

*Direct imaging for exoplanet atmosphere studies.*

Any directly imaged planet, by definition well separated from its host star, can also be observed spectroscopically given the right instrumentation and enough detected photons. Indeed, once the starlight has been suppressed to the appropriate level, conventional spectroscopic dispersion can be used to obtain a spectrum. Unlike transit imaging, direct imaging can be used to observe planets in a wide range of viewing geometries and, unlike the template-matching Doppler technique, direct imaging is not restricted to planets with special properties such as high orbital velocities. In principle, all planets that orbit outside the minimum angular resolution of the telescope and starlight-suppression scheme can have their atmospheres characterized. In the last few years, the first direct spectra have been observed in near-IR bands from the ground for hot giant planets typically at tens of AU from their host star.

Spectroscopic characterization of habitable-zone exoEarths will be exceptionally challenging for ground-based telescopes, even considering the upcoming ELT facilities. The challenges are due to residual starlight (from starlight suppression) induced by atmospheric turbulence and to Earth's atmospheric absorption, which blocks strong water-absorption features. Ground-based spectroscopy of habitable planets will therefore require significant advances in high-contrast imaging calibration techniques. The number of host stars around which exoEarths can be found by ground-based direct imaging is limited (to dozens or fewer M stars and only a few Sun-like stars, as emphasized in Section 3.2), which further limits the number of exoplanets for which spectroscopy can even be attempted.

The long-term future direction for direct imaging is to detect and characterize habitable planets, to go beyond planets orbiting M-dwarf stars to focus on Solar-type stars. Therefore, although space-based and ground-based facilities will offer complementary capabilities, *only HDST, a space-based, large-aperture telescope, will be able to search hundreds of stars to discover dozens of Earth-like planets orbiting Sun-like stars, and to characterize their atmospheres.*

### 3.3.3 Goals for Characterization of Rocky Exoplanets

The overarching goal for characterization of small exoplanets is to develop a comparative planetology that answers compelling questions. Do planets similar to Earth exist? Are they common? Do any have signs of life? We would most like to know which rocky planets are made habitable by the presence of liquid surface water, and if any show the presence of biologically significant gas molecules.

Rocky planets with thin atmospheres are the focus for the search for life—not just because they resemble our own Earth, but also because atmospheric biosignature gases will be detectable. Giant planets have no solid surfaces and are too hot for life beneath their atmospheres, due to trapped residual thermal radiation dissipating from the planet core.

A suitable starting point to understand what an exoEarth might look like is viewing Earth itself as an exoplanet. The spectrum of Earth's spatially unresolved, globally averaged atmosphere in reflected light can be measured by observing Earthshine reflected off the Moon and from observations with interplanetary spacecraft that look back at the unresolved Earth. Earth's spectroscopically visible, globally mixed gases in reflected light are shown in Figure 3-8. Water vapor, which shouldn't occur naturally in a small rocky planet's atmosphere, is indicative of surface liquid water. Carbon dioxide ($CO_2$) is indicative of a terrestrial exoplanet, since it is the dominant form of carbon in an oxidized atmosphere. Oxygen ($O_2$, and its photochemical byproduct ozone [$O_3$]) is the most robust biosignature gas because it is so reactive that it must be continuously replenished, most likely by life. The possibility of false positives can be assessed by considering the atmospheric context, that is, abundances of other gases in the atmosphere.

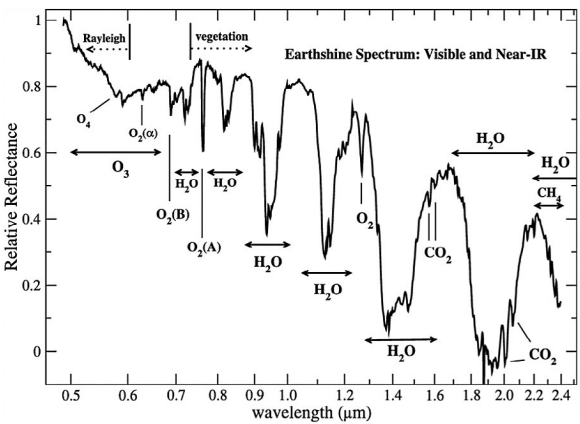

Figure 3-8: Earth's observed reflectance spectrum at visible and near-IR wavelengths. From Earthshine measurements (Turnbull et al. 2006), the spectrum is disk-integrated, i.e., spatially unresolved and shows what Earth would look like to a distant observer. The reflectance is normalized to one; the relatively high continuum and the presence of water vapor, oxygen, and ozone at visible wavelengths are relevant.

The diversity of exoplanets in size and orbit (Section 3.1) is expected to extend to exoplanet atmospheres, which may vary widely depending on the greenhouse properties of the atmospheric components. Planetary theorists are actively working to predict the ranges of potential atmospheres and their spectroscopically observable gases that might exist, and the optimal ways in which to categorize them. Figure 3-9 shows some of these model spectra. There might be a limited number of broad atmospheric scenarios because photochemistry, chemistry, and allowable temperatures are finite in scope for planets over a small range in size orbiting in their habitable zones. Or, small exoplanets may show an extensive diversity. The habitable zone itself depends somewhat on planetary properties: it may extend inwards for a dry exoplanet with minimal surface water, a low water vapor atmospheric abundance, and low atmospheric relative humidity. The habitable zone may extend outwards for planets that are massive and cold enough to hold onto hydrogen. Molecular hydrogen is a potent greenhouse gas owing to its continuous opacity induced by collisions at high pressures (Figure 3-4). An awareness that planets can be habitable, but different from Earth, or can lie outside of the traditional habitable-zone boundaries, should inform observational strategies and will increase our chances of finding habitable worlds.

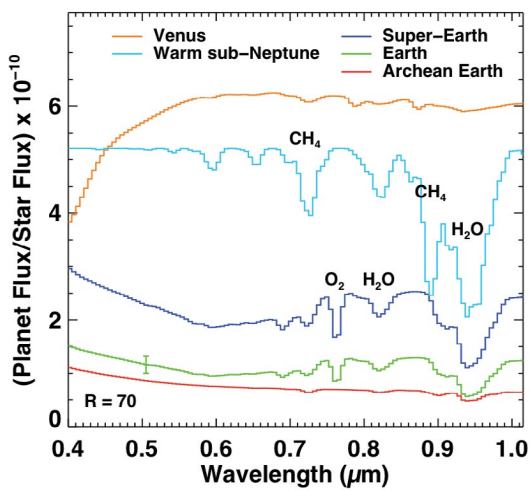

Figure 3-9: Simulated spectra of small planets orbiting a Sun-like star. The Earth, Venus, Archean Earth, and super-Earth models are from the Virtual Planet Laboratory. The sub-Neptune model is from R. Hu (personal comm.). The Earth, super-Earth, and Archean Earth are at 1 AU. Venus is at its 0.75 AU. The sub Neptune is at 2 AU. The spectra have been convolved to R = 70 spectral resolution and re-binned onto a wavelength grid with 11 nanometer bins. Figure courtesy of A. Roberge.

Once evidence that a planet is potentially habitable (with liquid water oceans and thin atmospheres) is in hand, the next challenge will be to search for the presence of biosignature gases. On Earth, plants and photosynthetic bacteria produce $O_2$; without them, Earth would have nearly 10 orders of magnitude lower $O_2$. While some abiotic production of $O_2$ has been hypothesized, studying the atmospheric context (the measured abundances of other gases) should help discriminate between geophysical and biological origins. Earth has other notable biosignature gases—methane ($CH_4$) and nitrous oxide ($N_2O$)—although these and other gases also have abiotic production routes. Earth also has dozens to hundreds of gases produced by life in quantities too small to be detectable remotely, but some of which nevertheless may rise to detectable levels on exoEarths with environments different from Earth's. Studied examples are dimethyl sulfide and methyl chloride. While the vegetation "red edge" and other spectral features due to pigments in vegetation or bacteria have been studied as biosignatures in reflected light, their signals are weak (partly due to limited surface coverage) and diminished by clouds, and are therefore generally not considered promising biosignatures.

Even with superb data, there is no single "smoking gun" biosignature gas. There are false positive scenarios, where the ambiguity between a biotic and an abiotic origin for biosignature gases simply cannot be resolved. This means that aiming for a robust detection of biosignature gases on a single planet may not be enough. To establish the presence of life beyond the Solar System, a large number of candidate exoEarths with detected biosignature gases will be needed.

### 3.4 The Motivation to Discover Dozens of ExoEarths

The main question for an exoEarth survey telescope is how large a sample of exoEarths is large enough? The motivation to discover dozens of exoEarths is to both understand the diversity of terrestrial planets and to increase the chance that biosignature gases will be detected.

Exoplanets are incredibly diverse and exist for nearly every mass, size, and orbit that is physically plausible. The discovery of dozens of terrestrial planets will tell us not only if this diversity extends to terrestrial-type planets, as anticipated, but also what kind of terrestrial planets are out there. Exoplanet atmospheres may be quite diverse even over the narrow range of small rocky worlds. This potential diversity— and the different warming potential of atmospheres of different masses and compositions—creates a compelling motivation to locate dozens of exoEarths to understand what fraction of terrestrial planets may actually have habitable conditions (where, recall, habitable refers to a surface temperature suitable for liquid water).

A large pool of exoEarths will increase the chances of finding a single exoplanet with an atmosphere containing a readily identified biosignature gas. Even one such planet with a suspected atmospheric biosignature gas would be a huge scientific advance and generate an enormous amount of astrobiological followup study.

The possibility of abiotic production of biosignature gases for most cases, however, means that the inference life at the 100% certainty level is not achievable for an individual planet with an identified atmospheric biosignature gas. The search for evidence of life beyond Earth therefore motivates a large sample of exoEarths for a second reason. That is the assessment of the incidence of life by considering individual planets with suspected biosignature gas detections in the context of planets with nondetections, and in conjunction with the planet's orbit, incidence of stellar energy, and other spectral properties.

An additional key motivation for aiming at a sample of dozens of exoEarths is to ensure against a low rate of incidence relative to present estimates, which may be biased by presently unknown systematic errors (see Section 3.5.1). Even if exoEarths are far more rare than anticipated (say, ten times less common than present limits), a large survey will still provide a useful null result to define the challenge faced by the next generation of astronomers in the search for true, habitable exoEarths.

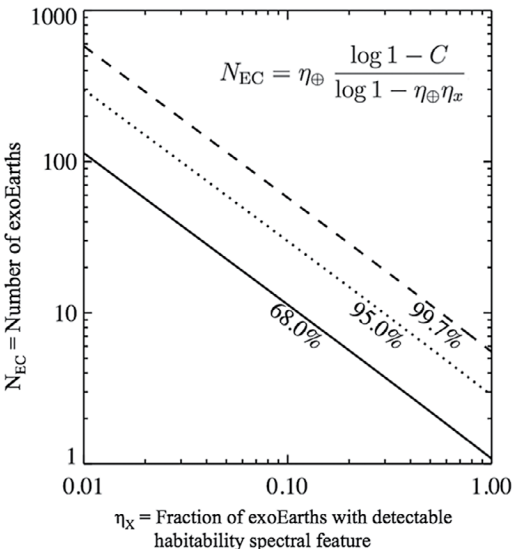

Figure 3-10: The exoEarth sample size, $N_{EC}$, required to ensure detection of at least one exoEarth with evidence for habitability (or even biosignature gases) as a function of the fraction of exoEarths with such detectable features. These results are computed using the binomial theorem (see equation in figure) where $C$ is the probability of detecting at least one life-bearing exoEarth in the sample, $\eta_\oplus$ is the eta Earth value (taken to be 0.1 for this computation), and $\eta_x$ is the fraction of exoEarths with detectable biosignature gases. Shown are the lines for three commonly used probabilities: 68%, 95%, and 99.7%. Figure courtesy of C. Stark.

$$N_{EC} = \eta_\oplus \frac{\log 1 - C}{\log 1 - \eta_\oplus \eta_x}$$

Figure 3-10 shows the basic statistical argument for a large sample of exoEarths. The plot shows results from using the binomial theorem to calculate the number of exoEarth planets that must be searched to have a given probability of detecting at least one that has a detectable habitability or biosignature gas feature in its spectrum. For example, if only 10% of exoEarths have detectable biosignature gases, then one needs a sample of ~30 exoEarths to have a 95% probability of detecting at least one inhabited world. If, in that same sample, no exoEarth atmospheres with biosignature gases are detected then one has determined, at a 95% confidence limit, that the fraction of exoEarths with detectable biosignature gases is less than 10%. In contrast, if exoEarths are more common than anticipated, surveys that precede *HDST* (see Section 3.2) will find a small number of exoEarths and *HDST* itself will reach a larger-than-anticipated sample, enabling an even more definitive result concerning life beyond Earth.

The *HDST* report recommendation is to seek dozens of exoEarths for detailed atmospheric characterization. The motivation is threefold: (i) comparative planetology for many terrestrial exoplanets; (ii) discover signs of life by way of biosignature gases; and (iii) infer the presence of life by detections or nondetections in context of a large sample of terrestrial worlds.

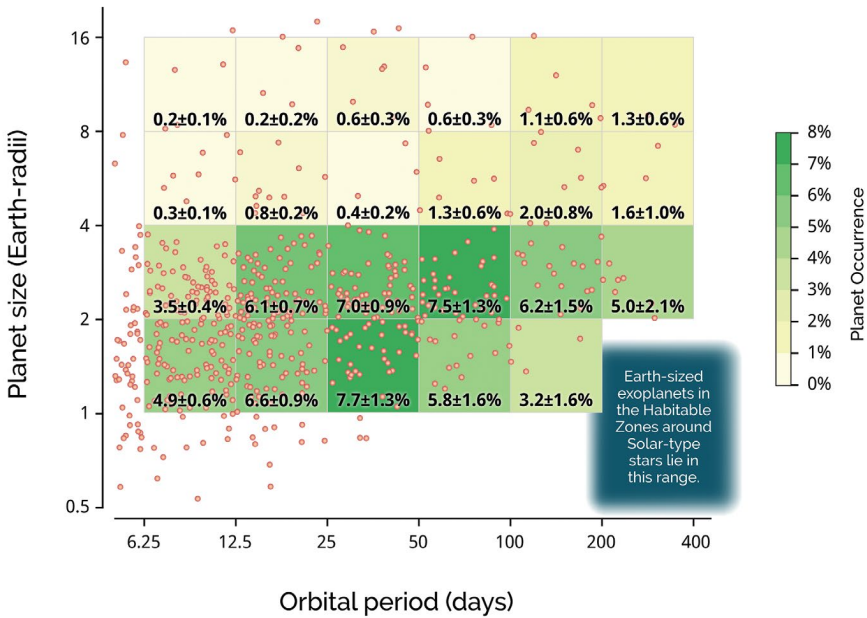

**Figure 3-11:** Planet occurrence as a function of planet size and planet orbital period. *Kepler*-detected small planets are shown by red circles. While the detected planets show the transit discovery technique's bias towards larger, shorter-period planets, the occurrence rates provided in each bin are corrected for biases. Figure adapted from Petigura et al. (2013).

## 3.5 ExoEarth Yield and the Design of Future ExoEarth Surveys

This section describes the number of exoplanets that can be discovered by a survey based on a fixed set of input assumptions, where the assumptions are related to the telescope aperture and other requirements. "ExoEarth yield" calculations connect telescope aperture, starlight-suppression performance, and other instrument properties to the number of Earth-like worlds that may be discovered and characterized. Here the focus is on visible-wavelength photometric discovery and follow-up spectroscopy using space-based direct imaging of exoEarths shining in reflected light and orbiting nearby bright main-sequence stars. This is the sample of stars and planets that *HDST* will specialize in, and is least accessible to ground-based measurements (even considering their future development).

We show that a telescope with *HDST*'s large aperture and strong starlight suppression down to small angular resolutions has the capabilities needed for discovering dozens of exoEarths.

### 3.5.1 Astrophysical Inputs to Planet Yield

While there are many possible astrophysical factors that will influence the number of habitable planets that could be discovered in a given survey, studies have found that two factors in particular dominate the eventual yield. Out of these several astrophysical inputs, the two most salient ones—the occurrence rate of Earth-size exoplanets and the brightness of exozodiacal dust—are discussed here.

*The Occurrence Rate of Earth-Size Exoplanets:* A main input to calculating the exoEarth yield is the occurrence rate, per star, of Earth-size exoplanets in the habitable zone, denoted $\eta_{Earth}$. For Sun-like and larger stars, $\eta_{Earth}$ is not measured, but is extrapolated from existing data. Specifically, the *Kepler Space Telescope* has not yet found any planets of 1 Earth radius in the habitable zone of Sun-like stars, but has found many planets larger than Earth in their host star's habitable zone, as well as planets of Earth size, but interior to the conventional inner edge of the habitable zone (Figure 3-11). This is an observational bias: both types of planets are easier for *Kepler* to detect than Earth-size planets in Earth-like orbits about Sun-sized stars.

A number of studies quantitatively assessed the completeness of the *Kepler* survey based on the photometric precision acquired per star (for tens of thousands of stars; see Table 3-1 and references therein). These studies considered sensitivity to Earth-size planets per star, to carefully derive $\eta_{Earth}$ for orbital periods where *Kepler* Earth-size planets are found. Habitable-zone Earth-size planets will have longer periods than currently constrained by *Kepler*, and thus extrapolations to longer orbital periods must be based on assumptions about the change in planet occurrence rate as a function of period (e.g., flat in log period or modeled assuming a smooth functional form). These studies produce indirect, somewhat model dependent, estimates of $\eta_{Earth}$ for *Kepler*'s parent population of stars.

The studies report values of $\eta_{Earth}$ ranging from 0.02 to 0.2 with large uncertainty and with some sensitivity on the details of the habitable-zone definition (i.e., orbital period or semi-major axis range) and the range of planet radii adopted as Earth-like (see Table 3-1). These reported $\eta_{Earth}$ values typically have large uncertainties about half as large as the value, owing to uncertainties in the extrapolation to the ExoEarth regime. All of these numbers may be revised as the *Kepler* data pipeline is improved, differences amongst pipelines are understood, and habitable-zone boundaries are agreed upon. There is no planned mission, however, that will directly measure $\eta_{Earth}$ for Sun-like and larger stars. In later sections, $\eta_{Earth}$ = 0.05–0.2 is assumed for Sun-like stars in estimates of total yield from different types of observatories.

The occurrence rate of Earth-sized planets in the habitable zones of M-dwarf stars is less relevant to a future space-based telescope, and more relevant to planet characterization methods that focus on M-dwarf stars, particularly from the ground (see Section 3.3). Reported numbers for M dwarfs based on the *Kepler* data set range from about 0.15 to 0.5 with large uncertainties. The value is $0.16^{+0.15}_{-0.07}$ for Earth-sized planets (defined as 1–1.5 $R_{Earth}$) and a habitable zone definition scaled to M stars (and equivalent of 0.99 to 1.7 AU for the Sun) and increases to $0.24^{+0.13}_{-0.03}$ for a broader habitable zone encompassing Venus and Mars (Dressing & Charbonneau 2015). M-dwarf $\eta_{Earth}$ estimates derived from radial velocity measurements are at the high end of this range, but broadly consistent with the values derived from *Kepler* data.

| $\eta_{Earth}$ (%) | $a_{inner}$ (AU) | $a_{outer}$ (AU) | Size ($R_{Earth}$) | HZ Definition | Reference |
|---|---|---|---|---|---|
| 22 | 0.5 | 2 | 1–2 | Simple | Petigura et al. (2013) |
| 5.8 | 0.95 | 1.37 | 1–2 | Kasting (1993) | Petigura et al. (2013) |
| 8.6 | 0.99 | 1.70 | 1–2 | Kopparapu et al. (2013) | Petigura et al. (2013) |
| $6.4^{+3.4}_{-1.1}$ | 0.99 | 1.70 | 1–2 | Kopparapu et al. (2013) | Silburt et al. (2014) |
| 14 ±5 | 0.7 | 1.5 | 0.66–1.5 | Brown (2005) | C. Stark scaled from Petigura et al. (2013) |
| 16 ±6 | 0.75 | 1.77 | 0.66–1.5 | Kopparapu et al. (2013) | C. Stark scaled from Petigura et al. (2013) |
| 10 ±4 | 0.99 | 1.67 | 0.66–1.5 | Kopparapu et al. (2013) | C. Stark scaled from Petigura et al. (2013) |
| $1.9^{+1}_{-0.8}$ | 0.67 (200 d) | 1.06 (400 d) | 1–2 | Simple | Foreman-Mackey et al. (2014) |

Table 3-1: Values of $\eta_{Earth}$ for Sun-like stars under different HZ and Earth-size radius assumptions. Note that functional form extrapolations for planet occurrence rates as a function of orbital period may also differ.

*Exozodiacal dust:* Starlight reflected by small dust particles is expected to be the main astrophysical noise source for space-based exoplanet direct imaging at visible wavelengths. The light reflected by the dust is referred to as exozodiacal light or exozodi for short. This dust is generated from evaporation of comet-like bodies and/or collisions among small bodies in an asteroid-type belt. The unit of "zodi" is assigned to be the dust-cloud brightness that is comparable to the estimated brightness of the Solar System's zodiacal cloud. With this definition, an exozodi value of 3 means that, assuming dust properties to be the same as in the Solar System, *HDST* would observe 3× the background of the Solar System zodi cloud (where the observational line of sight goes through 1/2 of the Solar System zodi *and* the entire thickness of the exozodi cloud. In other words, 1 exozodi is equivalent to 2× the average local zodi brightness (2 × 23 mag per square arcsec).

Although it is not known how much exozodiacal dust exists in exoplanetary systems, ongoing observational programs (e.g., the LBTI HOSTS survey, targeting several tens of stars) will be able to assess whether high levels of exozodi (10 to 100 zodi or higher) are common. Observational constraints on the median exozodi level should continue to be improved. In the exoEarth yield calculations, values of 3 to 100 zodi are used.

### 3.5.2 ExoEarth Yield Model

ExoEarths orbiting nearby Sun-like stars are faint, fainter than 30th magnitude. This is as dim, or dimmer, than the faintest galaxies ever observed by the *Hubble Space Telescope*. Yet exoplanet direct imaging is far more challenging than photometry of isolated objects at this magnitude because an exoEarth would be about a tenth of an arcsecond away from a host star that is about 10 billion times brighter. Achieving starlight suppression and angular resolution at these limits is so very challenging that a space telescope cannot detect a planet at all phases of its orbit.

The two observational limitations—photometric and obscurational bias—are illustrated in Figure 3-12. First, the planet goes through reflected light illumination phases that reduce its total brightness with respect to its "full" phase ("photometric incompleteness"). Second, as the planet orbits the star, the projected planet-star separation on the sky is changing, depending on the planet's orbital inclination. Often the planet is too close to the star as projected on the sky to be spatially resolved from the star by telescope. In other words, the star obscures the planet ("obscurational completeness"). Calculating the expected exoEarth yield requires consideration of both photometric and obscurational completeness.

The method for calculating exoplanet yield simulates an optimized and strategic observing plan to detect as many exoEarths as possible, given a finite total search time (Stark et al. 2014; see Figure 3-13). Rather than optimizing for the number of *stars* exhaustively observed, the observational strategy optimizes for exoEarth yield by considering a prioritized target list based on the likelihood of finding an exoEarth for each target.

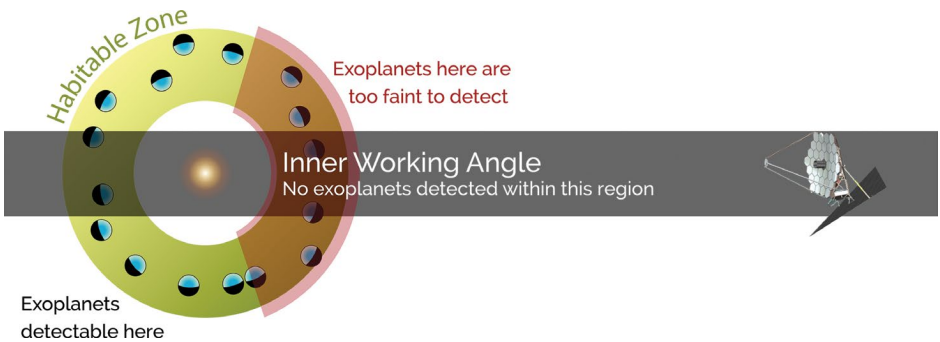

Figure 3-12: Illustration of the origins of obscurational and photometric incompleteness in direct imaging surveys. Figure courtesy of C. Stark.

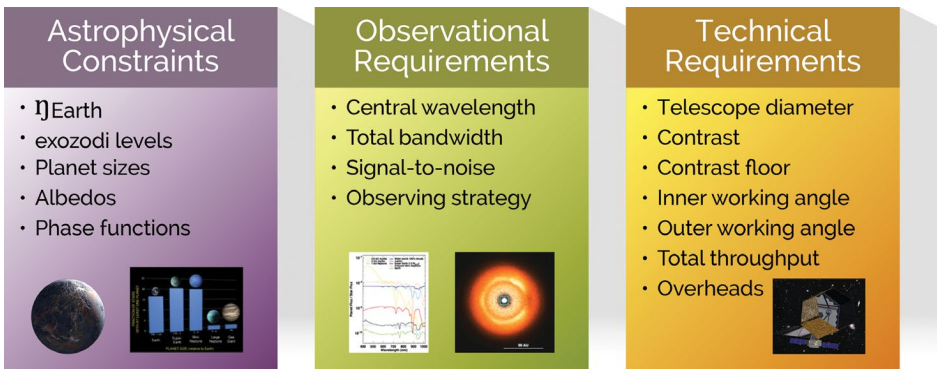

Figure 3-13: Schematic description of the input ingredients to the exoEarth yield calculations. Figure courtesy of C. Stark.

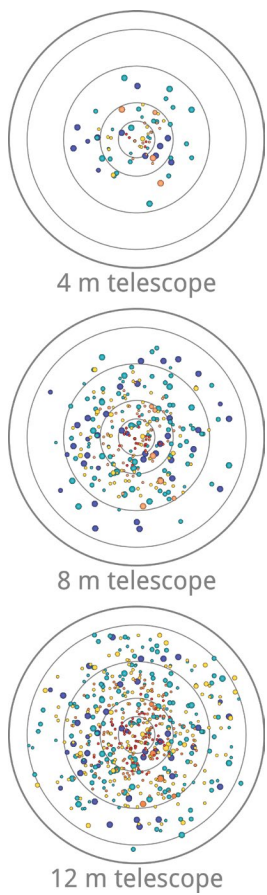

Figure 3-14: Star positions and properties for an exoplanet survey considering visible wavelengths and an internal coronagraph. Telescope and coronagraph values are consistent with those described in the text for the exoEarth yield simulations. Plotted are: stars surveyed to optimize exoEarth yield for a 4 m aperture (74 stars, top); 8 m aperture (291 stars, middle); and 12 m aperture (582 stars, bottom) telescope given 1 total year of integration time (with 100% overheads) and including spectral characterization. Stars are plotted on a projection of a sphere with radius 35 parsecs. Star colors correspond to their *B–V* color and star size corresponds to the stellar luminosity with respect to Solar.

In the simulations, the number of repeat observations of each star (i.e., visits) is optimized to maximize the exoEarth yield.

To yield dozens of Earths, the number of stars observed at least once must be on the order of hundreds (as shown in Figure 3-14), depending on the observing strategy. Assumptions for the results shown here assume a total of one year of on-target exposure time, which may translate into two total years of total observatory time (an estimate, as there is no detailed observatory efficiency model included here). The exoplanet search time is envisioned as being spread throughout the mission lifetime and interleaved with astrophysics observing programs (such as those described in Chapter 4).

### 3.5.3 ExoEarth Yield Results

The computed exoEarth yield for a large space telescope is based on a given set of assumptions about the true underlying distribution of planet properties, a variable set of telescope and instrument parameters, and the observational bias effects described above in Section 3.5.2. The calculations summarized here (Stark et al. 2014; C. Stark private comm.) adopt $\eta_{Earth}$ of varying values (5% to 20%), assume a habitable zone in the range of 0.75–1.77 AU (scaled by the stellar luminosity $(L_*/L_{sun})^{1/2}$), and assume planets are Earth-sized. The Earth's planetary geometric albedo of 0.2 is used, because the exoEarth yield is concerned with planets with Earth-like properties. An exozodi brightness of 3× the Solar System value is taken as nominal with values up to 100 also used.

The total ExoEarth yield is shown in Figure 3-15. Based on the calculations described above, for reasonable assumptions about $\eta_{Earth}$, exozodi levels, and technical performance, the desired yield of dozens of exoEarths corresponds to a telescope diameter of 9 to 12 m, or even larger. The upper end of this range would provide resilience against the possibility that one of the key input assumptions was overly optimistic. The results are shown graphically for varying assumptions of the driving astrophysical inputs, $\eta_{Earth}$ and exozodi levels. The occurrence rate of exoEarths, $\eta_{Earth}$, is estimated by extrapolation from the *Kepler* results, from larger planets or planets closer to the star (see Section 3.5.2).

The sensitivity of the exoEarth yield varies with specific telescope properties. Taking into account potential improvements for these properties, the yield is most greatly impacted by changes to telescope diameter, followed by coronagraph inner working angle and contrast, followed next by throughput and bandwidth. The yield is most sensitive to telescope diameter because a larger aperture provides three benefits: a greater collecting area, smaller physical inner working angle for a given coronagraph, and a smaller planet PSF that reduces the amount of background flux. The yield is also quite sensitive to small changes in the coronagraph's IWA, because decreasing the IWA increases the number of accessible habitable zones and enables the detection of bright and gibbous phase planets in already accessible habitable zones.

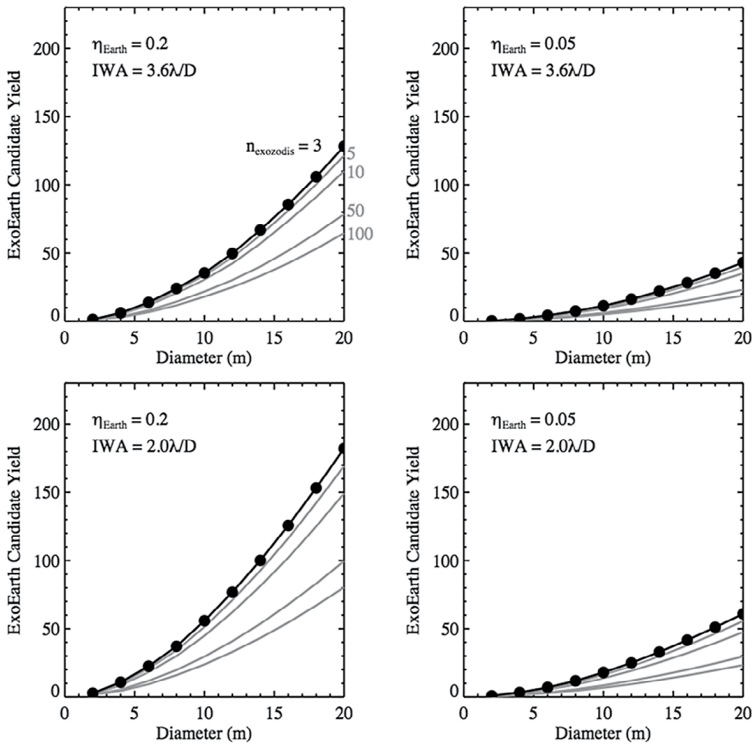

Figure 3-15: The number of candidate exoEarths that can be detected in the habitable zones of Sun-like stars as a function of telescope aperture diameter, assuming one year of on-sky observations, and including time to acquire R = 70 spectra centered at 550 nm. Results are shown for two IWA values and for two values of $\eta_{Earth}$. Black curves are for an assumed exozodi background of 3 zodi. Gray curves show the impact of increasing exozodi levels to 5, 10, 50, and 100 zodi. If the same one year of observing time is spent only on discovery and not spectra, the yields may increase by as much as 40%. Figure courtesy of C. Stark.

The planet yield also is only weakly dependent on the median exozodi level. There are two reasons for this. The first, and likely dominant effect, is a selection effect assuming a prioritized observation plan—as the median exozodi level increases, exposure times also increase, so the telescope cannot perform as many observations during the mission lifetime. The targets not reached because of fewer observations were low on the priority list, and have a weak impact on planet yield. The second reason is that the impact of a higher exozodi level can be mitigated by changing the observation plan—for higher exozodi levels, observations will not be pushed as far into the crescent phase as for planetary systems with lower levels of exozodi. Throughput and bandwidth, which cannot improve by an order of magnitude over assumed values, directly control the number of observations possible over the mission lifetime and thus result in the same sort of selection effect described above for exozodi. It is worth noting that if it were established observationally that median exozodi levels are closer to 3 zodi than the current upper limit of about 60 zodi, then the exoEarth yield calculation results would double for a given mission design.

An approximate scaling relation for the exoEarth yield calculations is as follows (and valid as long as deviations from the exponent values are small; Stark et al. 2015),

$$\text{Yield} \approx 25 \left[\frac{D}{10 \text{ m}}\right]^{1.97} \times \left[\frac{T_{\exp}}{1 \text{ yr}}\right]^{0.32} \times \left[\frac{\text{IWA}}{3.5 \; \lambda/D}\right]^{-0.98} \times \left[\frac{\text{Throughput}}{0.20}\right]^{0.35}$$
$$\times \left[\frac{\Delta\lambda}{0.10\mu}\right]^{0.30} \times \left[\frac{\text{Contrast}}{10^{-10}}\right]^{-0.10} \times \left[\frac{\eta_{Earth}}{0.10}\right]^{0.89} \times \left[\frac{\text{Bkgd}}{3.0 \text{ zodi}}\right]^{-0.23}.$$

### 3.5.4 ExoEarth Spectra

The exoEarth yield results above are for exoplanet photometric discovery and, therefore, set the number of planets that can then be selected for follow-up spectroscopy and atmospheric characterization. How planets will be selected for follow-up is beyond the scope of the present study; it may be optimal to select the brightest candidates, or to optimize for a uniform distribution across planet mass or orbital properties.

Spectroscopy spanning from visible wavelengths to 2.4 microns is desired so that multiple spectral bands of each molecular gas species can be observed to robustly identify the molecular species. A large spectral coverage also favors the detection and identification of unfamiliar or unknown gases. In particular, 2.4 microns would reach strong methane ($CH_4$) features. For an internal coronagraph, it is likely that a sequence of observations each covering a 10–20% bandpass would cover the full desired range.

Reaching to near-IR wavelengths presents challenges because the IWA of coronagraph-based starlight suppression scales as $\lambda/D$, implying that planets detected near the optical IWA will be inside it at longer wavelengths and so inaccessible for spectroscopy. The yield of planets with near-IR spectroscopy will therefore be lower than for optical spectroscopy, a natural consequence of the dependence of the IWA on wavelength. This reality means that spectra in the full desired range can only be taken for a subset of discovered candidate exoEarths. The number of exoEarths (out of the exoEarth yield shown in Figure 3-15) observable at different wavelengths is shown in Figure 3-16. Since it is harder to reach $2\,\lambda/D$ than $3.6\,\lambda/D$, a contrast level of $10^{-9}$ was assumed for the former while a contrast of $10^{-10}$ was assumed for the latter. The total observing time for spectroscopy at the full range of wavelengths shown in Figure 3-16 is not estimated here but would be in addition to the observing time needed for discovery and spectroscopic characterization at 550 nm.

## 3.6 Planetary Systems and Comparative Planetology

The search for ExoEarths is the primary focus of the *HDST* science case because of the unparalleled potential for advancing our understanding of our place in the cosmos. Many hundreds of stars will be searched, with the aim of uncovering dozens of exoEarths. A space mission capable of exoEarth discovery for hundreds of stars will therefore enable observational study of planetary systems in general. Newly discovered planets of all sizes, and their potential debris disks, will not only place detected exoEarths in context within their planetary systems, but are also interesting in their own right.

For example*, HDST* will be able to access cold outer planets including Jupiter and Saturn (see Figures 3-17, 3-18, and 3-19). The ability to observe multiple planets in the same planetary system is a major benefit for constraining the formation and orbital evolution mechanisms for planets. Spectroscopic characterization of planets in the same planetary system, and their comparison with planets in other systems, may yield useful correlations in planet atmosphere characteristics with planet orbital periods, and stellar types and metallicities. Planetary systems are expected to be naturally diverse, for reasons including the stochasticity of formation processes and the time-dependent and contingent nature of stellar evolution, planetary atmospheric growth and loss, and other still-unknown processes. The four-planet system HR 8799 is a good example of a multiple planetary system that was both unanticipated and informative for planet formation theories.

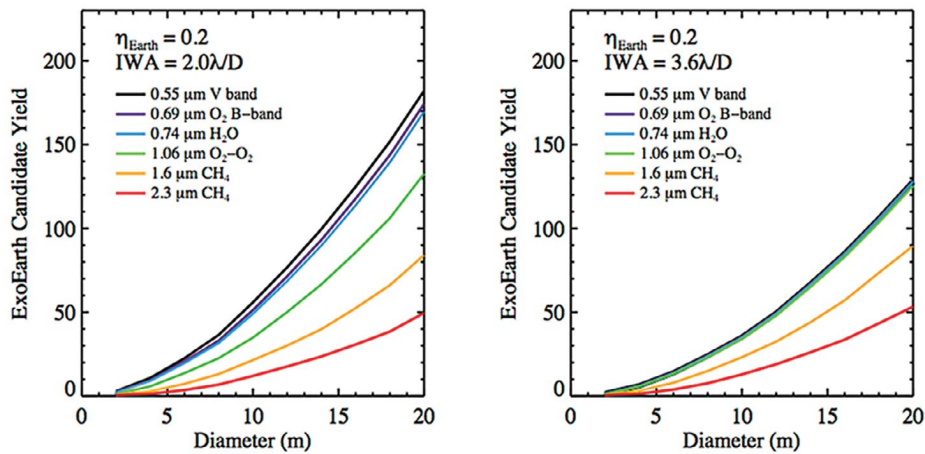

Figure 3-16: The number of planets discovered at 500 nm (black curve) that could also be detected at longer wavelengths (colored curves), as desired for spectroscopic observations, as a function of telescope aperture size and for two different assumed inner working angles. In other words, these plots show the number of planets that are visible at some point during their orbit, for a range of different wavelengths. Figure courtesy of C. Stark.

Dust belts and debris disks (from asteroids and comets) are expected to be present in planetary systems. Observations may point to unseen planets below the mission's direct-detection thresholds, whose presence can be inferred through patterns in the disk shaped by their gravitational influence. Planets induce radial structures (narrow rings, cleared gaps) in disks, vertical structures (midplane warps seen in edge-on systems), and azimuthal structures (eccentric rings, resonant clumps). Circumstellar dust can also be studied in the context of other planets in the same system and together with planet atmosphere spectroscopy may inform planetary system history by way of giant planets associated with dust belts and contribution of water and volatile inventory on any planets in the system.

Comparative exoplanetology based on direct-imaged reflected-light spectroscopy for a large number of planets in different categories is also a highly anticipated field of study. Sub-Neptune planets—planets between two to three Earth radii that have no Solar System counterparts—are a good example of a mystery that might be solved with atmosphere observations. Are these planets indeed all like Neptune with metal-rich but hydrogen- and helium-dominated envelopes? Are some of them water worlds (~50% by mass water in the interior) with thick steam atmospheres? Are some massive, rocky worlds with thin H atmospheres? Or, if all of the above and more are true, how are different classes of sub-Neptune planets populated? Spectroscopy of cold giant planets is another comparative planetology example. One expectation is that the cold giant planets will resemble each other, with atmospheres dominated by $CH_4$, $NH_3$, and $H_2O$ (depending on the presence of clouds and whether gas depletes into clouds). It may be that the giant-planet atmospheric C/H/N/O ratios are highly correlated to the host star's or the system's overall composition. In this case, we would have a general framework in which to understand the giant planets in any system. In contrast, the giant planets as seen by their atmospheres may be much more diverse than expected, and the story would change. Each new giant planet discovered is likely to be of interest for detailed observational study on its own.

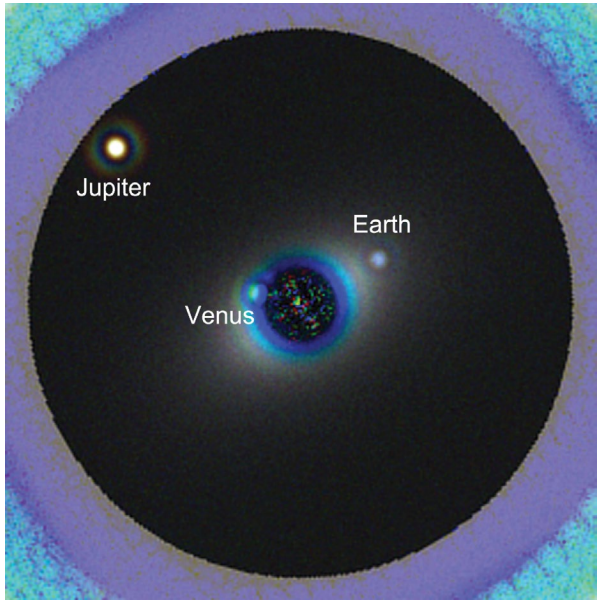

Figure 3-17: Simulation of a Solar System twin at a distance of 13.5 pc as seen with *HDST* (12 m space telescope) and a binary apodized-pupil coronagraph, optimized using the methods described in N'Diaye et al. (2015). The image here simulates a 40-hour exposure in 3 filters with 10% bandwidths centered at 400, 500, and 600 nm. The inner and outer working angles used in this simulation are 4 $\lambda$/D and 30 $\lambda$/D, respectively. The coronagraph design can support smaller inner working angles and larger outer working angles. Perfect PSF subtraction has been assumed (i.e., no wavefront drifts between target star and calibrator star). The Earth and its blue color are easily detected. The color of Venus is biased because that planet lies inside the inner working angle in the reddest exposure. The image employs a linear stretch to the outer working angle and a logarithmic stretch beyond that (where the purple-colored ring begins). Image credit: L. Pueyo, M. N'Diaye.

Transit and secondary eclipse spectroscopy is a possibility for *HDST*, provided high spectral resolution ($R$ ~1000s) instrumentation is included. Because this topic is a major focus for *JWST*, details are not presented here except to point out that *HDST*'s larger collecting area would allow an order of magnitude shorter integration times, allowing characterization of planets at the very edge of the *JWST* capabilities. As an example, to detect the presence of water vapor and $CO_2$ for even a potentially rocky transiting super-Earth-sized planet ($R < 1.5$ $R_{Earth}$) in the habitable zone or a nearby M star, *JWST* may require more than 50 combined transits (spread out over one to two years to catch enough transits), whereas *HDST* could accomplish the observations in 5–10 combined transits, spanning only a couple of months. With UV capability the ozone cutoff at 300 nm could potentially be observed. *HDST*'s transit spectroscopic capabilities will also enable new precision in mapping the 2D compositional and thermal properties of the atmospheres of larger and hotter planets, via multiple measurements of the secondary occultation.

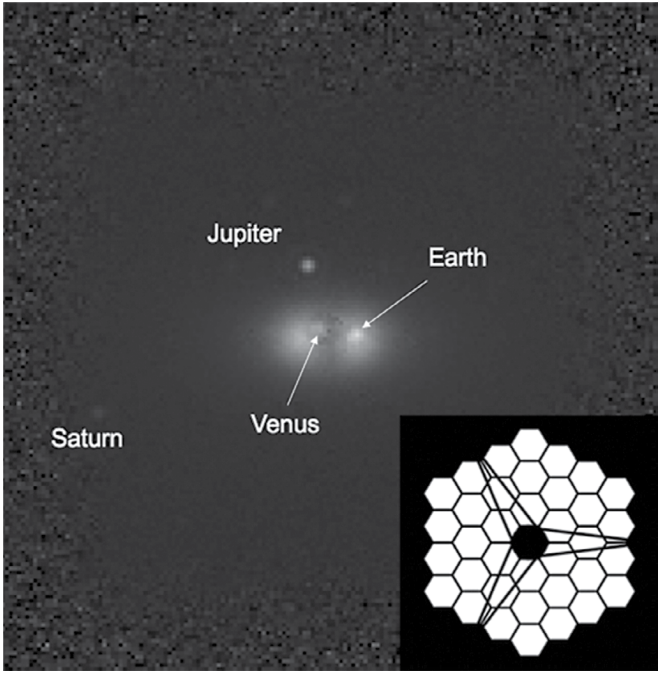

**Figure 3-18:** Simulated near-IR (1.6 µm, 20% band) image of a solar system twin at a distance of 13.5 pc as seen by *HDST* (12 m space telescope) with a 2 day exposure. The pupil geometry adopted for this simulation is shown in the lower right. A Phase-Induced Amplitude Apodization Complex Mask Coronagraph (PIAACMC), offering small IWA (1.25 λ/D), is used here to overcome the larger angular resolution at longer wavelength. Earth, at 2.65 λ/D separation, is largely unattenuated, while Venus, at 1.22 λ/D, is partially attenuated by the coronagraph mask. At this wavelength, the wavefront control system (assumed here to use 64 x 64 actuator deformable mirrors) offers a larger high contrast field of view, allowing Saturn to be imaged in reflected light. This simulation assumes PSF subtraction to photon noise sensitivity. In the stellar image prior to PSF subtraction, the largest light contribution near the coronagraph IWA is due to finite stellar angular size (0.77 mas diameter stellar disk).

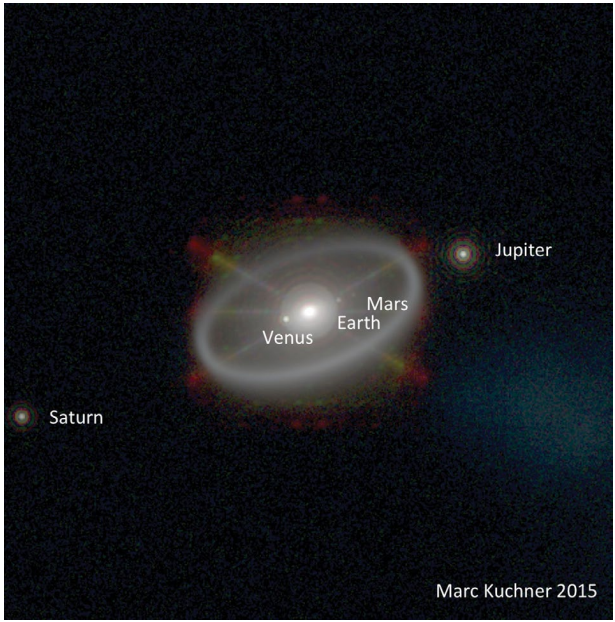

Figure 3-19: Simulated false-color (450–850 nm) image of a planetary system around a nearby G star (Beta Cvn) seen by a 12 m optical space telescope equipped with a free-flying ~100 m diameter starshade. Imperfections in the starshade scatter some starlight and sunlight into the center of the image, yielding a contrast of $4 \times 10^{-11}$ at 1 AU, but twins of Venus, Earth, Jupiter and Saturn stand out clearly from the PSF wings on either side of the starshade. A Mars twin may also be detectable in one-day exposures like this one with some careful calibration. The large aperture easily separates planets from background galaxies and local and external zodiacal dust (all included), but still senses the narrow dust ring at 3.5 AU in this model, 10× fainter than any other current or upcoming missions can reach. Credit: M. Kuchner.

## 3.7 The Role of Coronagraphs and Starshades in the Search for Habitable Worlds

Out of a number of different starlight-suppression options, this report recommends the internal coronagraph as the *HDST*'s baseline starlight-suppression method. *HDST* requires efficiency for searching hundreds or more stars for exoEarths, and the internal coronagraph poses little additional constraints on telescope slewing for repointing. A starshade to go with the *HDST* aperture is operationally challenging for a survey of hundreds of target stars, because realignment of the starshade and telescope for each new target star is cumbersome. Either the starshade or the telescope has to move across the sky, taking days to weeks for each retargeting maneuver, depending on the starshade size, separation from the telescope, and propulsion method.

While the internal coronagraph is the instrument of choice for the discovery phase of the mission, the starshade would perhaps eventually be added as a complement to the *HDST* mission. A starshade employed in a second part of the mission would enable more detailed characterization of a subset of interesting planetary systems already detected photometrically with the coronagraph. The starshade offers a high system throughput (on the order of 20 to 40 %) and a broad bandpass (a few hundred nm) enabling higher SNR spectra than the coronagraph system, for which total throughput may be significantly less. The starshade has flexibility in the wavelength range on the IWA, by changing the starshade–telescope separation for an appropriately designed starshade. This is beneficial for longer red wavelengths or even possibly near-IR observations. The starshade will enable the study of outer planetary systems (the starshade has no OWA), limited only by detector size. This is in contrast to typical OWA for internal coronagraphs (see Table 6.1; current lab demonstrations are at 10 to 20 $\lambda/D$ and *HDST* has a goal of > 75 $\lambda/D$).

Simulated *HDST* coronagraphic images of an Earth-like planet in a star system like our own are shown in Figures 3-17 and 3-18. A simulated image that might be obtained with a large starshade flying in formation with *HDST*, and for a similar planetary system, is shown in Figure 3-19.

The technical maturity and challenges for both the internal coronagraph and the starshade are discussed in Chapter 5. Here it is worth noting that both starlight-suppression techniques have major challenges to overcome, motivating the need to consider the starshade as a supplemental technological path for *HDST*. For the starshade, deployment and maintenance of the starshade shape has yet to be demonstrated to the required tolerances in a flight-like environment and precision formation flying at tens of thousands of km separation is under development. Yet, with those challenges overcome, the starshade has unique strengths, in that the starshade, and not the telescope, sets the IWA and achievable planet-to-star flux contrast. The starshade enables a telescope to detect planets as close as 1 $\lambda/D$ to their host stars, whereas most coronagraphs are being designed to work down to the 2 to 4 $\lambda/D$ IWA range. As illustrated in Figure 3-18, coronagraphs could potentially work as close as IWA ~1 λ/D, but light leaks due to finite stellar angular size are then brighter than the planet signal, requiring accurate subtraction, and reducing sensitivity due to residual photon noise. With a starshade, only planet light enters the telescope, meaning specialized wavefront control required for coronagraphs (on the picometer level for wavefront stability) is not needed. A significant disadvantage of a large starshade is that it is not a viable method for performing a survey of the hundreds of star systems needed in the search for Earth-like worlds, due to the lengthy transit times between targets. An internal coronagraph will thus need to be part of the instrument complement of *HDST* as the option that most effectively enables a large exoplanet survey.

### 3.8 Summary

*HDST*'s primary goal is to detect and spectroscopically examine dozens of exoEarths in their stars' habitable zones. To find these gems, the telescope will survey many hundreds of nearby main-sequence stars. Only a mission at the scale of *HDST* can deliver a high yield of exoEarths and such a rich database of information about all kinds of planetary systems at no additional cost in observing time.

Based on ExoEarth yield calculations, for reasonable assumptions about $\eta_{Earth}$, exozodi levels, and technical performance, the desired yield of dozens of exoEarths requires a telescope diameter of 9–12 m or larger. The upper end of this range would provide resilience against the possibility that one of our key assumptions was overly optimistic. The telescope diameter of 12 m is recommended based on the exoEarth yield, and also on general astrophysics science (Chapter 4) and technical considerations (Chapters 5 and 6).

The question repeatedly arises about ground vs. space-based capability. It must be emphasized that while existing and planned assets have the capability for small planet discovery in habitable zones, these potential discoveries are limited to a small number of target stars. In this regime, finding a habitable-zone planet will require a degree of good fortune. Moreover, the stars that can be searched are almost exclusively M dwarfs. Only space-based direct imaging with a large-aperture telescope can search hundreds of Sun-like stars to yield dozens of rocky planets, and to have the capability to characterize exoplanet atmospheres.

The *HDST* prime science driver is to find and identify exoEarths orbiting Sun-like stars. The search for biosignature gases is a main motivation for identifying an exoEarth, but with sobering caveats. Even with superb data, there is no single "smoking gun" biosignature gas, although oxygen is considered close to one. There are false-positive scenarios, where the ambiguity between a biotic and an abiotic origin for biosignature gases simply cannot be resolved. This means that aiming for a robust detection of biosignature gases on a single planet may not be enough. To establish the presence of life beyond the Solar System, biosignature gases in a number of planets may be needed, further motivating the desire to find a large number of candidate exoEarths.

# Chapter 4 Cosmic Birth Yields Living Earths

T<small>HE STORY OF LIFE</small> begins with the dark matter seeds of galaxies, which draw together gas from the diffuse cosmic reservoir created in the Big Bang. Galaxies form the stellar nurseries that make stars, without which the cosmos would forever lack the heavy elements and radiation that form and feed life. Stars with a broad spectrum of masses play multiple roles: low-mass stars provide an abundant source of energy for life, while the intermediate and massive stars create and disperse the heavy elements that form life's raw ingredients. Finally, those heavy elements find their way back into clouds that condense to form low-mass stars and their planets, on which the seeds of life can flourish.

All these steps preceded the formation of the Solar System and the birth of life on Earth. This origin story will be shared by any other life in the Galaxy, whether it is common or rare. Each generation of astronomers has filled in the chapters of this story, but it is still far from complete. Humanity does not yet fully understand the cosmic history leading to our own, so far unique, existence.

With its large collecting area, high spatial resolution, panchromatic UVOIR sensitivity, exquisite stability, and low sky backgrounds, *HDST* offers a radically improved vision to see what cannot yet be seen: planets lost in the glare of their stars, dim stars buried in crowded fields, and faint light and shadows from the nearly invisible cosmic web. Like each previous leap in capability, it promises to revolutionize understanding of the cosmos and our place in it.

The name of *HDST* uses the metaphor "high-definition" because a telescope of 12 meters in aperture will deliver stable diffraction-limited imaging to 10 mas in the visible. This simple number carries enormous promise for new discoveries: *HDST* can resolve objects down to a physical scale of 1 AU in the Solar neighborhood, 100 AU everywhere in the Milky Way, 0.1 pc everywhere in the Local Group, and 100 pc out to the edge of the observable Universe (Figure 4-1). These resolution thresholds will be reached at optical and ultraviolet wavelengths where they are difficult or impossible to achieve from the ground. Combined with the extreme depths (AB ~32 in ~1 hour) that can be reached with *HDST*'s low background and stable point-spread function (PSF), these physical thresholds bring transformative vision to the astronomical objects at each epoch. This sharp vision, together with efficient and highly multiplexed forms of UV and optical spectroscopy, gives a historic opportunity to make the journey "from cosmic birth to living Earths" in the next two decades.

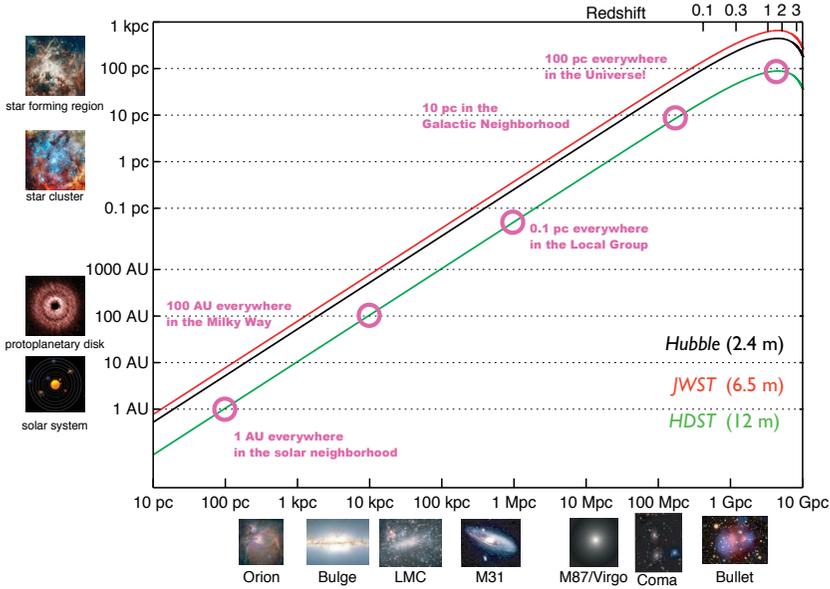

Figure 4-1: Telescopes imaging at their diffraction limit offer high-definition views of the Universe across cosmic distances. Major resolution thresholds reached by *HDST* are marked in magenta and illustrated at left. For a 12 m aperture, the resolution element at *HDST*'s 0.5-micron diffraction limit corresponds to 100 parsecs or less at all cosmological distances in the observable Universe. The D = 2.4 m *Hubble* (black line) is diffraction limited at $\lambda$ = 0.6 micron, and the 6.5 m *JWST* (red line) is diffraction limited at $\lambda$ = 2 microns, such that their physical resolution elements (proportional to $\lambda$/D) are about the same.

The following exploration of how this facility will transform astronomy proceeds through cosmic history (Figure 4-4), from the earliest seeds of galaxies (> 10 billion years ago), to the growth of galaxies like the Milky Way (6–10 billion years ago), to the birth of stars like the Sun in the Milky Way (5 billion years ago), culminating in the formation of planetary systems like the Solar System (now). At each epoch, the radically sharp vision, sensitivity, and wavelength coverage of *HDST* will reveal things previously unseen. What are the questions remaining about the cosmic origins of galaxies, stars, planets, and life? What major problems remain to be solved? And how will *HDST*, in concert with other frontline facilities in its era, address or solve these problems? The answers lie in the revolutionary capabilities this observatory brings to bear, what possibilities it opens with novel technologies and instrument modes, and how its vision is complementary to facilities operating in its era, across the electromagnetic spectrum in space and on the ground.

## 4.1 The Epoch of Galaxy Formation

Galaxies are the ultimate engines of creation, the factories that make stars, which in turn provide life with its basic ingredients and energy input. Over billions of years these galactic factories grow inside their dark-matter halos by acquiring gas from the intergalactic medium (IGM) and by merging with other galaxies, large and small. Telling the complete story of galaxy formation focuses on two main chapters: detecting and resolving galaxies' building blocks at any time, and tracing galaxies' evolution with significant samples over time. The first requirement demands high physical resolution and broad wavelength coverage, and the second demands high "mapping speed"—the ability to efficiently cover large areas of sky at deep limiting magnitudes.

### 4.1.1 Galaxies and their Building Blocks in High Definition

Current data indicates that the overall cosmic star-formation history peaked about ~10 billion years ago and has since declined, and that the largest and smallest galaxies known today essentially ceased forming stars early on, while galaxies like the Milky Way continued to do so roughly in proportion to their mass. In *Hubble*'s deepest images, the building blocks of galaxies—massive star-forming regions in disks and accreting satellites—look different at the time the Milky Way formed than they do today. Yet *Hubble* poses major questions that it lacks the resolution to answer: Why are early galaxies so small and dense, sometimes packing the entire mass of the Milky Way into less than 1/100th its volume? How are galactic disks assembled—through slow accretion of gas, or through many mergers with smaller galaxies? What are the massive, unresolved clumps dominating early disks? Do they form bulges? What causes the quenching of galaxy star formation? Is it feedback from supermassive black holes, dynamical interactions, or both? Is there a link between nuclear star formation and supermassive black hole growth?

> **Big Questions**
>
> *How was the Milky Way assembled?*
>
> *How did it get its bulge, disk, and satellites?*
>
> *When and how do galaxies stop forming stars?*

## *HDST* and the Extremely Large Telescopes: Better Together

In the 2030s, *HDST* will be operating alongside a new generation of ground-based telescopes that are set to begin operations in the 2020s. These telescopes' large 20–40 m apertures will offer unprecedented light-collecting ability, and superb resolution when operated with adaptive optics. Much like the *Hubble* and its 8–10 m ground-based contemporaries, *HDST* and the 20–40 m telescopes will leverage their respective capabilities and work together to push back the scientific frontiers.

Ground-based telescopes excel at spectroscopy to deep limits (currently AB ~28) or high spectral resolution (R ~10,000–200,000). They also provide superb wide-field coverage (up to > 1 deg) and spectroscopic multiplexing (100–1000 sources per field). These capabilities benefit science programs needing large samples of sources over large areas of sky and/or high spectral resolution in the optical and near-IR. Ground-based telescopes equipped with adaptive optics systems have also made great advances in high-resolution near-IR imaging where > 1 arcmin fields are not required.

Space-based telescopes' strengths are complementary. Due to backgrounds that are 10–100× darker than from the ground, and a stable operating environment, space observations reach the same broadband imaging depths up to 100× faster, achieve fainter ultimate limits (AB ~32–34 mag), and extend stable diffraction-limited PSFs to larger fields of view (3–4 arcmin) and bluer wavelengths. These advantages favor science programs requiring high-resolution imaging of faint objects, < 0.0001 mag photometric precision, low-resolution and/or slitless spectroscopy, imaging in crowded fields, and long-term variability and proper motion studies. Space telescopes also uniquely cover the information-rich UV spectrum that is otherwise blocked by Earth's atmosphere. These complementary strengths will likely persist into the *HDST*/ELT era. Even if the ELTs achieve diffraction-limited imaging at wavelengths longer than 1 or 2 microns, *HDST* will still be necessary to extend ~10 mas resolution from the near-IR into the UV, given that no other currently planned facility will achieve < 40 mas resolutions at wavelengths less than 1 micron. The combination of *HDST* in the UV and optical, the ELTs in the near-IR, and ALMA in the submillimeter, will provide a truly panchromatic view of the high-definition Universe (Figure 4-2).

Despite its smaller aperture, *HDST* provides key advantages over the ELTs in the time required to reach a given SNR for broadband imaging and low-resolution spectroscopy. Figure 4-3 compares 10-$\sigma$ point-source limits for broadband imaging (R = 5) between *HDST* and the European ELT (E-ELT) as the largest limiting case. For broadband imaging, the ground-based facilities will hit a "wall" at around AB ~30–31 imposed by the sky background of light scattered by Earth's atmosphere. *HDST* is needed to open a window into the "ultra-faint" universe below 1 nJy (AB > 31) in the UV-optical regime, where *HDST* will reach rocky exoplanets around Solar-type stars at > 10 parsecs from Earth and individual main-sequence stars out to 10 Mpc. The advantage of *HDST* over E-ELT declines at higher spectral resolution, but *HDST* still offers significant gains at R = 100 at all optical and near-IR wavelengths and at R = 2000 in the visible (Figure 4-17). The combination of *HDST*'s deep, precise photometry and low-resolution spectroscopy with the deep, wide-field, and multiobject spectroscopy and high-resolution near-IR imaging will enable key advances in the science cases discussed in this Chapter.

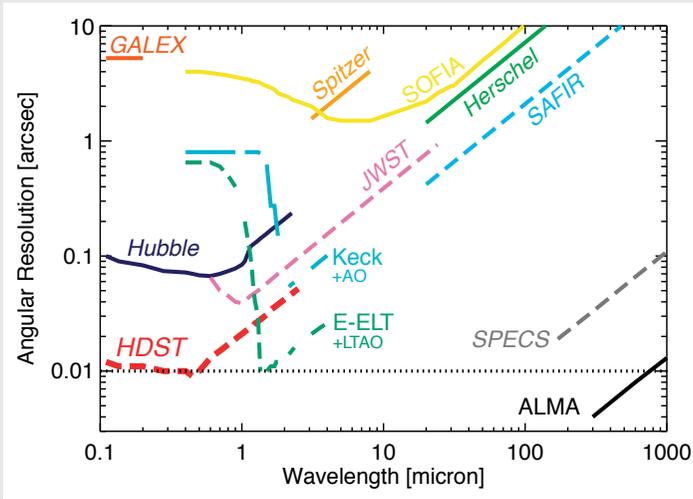

Figure 4-2: Angular resolution as a function of wavelength for current, planned, and possible future astronomical facilities. The *HDST* resolution is shown assuming a 12 m aperture, diffraction limited at 500 nm. The expected performances from facilities that are still in pre-construction or conceptual phases are shown as dashed lines. Some ground-based facilities have non-continuous wavelength coverage due to atmospheric absorption features.

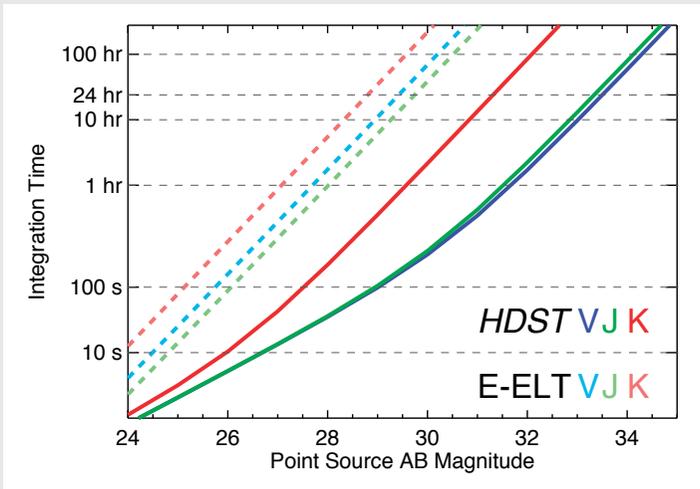

Figure 4-3: The total integration time needed to reach a 10-sigma point-source limiting magnitude for the 38 m E-ELT and a 12 m *HDST* (assumed here to operate at 270 K). Computations are done for 3 different passbands: *V*, *J*, and *K*. Total system throughput and instrument characteristics (read noise, dark current) are adopted from the ESO E-ELT ETC website. The E-ELT is assumed to perform at its diffraction limit at wavelengths longer than 1 micron and seeing-limited (0.6 arcsec) for wavelengths below 1 micron.

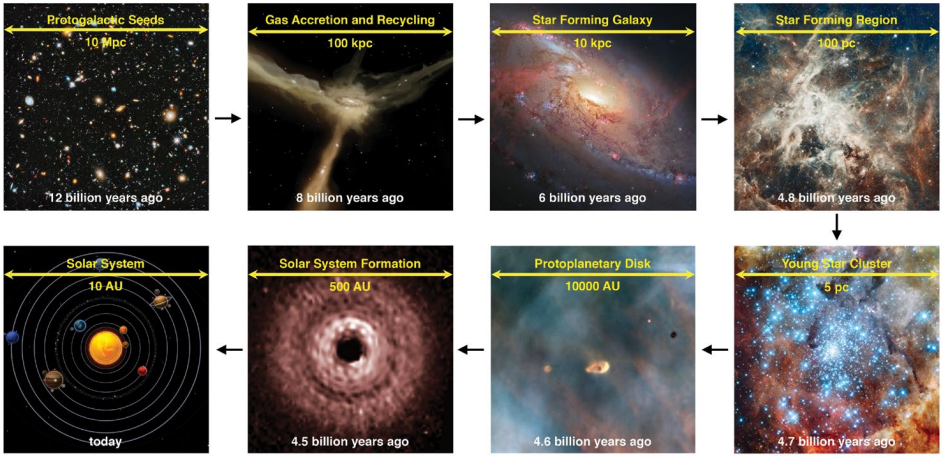

Figure 4-4: The path from Cosmic Birth to Living Earth, witnessed at a range of physical scales. Stars in the first galaxies produce the first heavy elements, which leave those galaxies and recycle into the larger galaxies they become. Mature galactic disks form later, composed of many star-forming regions at 50–100 parsec in size. Inside these regions form multiple star clusters, each of which includes massive stars that forge more heavy elements and low-mass stars that host protoplanetary disks. At 4.5 billion years ago, one such disk in our Milky Way formed the Earth and its siblings in the Solar System.

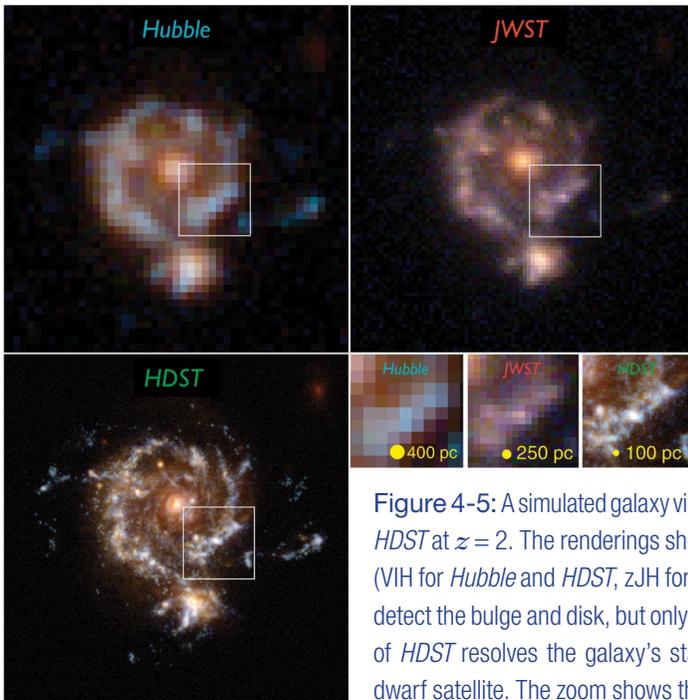

Figure 4-5: A simulated galaxy viewed by *Hubble*, *JWST*, and *HDST* at $z = 2$. The renderings show a one-hour observation (VIH for *Hubble* and *HDST*, zJH for *JWST*). *Hubble* and *JWST* detect the bulge and disk, but only the exquisite image quality of *HDST* resolves the galaxy's star-forming regions and its dwarf satellite. The zoom shows the inner disk region, where only *HDST* can resolve the star-forming regions and separate them from the redder, more distributed old stellar population. Image credit: D. Ceverino, C. Moody, and G. Snyder.

*JWST* will address these questions, pushing to deeper limits and higher mapping speed than *Hubble*, but with comparable physical resolution at their respective diffraction-limited wavelengths (Figure 4-1). In the near-IR, *JWST* will see rest-frame visible light from galaxies at $z$ = 1–4, not their star-forming regions and satellites shining brightly in UV light from young massive stars. These UV measurements would detect stars that have emerged from their birth environment, and complement the mid-IR and Atacama Large Millimeter Array (ALMA) studies of dusty star-forming regions at similar physical resolution. Because of their limited spatial resolution and sensitivity, neither *Hubble* nor *JWST* resolve the formation of Milky-Way-like galaxies down to their smallest building blocks.

A great stride in observing the building blocks of galaxies requires the transformative capability of *HDST*. At 12 meters, *HDST* will pack 25× more pixels into the same area compared to *Hubble*, resolving galaxies at all cosmic epochs to scales of 80–150 parsecs in the visible bands. At this threshold, the internal structure of galaxies and their individual star-forming regions will snap into view where *Hubble* now sees indistinct smudges and only the very brightest satellites.

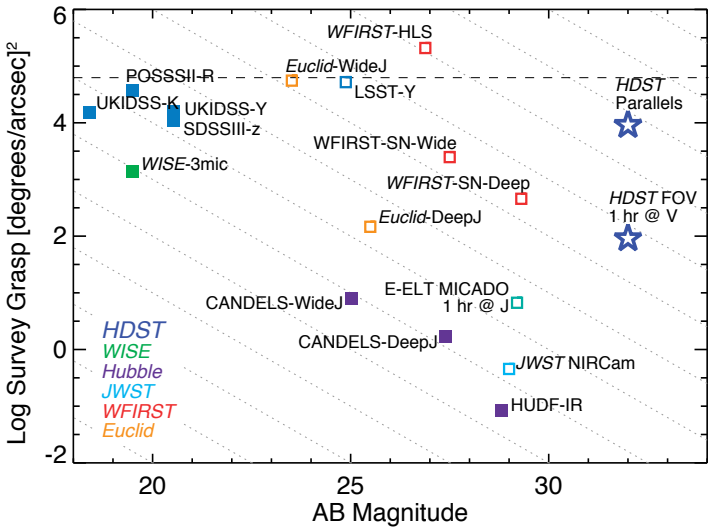

Figure 4-6: A comparison of present and future observatories and sky surveys in terms of "information content," defined as the number of spatial resolution elements in the observed field. This figure of merit increases toward the upper right, perpendicular to the diagonal lines. Current facilities are marked with filled symbols and future facilities with open symbols. All-sky ground-based surveys cluster around the horizontal line, which marks the entire sky observed at 1" resolution. Space-based narrow/deep fields cluster at lower right. Large ground-based telescopes achieve high resolution, but over relatively small fields of view (~1 arcmin), and are limited in depth by sky backgrounds. *WFIRST/AFTA* will expand *Hubble*-like imaging and depth to much larger areas, moving toward the upper right. Though it observes smaller fields, its exquisite spatial resolution means that *HDST* will surpass *WFIRST/AFTA* in information content on galaxies.

These gains are plain in Figure 4-5, which compares mock *Hubble*, *JWST*, and *HDST* images of a simulated $z = 2$ disk galaxy. The average distance between bright star-forming knots (a few hundred parsecs) is not resolved by *Hubble* and *JWST*; to them, this is a single supermassive star-forming region. By contrast, *HDST* can see the individual star-forming regions, resolve some nearby satellites, cleanly separate and age-date bulges and disks while both are in formation, distinguish minor mergers from disk instabilities, count star clusters, and see inside the compact star-forming and passive galaxies, including those in transition. This sharp vision of galaxy formation at all cosmic epochs will transform our understanding of the birth of our own Galaxy and others like it.

This epoch is also the peak era for luminous active galactic nuclei (AGN). *HDST* will intensively study the relationships between AGN and their host galaxies, resolve the reaction of galaxies to their AGN at 100 pc scales (both through imaging and spectroscopy), and examine the relationship between AGN and galaxies at low-galaxy mass. It is increasingly clear that the evolution of AGN, galaxies, and supermassive black holes are tightly coupled, with the properties of each influencing the other. *HDST* will carefully dissect the properties of each component, revealing key ingredients of the underlying astrophysics that shapes AGN feedback and supermassive black hole growth.

*HDST* also promises dramatic gains in terms of the cosmological volumes, spatial resolution, and depth of galaxy surveys, especially if it can operate its wide-field imaging cameras in parallel during long, staring exoplanet observations (Chapter 3). In approximately 500 observations spread over a year of exoplanet observations, deep images taken in parallel could accumulate a total of ~1 $deg^2$ of sky with limiting magnitudes of AB ~31–32; this is 15–40× deeper and 30–60× larger area than *Hubble*'s state-of-the-art CANDELS-Deep survey. Because these *HDST* "Deep Parallels" will be sampled at the diffraction-limited resolution of a 12 m class telescope by the wide-field imaging camera (Chapter 5), their information content will exceed any presently contemplated survey, even the much larger *WFIRST* high-latitude surveys (Figure 4-6). The cosmological volumes covered are impressive: from $z = 2$–3 alone, the comoving volume is 10–20 million $Mpc^3$, comparable to the Sloan Digital Sky Survey (SDSS) out to the distance at which SDSS's (seeing-limited) spatial resolution is 1 kpc or better. The *HDST* Deep Parallels can slice the high-$z$ cosmos into multiple timesteps comparable to the entire SDSS, but at 10× better physical resolution. Because these observations will occur during observations of bright exoplanet host stars by the on-axis coronagraph, they require an off-axis imaging camera and good control of internal scattered light, both readily achievable (see Chapter 5). *HDST* will have the sensitivity, spatial resolution, and operational efficiency needed to survey massive cosmological volumes at 100 pc or better spatial resolution, allowing us to directly relate modern galaxies to their early progenitors to tell the full story of our galactic origins.

### 4.1.2 The First Galaxies at the Dawn of Cosmic Time

*HDST* will continue the search for the youngest galaxies in the post-*JWST* era. Near-infrared imaging with the *Hubble Space Telescope* has already traced the earliest detected seeds of modern galaxies to within a few hundred million years of the Big Bang. This process started with the Deep Field in 1995 and continued through each new generation of *Hubble* instruments, allowing *Hubble* to push back the redshift frontier and identify the brightest star-forming galaxies at redshift $z \sim 10$, when the Universe was only 400 million years old. The on-going *Hubble* "Frontier Fields" are carrying this search to potentially higher redshifts and lower galaxy luminosities, by using gravitational lensing to magnify the more distant universe in select regions where the lensing effect is particularly strong. *HDST*'s higher resolution and greater sensitivity match the typical increase in resolution and depth offered by gravitational lensing, and thus *HDST* will effectively carry out the same search as the Frontier Fields, but anywhere in the Universe it chooses to point. These searches will benefit directly from the power of the next generation of ELTs, which can provide the needed long-wavelength constraints for these searches.

> **Big Questions**
>
> *What were the first seeds of the Milky Way?*
>
> *What and when were the first cosmic explosions?*

In the immediate future, well before the launch of *HDST*, this effort will be continued by *JWST*, which is optimized for sensitivity at near- to mid-IR wavelengths (0.6–28 µm) where it can probe the rest-frame UV and visible signatures of star formation and stellar explosions in the first galaxies. *JWST* will also have superb image quality (60 mas at 2 µm), comparable to *Hubble*'s in the optical. *JWST* will surely make breakthroughs that cannot be anticipated, and thus it is natural to ask: how might *HDST* build on, or complement the achievements of its predecessor in the high redshift Universe?

> **Big Questions**
>
> *How do galaxies acquire, expel, and recycle their gas?*
>
> *How do massive black holes influence their galaxies?*
>
> *What are the many ways stars can die?*

In terms of capability, the larger aperture of *HDST* will outperform *JWST* in image quality at near-infrared wavelengths below 2 microns, and dramatically exceed the resolution and sensitivity of *Hubble* and *JWST* at optical wavelengths. This capability alone will make *HDST* a partner with *JWST* in exploring high-redshift galaxies in a panchromatic fashion. The increased spatial resolution will improve the discovery and localization of high-redshift transients such as early supernovae, tidal disruption flares from supermassive black holes, and even gravitational-wave sources within their host galaxies.

In addition to *HDST*'s higher resolution, at the redshifts where *JWST* is expected to make its greatest contributions ($z > 6$), the information-rich rest-frame UV wavelengths are still observed at < 1 micron, where *HDST*'s performance is unmatched. These visible-band photons directly trace the radiation of the massive stars that drive the feedback of their galaxies and the reionization of the intergalactic medium (IGM). While *JWST* will use high spatial-resolution Integral Field Units (IFUs) to dissect the internal dynamics of early galaxies, it will do so in the rest-frame visible, and so miss many of the key physical diagnostics of star formation and gas dynamics that are available in the rest-frame UV (e.g., Lyα, C IV, O VI). In particular, star-formation rates derived from optical light are subject to significant age-metallicity degeneracies, but those derived from rest-UV light see the massive stars directly. These *HDST* measurements will also complement ALMA observations of the redshifted [C II] line, which only provides an indirect measure of the energy input into ISM. While *JWST* will certainly make unanticipated discoveries, it is likely that there will still be unanswered problems that only *HDST*, with its access to the UV and visible and spatial resolution, will be better suited to solve.

With a 2018 launch and a 10-year goal for full science operations, *JWST* will likely cease operating before *HDST* can be launched. Thus it will be *HDST*'s responsibility to carry low-background near-infrared capability in space into the post-*JWST* era. *HDST* will operate near 250–295 K to stabilize itself for exoplanet coronagraphy, so it will not be a cryogenic observatory. However, it could include a cryogenic instrument extending coverage out to > 2 microns with sensitivity limited by the thermal background of the telescope. This would give *HDST* panchromatic imaging even greater than *Hubble*'s current broad reach, 0.12–1.7 microns. Even without a fully cryogenic telescope, *HDST* could extend astronomy's infrared vision in space past the era of *JWST*, building on its successes and continuing to probe the depths of the Universe in high definition. Defining the exact capabilities that *HDST* must support in the near-IR and mid-IR will need to await the discoveries in these bands to come from *JWST*.

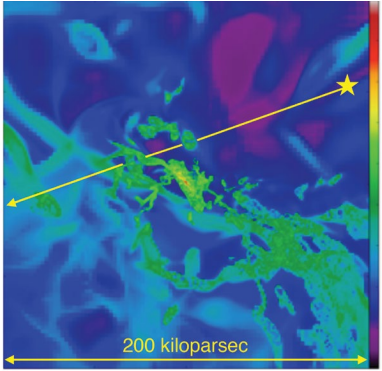

Figure 4-7: A visualization of the Circumgalactic Medium (CGM) gas fueling a Milky-Way-like galaxy. The color renders the strength of UV emission by triply ionized Carbon (C IV) in the CGM gas surrounding a Milky-Way-analog galaxy at $z$ ~0.25. *HDST* should be able to directly detect gas emission from infalling filaments (green regions) and absorption from C IV and many other species in absorption at all densities. Image credit: Joung, Fernandez, Bryan, Putman & Corlies (2012, 2015).

## 4.2 The Epoch of Solar System Formation

By four-and-a-half billion years ago ($z = 0.44$), the Milky Way had established the conditions for life on at least one rocky planet orbiting an otherwise ordinary G-type star. By then, the Galaxy likely had settled into an old bulge with a star-forming disk, and stars had ejected enough heavy elements into the interstellar gas to form dust, proto-planets, wet worlds, and life. At this epoch, all the main galactic components—disks and bulges, ISM and halo gas, satellite populations, and central black holes—were in place. This epoch is the "sweet spot" in redshift space, where the complex interactions and co-evolution of these components can finally be seen in detail from Earth, because the most powerful probes of galaxy gas flows have redshifted from the rest-frame far- and extreme-UV into the observed-frame UV and visible, at the same time that optimal stellar diagnostics are still accessible in the visible and near-IR. *HDST* will thus be able to witness, up close, how galaxies acquire, process, and recycle their gas and heavy elements and why they stop doing so (4.2.1), and to pinpoint the explosive events that announce the creation and dispersal of many of the chemical ingredients of life (4.2.2). Thanks to its unique sensitivity, spatial resolution, and multiplexing, *HDST* will bring the full picture of how galaxies help set the conditions for life into sharp focus.

### 4.2.1 Galaxy Fueling, Recycling, and Quenching

The birth of stars, planets, and life is fueled by the diffuse reservoir of cosmic gas that ultimately forms and feeds galaxies. Pictures of galaxies

> ## Essential Astrophysics in the Ultraviolet Band
>
> The UV capabilities of *HDST* provide information that is inaccessible in other wavelength bands. The UV covers thousands of atomic and molecular transitions that can be probed in emission or with absorption-line spectroscopy, yielding *direct, quantitative measures of the state of many astrophysically important elements* in the majority of their ionization states, down to exceptionally low column densities. These transitions provide powerful diagnostic tracers of hot main-sequence and white-dwarf stars, protostellar disks, diffuse gas in the ISM and IGM, black hole accretion disks including AGN, and star-forming galaxies.
>
> The increasing density per unit wavelength of diagnostic lines at <2000 Å makes the UV band a uniquely powerful probe of matter in the Universe. Figure 4-8 shows a representative sample of the available lines, categorized by their ionization species, as a function of redshift. The rest-frame 1000–2000 Å range encompasses Ly$\alpha$, the strongest lines from hot and cold ISM and IGM gas, strong rotational-vibrational bands of $H_2$ and CO, the peak emission from hot stellar atmospheres, and AGN outflows. This information-rich spectral range remains in the space UV for all redshifts $z \lesssim 1$–2, and at $\lesssim 1$ micron for all redshifts $z < 6$, covering the last 12 Gyr of the Universe's 14 Gyr history. Thus the UV is an essential window into vitally important astrophysical environments over most of cosmic time.
>
> At sufficiently large redshift, the UV reaches extremely high-ionization species that arise in hot, diffuse gas and the ejecta of supermassive black holes. Species such as Ne XIII, Mg X, and Na IX form at ~1–2 million K and exhibit lines at ~600–800 Å in the rest frame that become visible above the 1000 Å cutoff of *HDST* at $z \gtrsim 0.3$. This makes the UV the natural high-redshift partner of X-ray observations at low redshift, opening up the possibility of evolutionary studies with Athena and other X-ray facilities.
>
> *HDST* will deploy these unique diagnostics at 100-pc spatial resolution and with 50–100× the UV point source sensitivity of *Hubble*. Still more discovery space will be opened by novel UV multi-object and integral field capabilities that support the *HDST* UV drivers in this Chapter. When over >4 arcminute fields of view, and together with the complementary optical and near-IR powers of ELT peers on the ground, *HDST*'s unique UV capabilities will deliver compelling scientific returns that cannot be reached in any other way.

deceive even the best cameras: in fact, galaxies as seen in starlight are only the brightly lit core of a large gaseous structure—the circumgalactic medium (CGM)—that spans 30× the radius and 10,000× the volume of the visible stellar disk, and which may contain about as much mass as the stellar disk (Figure 4-7). Beyond the CGM lies the intergalactic medium (IGM), the Universe's nearest thing to empty space (around one atom per cubic meter). If life is made of star stuff, this is the stuff of which stars are made.

Characterizing the CGM and the IGM are fundamental goals that require UV access, because as diffuse gas moves in or out of galaxies the majority of its energy is emitted or absorbed in rest-frame UV-band transitions of hydrogen, carbon, oxygen, neon, and other heavy elements ($\lambda_0$ = 500–2000 Å). Even when redshifted by the expansion of the Universe, these lines still appear in the UV over the last 8–10 Gyr of cosmic time ($z \lesssim 2$; Figure 4-8). These transitions are readily seen in absorption in the spectra of QSOs that happen to lie behind the halos of individual galaxies; repeating the observations for many galaxies builds a "statistical map" of CGM absorption for typical galaxies. Using these UV diagnostics and this QSO absorption-line technique, *Hubble* has shown that the IGM and the CGM contain more of the baryonic matter in the cosmos than do galaxies themselves, and that the CGM acts as the fuel tank, waste dump, and recycling center for galaxies.

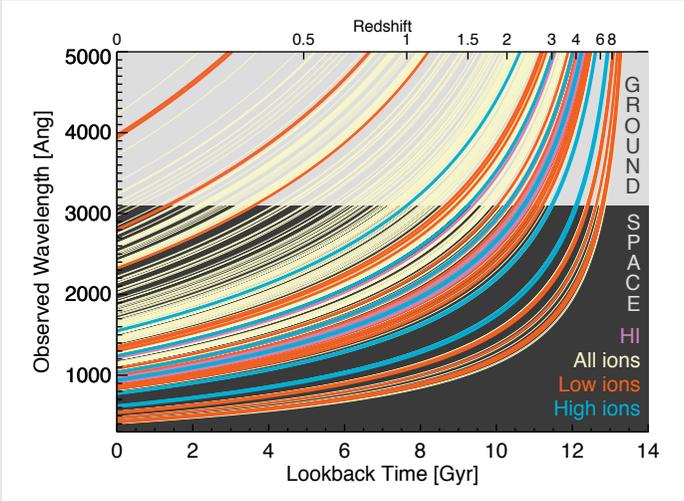

Figure 4-8: The availability of physical line diagnostics versus lookback time and redshift. Rest wavelengths are marked where lines touch the axis at left ($z = 0$). The line density of available diagnostics increases sharply toward the rest-frame UV. All the important diagnostics of H I (magenta) and hot and/or high ionization gas (blue) lie in the UV at 8–12 Gyr lookback time, and do not cross the 3100 Å atmospheric cutoff until $z \gtrsim$ 1.5–2. While a few low-ionization lines (orange) are available from the ground at low redshift, over most of cosmic time the measurement of physical conditions and elemental abundances in multiphase gas and stellar populations requires access to observed-frame UV wavelengths.

As insightful as these observations have been, the one-sightline-per-galaxy technique smears out the rich structure seen in simulations of the CGM, where gas accretion and feedback occupy distinct physical regions in galaxy halos and have a widely varying metal content and kinematics. Critical questions remain concerning how this gas enters and leaves galaxies. It appears that galactic star-formation rates and SMBH growth may be limited by the rate at which they can acquire gas from their surroundings, and the rate at which they accumulate heavy elements is limited by how much they eject in outflows, but the importance of these processes and their astrophysical drivers are not quantitatively understood. Without knowledge of how galaxies acquire, process, and recycle their gas, much of the story of how galaxies form and set the conditions for stars, planets, and life will remain unknown. *Hubble* has nearly reached the limits of its ability to answer these questions because of its limited sensitivity and the small number of UV-bright QSOs from which to make statistical maps.

*HDST* will revolutionize the picture of galaxy gas flows by transforming what are now 1-D statistical maps based on *absorption* into richly detailed 2-D maps in *emission* from the diffuse CGM. Hydrodynamical simulations (see Figure 4-7) predict that the energy lost by cooling CGM gas is emitted in the same key UV diagnostic lines (such as Ly$\alpha$ 1216 Å, O VI 1032/1038 Å, and C IV 1548/1550 Å) that are now commonly used to detect the CGM in absorption. With its large collecting area and unique UV access, *HDST* will map the density, temperature, and mass flow rates of the CGM using this "the faintest light in the Universe" over most of cosmic time. Ground-based telescopes attempting to detect this radiation at redshifts $z > 2$–3, where it appears in the visible, struggle to perform the exquisite sky foreground subtraction needed to reveal the faint underlying CGM signal. Sky foregrounds will be considerably lower for *HDST* (by 10–100×), shortening required exposure times by an equivalent factor.

The redshift range of $z \lesssim 1$–2 where *HDST* can map the CGM in emission covers the last 8–10 Gyr of cosmic time (Figure 4-8), when most of the current stellar mass in the Universe was formed, and when the recognizable modern day shapes and sizes of galaxies were established. These imaging studies of the CGM can be coupled to far more sensitive absorption-line studies in the same galaxies. *HDST*'s moderate resolution (R ~5000), wide-field multi-object UV spectrograph can observe up to 50–100 background sources at a time—a feat only possible because of *HDST*'s large aperture, which enables high signal-to-noise spectroscopy of much fainter, more numerous UV background sources. These CGM absorption-line maps will target fields where deep *HDST* imaging identifies filaments in the large-scale structure, and where ground-based ELTs have performed deep redshift surveys to pinpoint the galactic structures and sources of metals to be seen in the CGM. The sensitivity of absorption-line studies will count up the heavy element content of the gas, trace the flows as they are ejected and recycled, and witness their fate when galaxies quench their star formation, all as a function of galaxy type and evolutionary state. The ability to examine the rich structure and composition of the CGM and to relate it to the detailed properties of individual galaxies resolved at ~100 pc or better promises to resolve some of the most compelling open questions about how modern galaxies came to be.

As but one example, *HDST* can address the long-standing mystery of why some galaxies cease to form stars ("quench"). The transition to "passive" evolution has occurred in the most massive galaxies at almost every epoch as far back as $z$ ~3. The number density of non-starforming galaxies increased 10-fold over the past 10 Gyr interval (since $z$ ~2). The process of quenching is likely to be tightly coupled to the feedback that all galaxies experience: the galactic superwinds driven by supernovae and stellar radiation, the hot plasma ejected by black holes lurking in galactic centers (which appears suppressed in quenched galaxies), and the violent mergers that transform galaxy shapes, while triggering the consumption or ejection of pre-existing gas. All of these feedback processes should leave clear kinematic, thermal, and chemical imprints on the surrounding CGM.

*HDST* will have a unique ability to examine the stars and gas in galaxies undergoing quenching. High-definition imaging will lay bare the transformation of star-forming disks to mostly spheroidal passive galaxies at 50–100 pc spatial resolution and closely examine the influence of AGN on this process, resolving the AGN from the underlying disk and/or bulge to unprecedented precision. Emission maps of the surrounding CGM will determine the fate of the gas that galaxies in the process of quenching must consume or eject, and will powerfully elucidate the physical mechanisms that trigger, and then maintain, quenching. Only *HDST*, with its high UV sensitivity and visible-band spatial resolution, can address the co-evolution of stars and gas in galaxies undergoing this transition.

*HDST* will also enable UV spectroscopy of AGN outflows, a primary candidate for both quenching and for coupling the mass of central black holes to the galactic bulges they inhabit. *HDST* will extend into the UV measurements from X-ray observatories (e.g., *Athena*) to provide a global picture of outflow kinetic energy and mass flux across a broad range of gas temperatures and densities. At redshifts of $z$ ~0.2–2, spanning most of cosmic time, *HDST* will access rest-EUV lines from the highly ionized species (including Ne VIII, Mg X, and Si XII) that are likely to trace most of the kinetic energy and mass in AGN outflows. Some of these lines are density sensitive, and thus can be used to deduce the radial location of the absorbing material. This key parameter feeds into estimates of mass outflow rates and kinetic luminosity, but is often uncertain by 1–2 orders of magnitude. Currently only a small number of the rare luminous AGNs have been studied in these lines, but *HDST* will open such studies to the majority of the AGN population. As some of the transitions detected by *HDST* are thought to arise in the same gas traced by X-ray absorption, joint UV/X-ray measurements can be used to characterize in detail outflow dynamics, structure, column density, and radial distance. Outflow studies of AGN samples with well-characterized star formation, reaching out to the peak in cosmic star-formation rate, will provide a direct assessment of the effectiveness of outflows in modulating galactic star formation. This latter component will be revealed in *HDST*'s exquisite imaging, allowing for intensive study of the detailed relationships between galaxies and their central black holes.

Observing these hundreds of AGN to unravel their feedback effects has another benefit: *HDST* is uniquely positioned to detect the long-sought "missing baryons" thought to reside in a very diffuse, high temperature (T > 500,000 K) phase of the IGM. With single-object UV spectroscopy at 50–100× the sensitivity of *Hubble* for point sources (AB ~22 in the FUV), *HDST* will trigger another revolution. There will be > 100× more UV-bright quasi-stellar objects within its reach compared with the limited sample of < 1000 on the whole sky that *Hubble* can observe. *HDST* will even be able to use the much more numerous UV-bright galaxies as background sources. Using this large gain in sensitivity, *HDST* will be able to perform a complete census of gas and metals outside galaxies, to relate those metal budgets to the galaxy properties. *HDST* will again complement X-ray observatories as they too search for the missing baryons. *HDST* can reach weak UV lines of high-ionization species such as Ne VIII or Mg X (among the blue lines in Figure 4-8) that are thought to trace this hot phase if it is metal-enriched, or using broad lines of H I (Lyα) if the hot gas is metal-poor. While next generation X-ray observatories can reach metal-enriched gas at 1 million K, only UV measurements can detect those hot, missing baryons that are not yet enriched with metals, making the two approaches highly complementary. The hot IGM and AGN outflows are both UV and aperture drivers: with a massive gain in UV collecting area and novel spectroscopic capability, *HDST* should be able to map the gas flows from the IGM to the CGM to the ISM and back again, and to fully unravel all the processes by which galaxies acquire, process, and recycle their gas to form stars, planets, and life.

### 4.2.2 How Stars Disperse their Products: *HDST* and the Transient Sky

The death of stars is a critical part of the story of cosmic birth. All the elements of life were produced in stellar interiors and then dispersed by winds or supernova explosions. Even long after massive stars have reached their catastrophic demise through core-collapse, the compact remnants left over (white dwarfs, neutron stars, and black holes) can merge with one another to create other explosive phenomena observed as supernovae (SNe) and gamma-ray bursts (GRB). Understanding the physics of stellar explosions and nucleosynthesis requires relating these explosive events back to the stars that produced them. *Hubble* has played a crucial role in matching stellar explosions to their progenitor populations, by placing them in the context of their local star-forming environment (or lack thereof), and in fortuitous cases, pinpointing the ill-fated star itself in archival pre-explosion imaging. The UV sensitivity and spatial resolution of *HDST* will push this capability to higher redshifts, where the next generation of deep all-sky surveys will make the majority of SNe and GRB discoveries.

One of the most rapidly growing areas of astrophysics in the next decade will be multi-messenger time domain astronomy. By the mid-2020s, several powerful ground-based wide-field synoptic surveys will be in full operation, including the Large Synoptic Survey Telescope (LSST) at optical wavelengths, the Square Kilometer Array (SKA) at radio wavelengths, and the Advanced LIGO and Virgo detectors for gravitational waves. These surveys will explore a new discovery phase space (see Figure 4-9), populating it with new classes of transients and pushing to fainter limits (and thus higher redshifts) on known classes of transients. Prompt (~24 hr) target of opportunity (TOO) capabilities with *HDST* will enable the precise localization and follow-up of transients over the full range of luminosities and timescales shown in Figure 4-9. At the redshift limits shown for single visits by LSST (one night's observation), *HDST* will reach localization precision sufficient to identify the stellar populations giving rise to the explosive phenomena, and with its spectroscopic capabilities, measure the metallicity local to the explosion (an important driver of stellar evolution and mass loss). The rest-frame UV capability of *HDST* is particularly important for probing young (hot) SNe, and can be used as a sensitive diagnostic of their progenitor mass and radii, and explosion energies. The UV and optical diagnostics for the age and metallicity of the progenitor star and its host population are more effective than the near-IR diagnostics that would be used by ground-based telescopes, especially at young ages. The deep magnitude limits of *HDST* are not achievable from the ground, and *HDST* will be the only telescope capable of following the evolution of a faint optical/near-IR counterpart (postulated to be a kilonova) to a localized gravitational wave source for long periods after the event.

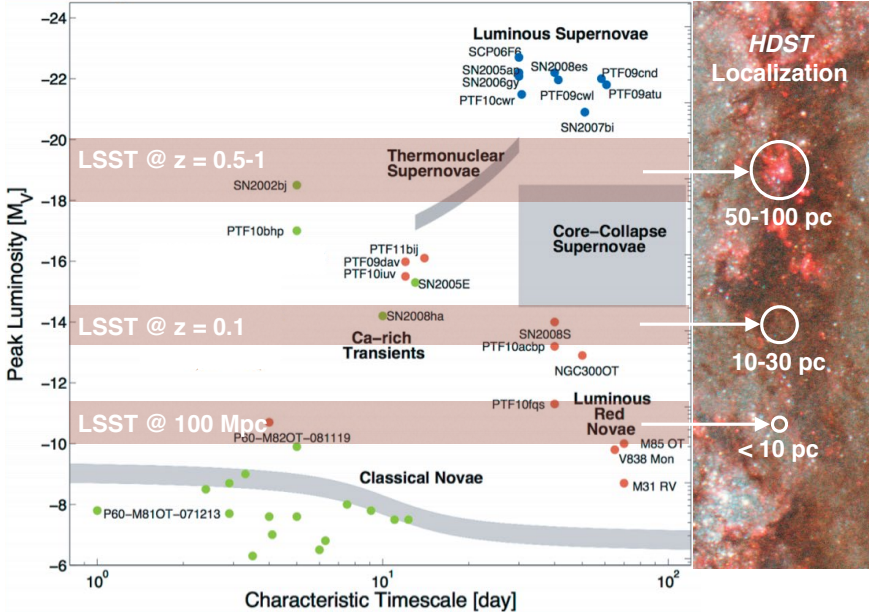

Figure 4-9: The energy/timescale parameter space for the transient sky. In single-night visits, LSST will reach events at the luminosity limits shown by the pink bands, e.g., events at $M_v \sim -19$ will be detected out to $z = 0.5–1$, and *HDST* will localize these events to within 50–100 parsec precision within their host galaxies. Localization becomes 10 parsec or better, the size of a single molecular cloud region, for events detected out to 100 Mpc. Credit: M. Kasliwal (Caltech).

The most common variable extragalactic source detected by time-domain surveys will be active galactic nuclei. Any variability associated with accretion activity onto supermassive black holes (SMBHs) is intrinsically bright in the UV, and will have the highest contrast to underlying starlight in the UV. High-resolution UV imaging with *HDST* can pinpoint the positions of transients relative to their host galaxy nuclei, and confirm rare and interesting SMBH configurations discovered by wide-field surveys, including flares from the tidal disruption and accretion of individual stars by previously dormant SMBHs, periodic modulations in accretion due to binary SMBH systems, or spatially offset AGN marking a recoiling SMBH from a binary merger.

## 4.3 The Epoch Imprinted in the Galactic Neighborhood

The story of Cosmic Birth does not end when the Sun forms. Processes that lead to life occur on scales smaller than can be seen at any significant redshift. Other more elusive processes can only be studied through the "fossil record" that they leave behind. To address these phases of Cosmic Birth, our telescopes must probe galaxies whose stars and gas can be best resolved.

Galaxies in the Galactic Neighborhood, out to about 100 Mpc, are in a comparable epoch to that of the Milky Way, and are sufficiently close that their constituent parts can be disentangled with ease. Observations can trace their internal motions, the relationships between their stars and their gas, and their population of faint satellites. This local volume contains a fair sampling of almost all astrophysical phenomena, allowing us a front-row view of the processes that shape galaxy evolution at every redshift.

Within the local neighborhood, *HDST* will resolve galaxies and their stellar populations down to ~10 pc at 100 Mpc, ~1 pc at 10 Mpc, and ~0.1 pc at 1 Mpc (i.e., everywhere in the Local Group). These thresholds will transform our ability to track the internal motions of galaxies, to reveal the detailed histories of their stellar populations, and to uncover the evolutionary relationships between galaxies' stars and their gas. These extraordinary gains in resolving power can take almost any study done in the Milky Way or its neighbors, and carry it out in systems with drastically different properties. For example, is the stellar Initial Mass Function (IMF) the same in galaxies with much more intense star formation than the Milky Way? How is the small-scale clustering of gas and stars shaped by the large-scale properties of the host galaxy? This section covers just a few of these possible investigations, many of which bear directly on how galaxies and stars set the ultimate conditions for life.

### 4.3.1 Building Galaxies from Stars and Clusters

The Sun, like most stars, formed in a stellar cluster that condensed from a single massive molecular cloud. Clusters contain from a few hundreds to hundreds of thousands of stars. The life-cycle of clusters is typically fast, forming quickly over a few million years, and then often dispersing on comparable timescales when their host molecular cloud dissipates under the force of ionizing radiation and explosions of newly formed massive stars. As the birthplace of stars, star clusters are a critical step on the pathway from cosmic gas to life.

Using a wide-field imaging camera to produce precise multiband photometry, and with ~1 parsec spatial resolution for thousands of galaxies, *HDST* imaging can precisely weigh and age-date star clusters down to a few thousand solar masses. These studies rely on the UV/optical coverage, wide

> **Big Questions**
>
> *How do stellar clusters build galaxies?*
>
> *How do massive stars influence their galaxies?*
>
> *How does the IMF vary with environment?*
>
> *How does dark matter drive galaxy evolution?*

fields of view, and stable PSFs that space telescopes can provide. *Hubble* has pushed these techniques to their limits to dissect all the young clusters visible in very nearby galaxies, such as M51 and the Antennae. *HDST* will enable comparable studies all the way out to 20 Mpc, spanning over 125× the volume that *Hubble* can probe. This is galaxy formation in action. For these same clusters, efficient spectrographs on ELTs will obtain detailed chemical abundances, and ALMA and SKA will probe the remaining molecular and atomic gas from which the stars formed, with comparable spatial resolution. With all these facilities working in concert, it should be possible to tell the complete story of how stellar clusters formed and dispersed their stars throughout their host galaxies.

*HDST*'s UV spectroscopic capabilities will also answer outstanding questions about hot, massive stars. These rare but important stars drive the mass flows that act as feedback on galaxies and provide a large share of the heavy elements from which life is built. Yet the ways in which massive stars vary in these behaviors over cosmic time and metal content are very poorly known, because the UV spectrographs used to study them cannot reach far beyond the Local Group, where the range of heavy element content is still narrow. While *Hubble* can examine individual massive stars in the low-metallicity Magellanic Clouds, a thorough mapping of massive star properties in chemically primitive galaxies will require a much greater reach out to individual massive stars in galaxies at 20–50 Mpc. Over this interval, *HDST* will have spatial resolution of ~1 pc in the UV, and will be able to collect intermediate resolution spectroscopy for hundreds of individual OB stars and clusters in such metal-poor galaxies as I Zw 18 (d = 18 Mpc), which has a metal content only a few percent of the solar ratio and serves as a local proxy for such primitive galaxies at early cosmic times. Only *HDST*'s great leap in UV spectroscopic capability will be able to map out the mass function, rotation velocities, stellar winds, and radiation fields of massive stars over a range of conditions expected throughout.

Complementing its UV spectroscopic capabilities, *HDST*'s panchromatic imaging of resolved stars in nearby galaxies will address galaxy quenching from another. The complexity of the observed phenomena is daunting: it appears that galaxies at both extreme ends of the mass function formed stars only early in cosmic time, and then stopped. This is true for the giant elliptical galaxies in the Galactic Neighborhood and for the smallest "Ultra-faint Dwarfs" orbiting the Milky Way, but the galaxies in the middle range continue to form stars for essentially all of cosmic time. Using color-magnitude decomposition of fully resolved UV, optical, and near-IR stellar populations, *HDST* will be able to map the full star-formation history of any galaxy out to ~10 Mpc, while fully accounting for dust and tracing the history of metal enrichment. This large volume includes the nearest giant elliptical galaxies. At the other extreme, *HDST* will easily measure the smallest and most ancient galaxies out to the 400 kpc edges of the Milky Way halo, where LSST is expected to find hundreds of additional, presently invisible dwarf satellites.

### 4.3.2 Galactic Feeding and Feedback in High Definition

For nearly two decades, the flow of gas into and out of galaxies has been recognized as one of the key physical processes shaping galaxies' evolution. As described in 4.3.1, *HDST* will provide a groundbreaking opportunity to trace the gas flows in the ISM and CGM in exquisite detail. However, *HDST* also opens up genuinely new possibilities for examining gas flows in a resolved fashion in nearby galaxies. In the *HDST* era, the orders of magnitude increase in UV sensitivity over *Hubble* will transform such studies from statistical exercises to true physical mapping of gas within individual galaxy halos. All of the galaxies within the Galactic Neighborhood will have 10 background QSOs that are bright enough to probe variations in gas density, temperature, metal enrichment, and kinematics from place to place within the galaxies' halos, which is currently only possible for the very closest massive galaxy, M31. *HDST* will open this kind of study to galaxies within 20 Mpc, which includes the exact same galaxies for which star-formation histories are measured from resolved and semi-resolved stellar populations, and for which *HDST* can measure AGN properties, and perhaps even proper motions. Thus, for the first time, *HDST* can link the time-resolved history of energy input from stars and AGN to the properties of the surrounding gaseous halo. Technically, this project requires sensitive UV point-source spectroscopy at wavelengths down to ~900–1000 Å to allow detection of the critical diagnostics of the Lyman series lines and the 1032, 1038 Å doublet of O VI. This will be the most detailed possible comparison between galactic star formation, AGN activity, and the gas that fuels it, providing superb constraints on the uncertain feedback included as ingredients in all modern galaxy-formation simulations.

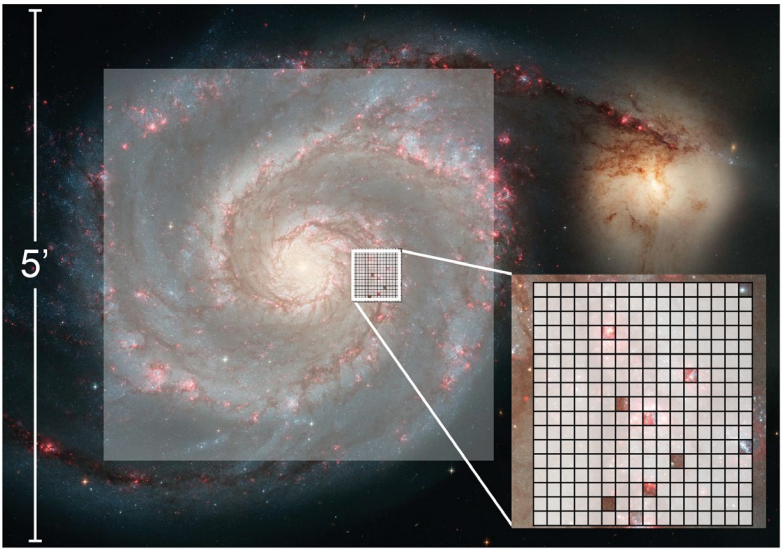

Figure 4-10: Dissecting galaxy outflow and inflow requires UV multi-object spectroscopy. A UV MOS mode with 50–100 objects observable at once in a 3–5' FOV will enable efficient studies of young stellar clusters, ISM gas, and energetic feedback for thousands of clusters in dozens of galaxies in the nearby Universe. Together with star-formation histories and halo gas, these observations will dissect the past evolution and current activity of nearby galaxies. This capability will also enable intensive studies of the circumgalactic medium, young stars in Galactic clusters, and many other frontier science problems.

Beyond mapping the connection between halo gas and star formation, *HDST* can peer into the individual sites where winds are thought to be launched. An efficient UV multi-object spectrograph (MOS) will provide revolutionary access to the outflows driven by young stellar clusters, giant star-forming regions, and AGN. Assuming that such an instrument can cover ~50 sources over 3–4 arcmin FOV (comparable to *JWST*'s NIRSpec), it could measure UV absorption towards hundreds of individual clusters or AGN within nearby galaxies in a matter of hours, using the wide range of physical and abundance diagnostics available in the rest-frame UV (Figure 4-10). These observations would measure the outflows of gas driven by the radiation pressure and supernovae from these young clusters, while simultaneously characterizing the stellar clusters' mass, age, metallicity, and connection with local ISM properties. Such measurements are possible for *Hubble*, but would require ≫100 orbits to map even a single galaxy with the same density of sources. In contrast, cluster and down-the-barrel outflow measurements of hundreds to thousands of clusters in nearby galaxies could be done in < 100 hours with *HDST*, making it possible to fully characterize the sources of heavy elements and galactic feedback for a representative sample of galaxies. Moreover, these measurements will be done for the same < 10 Mpc galaxies with measured star-formation histories and halo-gas measurements, thus completing the loop of galactic recycling from the disk, to the halo, and back, in a fully resolved way. These observations would take gas flows out of the regime where they are knobs for tuning in simulations, and instead would link them to measurements that constrain the fundamental energetics of this critical astrophysical process.

### 4.3.3 Stellar Populations and the Mass Function of Stars

The distribution of elements necessary for building life and the number of low-mass stars optimal for hosting habitable worlds are all controlled by the relative numbers of stars of different masses. Thus, the cosmic pathway leading to life runs straight through the stellar initial mass function (IMF), the mass distribution of newly born stars. The IMF of massive stars also controls energetic feedback within and outside galaxies at nearly all physical scales, and so plays a central role in astrophysical problems far beyond the scale of individual stars. It is not clear which physical processes interact to produce the IMF observed locally, nor is the extent to which this local IMF is universal, even within the Milky Way itself. To unravel this complex mixture of influences and outcomes, *HDST*'s superb image quality and stability can measure the IMF in a wide range of galactic environments, and over a long span of cosmic time. Specifically, *HDST* can: (1) measure the Milky Way disk IMF (and the masses of all dark stellar remnants) with microlensing; (2) extend star-count IMFs out to the edge of the Local Group, in new regimes of galaxy mass and metallicity; (3) constrain the IMF in distant galaxies using surface brightness fluctuations; and (4) measure the high-mass IMF in stellar clusters.

*The IMF through Astrometric Microlensing:* The high spatial-resolution imaging of *HDST* will enable a new microlensing technique for measuring the mass function of stars in the Milky Way down to the hydrogen-burning limit (0.08 $M_\odot$), and for simultaneously measuring the mass distribution of dark stellar remnants—black holes, neutron stars, and the coolest white dwarfs. Microlensing surveys watch for the sudden brightening of stars caused when foreground objects gravitationally lens a background star. The technique is sensitive to the mass of the foreground object, which need not be luminous for its influence to be detected. With its exquisite spatial resolution (~10 mas in the optical) and unprecedented stability, *HDST* will be able to break the typical degeneracy between the mass and the velocity of the lens by also measuring the astrometric deflection of the background source when the lens passes by. These astrometric shifts are tiny: 0.1 mas for a lens at the hydrogen-burning limit and less than 1 mas for every lens mass up to 10 $M_\odot$. Yet, as small as these shifts are, they are within reach of a 12 meter telescope with 3–5 hour integrations (V = 27–28 at S/N = 300). By monitoring dense stellar fields in the galactic bulge, *HDST* can detect hundreds of such events in a single monitoring campaign. Deep follow-up imaging will even be able to detect any lensing object that is a living star (down to 0.1 $M_\odot$) once it has moved away from the background source. If detected, the lensing star will have its luminosity tied directly to its mass (with a possible mapping to metallicity), and if it is not detected, the lens will be identified as a dark remnant—which have still not been unambiguously detected outside binary systems. Only *HDST*'s unique combination of precise astrometry with deep, wide-field imaging will allow these measurements of the mass function for living and dead stars.

***The IMF through Star Counts in Diverse Environments:*** The most reliable measures of the IMF come from simply counting stars in images in which they are fully resolved, and their brightness can be precisely measured and then mapped to stellar mass. With *Hubble*, star-count measurements of the low-mass IMF have been extended as far outside the Milky Way as the Magellanic Clouds and the Ultra-Faint Dwarfs, and already there are emerging signs of a dependence of the IMF on either galaxy metallicity or mass, with smaller, more metal-poor galaxies forming relatively fewer low-mass stars. Likewise, star-count measurements of the high-mass IMF were carried out in M31 and M33, but can go no further due to the inability of *Hubble* to resolve stars in the centers of young, massive clusters.

With its exquisite optical-image quality, one-hour exposures with *HDST* will be sufficient to detect the faintest M dwarfs (~0.1 $M_\odot$) to distances of 100 kpc from the Sun. In 10-hour integrations, *HDST* will reach down to 0.2 $M_\odot$ everywhere within ~400 kpc of the Milky Way, and to 0.5 $M_\odot$ anywhere in the Local Group (to ~1 Mpc). Likewise, with its dramatic improvement in resolution, the high-mass IMF can also be extended to resolve young cluster stars beyond the confines of the Local Group. Within these greatly expanded volumes reside the whole population of satellite galaxies of Andromeda and the Milky Way, with a diverse range of metallicity and of current and past star formation, and the much larger population of massive galaxies beyond. Multi-epoch observations will yield parallaxes, and hence distances to these same stars in the Local Group, anchoring their masses with secure measurements of their luminosities. By placing the IMFs of these objects within reach, *HDST* will finally and definitively determine if both the low- and high-mass IMF is truly universal, or if it varies with environment.

***Semi-resolved Stellar Populations:*** Observations with *Hubble* have revealed the potential of *semi-resolved stellar populations* where the stars are too crowded to resolve individual low-mass stars, but where a small number of rare, luminous giant stars produce pixel-to-pixel brightness variations on top of the smooth background of fainter stars. The power of semi-resolved stellar populations is just coming into its own. So far, it has been used to estimate distances to galaxies in the Galactic Neighborhood that depart significantly from the Hubble flow, and for which recession velocity is not a reliable measure of distance.

However, the semi-resolved universe offers us much more than a yardstick. In principle, one can statistically reconstruct a color-magnitude diagram of evolved stars in a distant galaxy. Even 30 m telescopes will be unable to resolve stars fainter than the upper giant branch in the Virgo cluster. Analysis of semi-resolved stellar populations will allow us to push to greater distances and into the cores of nearby massive ellipticals—providing new windows into some of the most extreme environments in the local universe.

Observations of this effect, often called "surface brightness fluctuations," with *Hubble* have been obtained for ~200 galaxies in the local volume. The semi-resolved signal scales as the number of stars per resolution element, so a factor of 5 increase in aperture size would enable a ~25× more sensitive measurement for the same galaxies, and would allow the same type of measurement afforded by *Hubble* in a volume ~125× larger than current samples, encompassing a cosmologically significant volume out to ~100 Mpc.

Using diffraction-limited spectroscopy to analyze Poisson fluctuations in the number of giant stars per pixel will be an extremely powerful combination. Analyzing pixels that contain few/no giants yields a relatively unobstructed view of the fainter main-sequence stars, whose characteristics are well understood. This would allow for accurate metallicity and age estimates of distant stellar populations, which are best derived from temperature-sensitive blue optical light. Spectroscopy of the warmer turnoff stars would provide abundance analysis of elements that are otherwise very difficult to obtain, such as the *s*- and *r*-process neutron-capture elements. Moderate-resolution spectroscopic analysis at the pixel-by-pixel level would also provide high-precision kinematics, as the finite number of giants per pixel entails that the usual line-of-sight velocity distribution would not convolve the observed spectrum. In this regime, the velocity dispersion of the system would not limit the velocity resolution. Pixels with few giants will also provide much more accurate measurements of the stellar IMF for a wide range of galaxy types.

### 4.3.4 Unraveling Galactic Dynamics with Proper Motions

The mysterious dark matter that dominates the mass of the Universe binds and organizes all large-scale cosmic structure. The first seeds of galaxies, stars, and ultimately life were pulled together by small concentrations of dark matter shortly after the Big Bang. Dark matter has continued to drive the dynamics of galaxies ever since.

The nature of the dark matter remains unknown, and yet its gravitational influence on the dynamics of stars and galaxies is clear. The motions of stars on the sky (their proper motion) are sculpted by the underlying dark-matter-dominated gravitational potential, and thus they reveal the true structure and movements of the galaxies those stars inhabit. *Hubble* has pushed this astrometric proper-motion technique to its limits, searching nearby stellar clusters for signs of black holes and tracking the orbits of galaxies in the Local Group. *JWST*'s comparable precision will allow this work to continue, yielding longer time baselines for fields previously studied with *Hubble*. While the forthcoming *Gaia* and *WFIRST* space missions will measure proper motions over huge fields, revealing internal Milky Way dynamics with billions of stars, even these major advances are not enough to reveal the true nature of dark matter or to resolve the detailed motions of distant galaxies.

In contrast, *HDST* has the spatial resolution at visible wavelengths to achieve an entirely new level of precision in measuring the effects of dark on normal matter. The key to these gains is its aperture: at 12 meters, with 10 mas spatial resolution, millipixel astrometry, and rapid high-S/N photometry, *HDST* will be able to measure the motions of stars on the sky to unprecedented low limits: 100 centimeters per second out to 100 parsec (a human walking pace), 0.1 kilometers per second (a race car) anywhere in the Milky Way, and 10 kilometers per second (an orbiting spaceship) anywhere in the Local Group. At the limits achievable in the Milky Way, virtually every star can be seen to move, making it possible for *HDST* to measure the dynamics of virtually any Galactic stellar population, young or old.

Measurements of this precision can directly constrain the nature of dark matter. In the ultra-faint dwarf satellites of the Milky Way (and the hundreds of additional dwarfs that LSST is likely to reveal), baryons contribute essentially nothing to the mass of the galaxy, making their stars ideal tracer particles of the dark matter potential. The micro-arcsec per year precision can reach the velocity dispersions of the dynamically coldest structures in the Milky Way system, such as the ultra-faints and stellar streams, in a way that is not possible with a significantly smaller telescope. Further afield, *HDST* can use proper motions to measure the mass of the black hole at the center of Andromeda—which will eventually merge with the Milky Way's own. And further still, *HDST* can measure the proper motions of entire galaxies, such as the giant elliptical galaxies in the Virgo Cluster (at 15 Mpc). These observations will lay bare the true dynamics of local galaxies, write the early formation history of our own Milky Way, and constrain the nature of dark matter with radically transformed vision.

> ### Capability and Serendipity Drive Discovery
>
> The great French chemist Louis Pasteur famously stated that "In the fields of observation chance favors only the prepared mind." Pasteur's aphorism captures a fundamental truth about discovery in science, but he could not have known that he was also telling the story of the great space observatories to be launched 150 years after his remark. The history of the *Hubble Space Telescope* amply illustrates that an observatory's greatest achievements can be unknowable at its conception, design, and launch. Although *Hubble*'s planned science goals—to measure the expansion rate of the Universe, study distant galaxies, and map the intergalactic medium— were met by its early Key Projects, any list of its greatest hits now includes discoveries that none of its planners could have imagined, and solutions to questions they could not even have posed. *Hubble* has proven that the expansion of the Universe is accelerating, found supermassive black holes lurking in the core of nearly every galaxy, traced the orbits of nearby galaxies, age-dated the oldest stars, and detected the atmospheres of exoplanets. These discoveries echo Pasteur's truth if the "prepared mind" is equated to the immensely capable instruments *Hubble* carried—all of which were improved many-fold over prior generations. Order-of-magnitude strides in instrumental capability lead to serendipitous discovery, even if the questions originally driving that capability fade and are replaced by new problems over time. Keep this lesson in mind: while *HDST* will deliver on the goals envisioned here, it may in the end be most remembered for things that cannot be foreseen. This is why each leap in capability must be a large one, and must be sustained over time—if it is designed only to solve narrow, foreseeable problems, an observatory will not prepare our minds to enjoy the favor of chance that leads to discovery.

## 4.4 Star and Planet Formation in the Milky Way

The stars themselves occupy center stage in the grand story of Cosmic Birth. For life as we know it to exist, stars must form over a wide range of mass, from less than 1 $M_\odot$ to > 10–100 $M_\odot$. Intermediate- and high-mass stars have been creating the carbon, nitrogen, oxygen, calcium, iron, and other elements that form the basic building blocks of life since the beginning of time. Massive stars and their supernovae drive many of the gas flows that make galaxies look as they do, and may govern whether or not galaxies continue to form stars. And of course, low-mass stars and their planets form together, stars bathe their planets in light and warmth to make them habitable and drive the energy gradients that give rise to life. Understanding why the stars have the masses they do is part of the story of how the materials making up all life came to be. *HDST* will make fundamentally new contributions to our understanding of how stars acquire their mass and generate the IMF (4.5.1), how protoplanetary disks form, evolve, and dissipate (4.5.2), and how stars can influence the detectable properties of their atmospheres in ways that are relevant for biosignature-gas detection (4.5.3).

### 4.4.1 How Stars Get Their Mass

Stars form in dark, self-gravitating clouds of gas and dust. Much of their evolution is hidden within highly reddened, dark cloud cores that are best studied in the far infrared or radio bands. At later stages, however, young stars emerge from their dusty cocoons, becoming visible in the UVOIR bands. These newly revealed stars often exhibit protoplanetary disks, debris disks, or young families of planets. This evolutionary phase is thought to be very fast, with only a few million years elapsing between dense molecular clouds, through protostellar cores, to thriving self-powered stars and planets. Major questions still remain about how stars acquire their final mass and how much mass remains to form planets. Still more complex are the questions about how environmental factors such as galaxy mass, gas metallicity, and other local conditions influence the final stellar-mass distribution and its influence on planets. The finding that planets are more frequent around stars of higher metal content highlights the question of how stars form in environments of different metallicity; regions that form high- and low-mass stars in different proportions may make planets of different compositions and varying chances of making life.

> **Big Questions**
>
> *How do stars get their mass?*
> *How do planets form in disks?*
> *How do Solar Systems vary in composition?*

The distribution of stellar masses is a perennial grand prize in astrophysics that can truly only be claimed by counting individual young stars at all masses directly in the environments of interest. The need to resolve individual stars in dense star clusters in a representative range of galactic environments in the Milky Way, its satellites, and the rest of the Local Group is a key aperture driver for *HDST*. The question of *why* the IMF varies is best addressed by measuring the rates at which stars acquire mass while they are still forming. Recent findings indicate that UV emission is an excellent proxy for accretion rate; as it moves from the protostellar disk to the star, the accreting material shocks, heats, and produces a characteristic continuum emission and broad UV emission lines. Observing these lines and continuum features is a key UV driver for *HDST*, which can measure the time-varying accretion rate onto the star self-consistently across the UVOIR with the present stellar mass, and thus explain how stars reach their final mass and planets form.

Progress in this field requires performing similar observations in younger clusters, and especially in the Magellanic Clouds, to explore star formation at the low metallicity comparable to what the Milky Way had ~10 Gyr ago (Figure 4-11). Pushing *Hubble* to its limits, it is possible to obtain broad-band photometry of ~0.2 $M_\odot$ pre-main-sequence stars in the LMC, but only if the stars are relatively isolated. In stellar clusters, where most stars are born, stellar crowding puts measurements of individual stellar masses and accretion rates far beyond present capabilities. For instance, the LH95 cluster in the LMC appears to contain about 100 stars per square arcsecond, a density that corresponds to only 4 pixels for every star observed with *Hubble*'s Advanced Camera for Surveys (ACS). *JWST* will deploy higher spatial resolution, but not in the critical ultraviolet range where stellar accretion signatures are most robust. At 12 meters in aperture, *HDST* should be able to observe these regions with a comfortable 100 pixels for every star, fully resolving all the stars, producing robust star-count IMFs, and measuring accretion rates down to ~0.1 $M_\odot$ throughout the Milky Way, the Magellanic Clouds, and beyond.

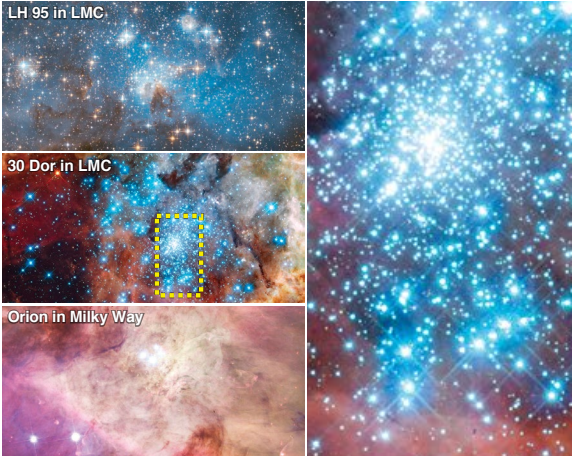

Figure 4-11: Most stars form in clusters like these. *Hubble* can resolve individual stars in the nearest massive star-formation region to Earth (Orion, at bottom). As shown in the zoom-in panel at right, dense clusters in the low-metallicity Magellanic Clouds, which better resemble conditions in the early history of the Milky Way, cannot be resolved into individual stars for star-count IMFs or UV-derived mass-accretion rates. *HDST* will fully resolve such clusters and directly measure the shape and causes of the IMF in these chemically primitive environments.

### 4.4.2 How Planets Form in Disks

Under the right conditions, the gas that stars do not acquire for themselves can settle into a protoplanetary disk, where the smallest dust grains adhere into progressively larger grains and pebbles until they form rocky planets and the cores of gas and ice giants. The enormous diversity in exoplanetary systems discovered so far shows that there are many paths to the formation of planets and many variations in their subsequent dynamical evolution, which may reflect diversity in the initial protoplanetary disk conditions. There are currently few constraints on the number, masses, and orbits of planets depend on the properties of the host star, the composition of the pre-stellar nebula, the dynamical or radiative influence of other nearby stars, or other undiscovered factors. Progress requires large suites of observations of proto-planetary and debris disks in a range of conditions.

For *HDST*, exploring young planetary systems is both an aperture driver and UV/optical driver. Nearby protoplanetary disks in the Milky Way span 10–50 AU and evolve on Keplerian timescales of ~1–100 years. At 12 meters, and with spatial resolution of < 1 AU out to 100 parsec and < 10 AU out to 1 kpc, *HDST* will resolve disks, see the gaps and instabilities introduced by nascent planets, and detect individual planets in their own right. Over 5–10-year timescales, *HDST* will be able to follow the motions of planets and disk features as they develop; these motions encode the current dynamical state of the system and help predict what it will look like when it is a mature planetary system. These optical and near-IR measurements will complement the maps of the remaining molecular gas and dust to be obtained with the fully operational ALMA at longer wavelengths, where it has comparable spatial resolution (~10 mas). *HDST*'s spatial resolution of ~1 AU for the nearby star-forming complexes will be able to trace conditions in warm and/or diffuse gas in the inner disk (≲5 AU). *HDST* and ALMA will therefore complement one another in the exploration of protoplanetary disks at all scales.

*HDST* will also relate the properties of planets to the properties of their disks. Using its unique UV high-resolution spectroscopic capability (R ~100,000, 0.1–0.3 micron), *HDST* can examine the molecular gas content of protoplanetary disks using direct tracers of the mass: molecular hydrogen ($H_2$) and carbon monoxide (CO), both of which have numerous emission and absorption bands in the far UV that are intrinsically far stronger than the molecular emission bands that *JWST* will access in the near-IR. These disks are thermally and kinematically cold, so high resolution is required to resolve these molecular emission lines (FHWM of a few km s$^{-1}$) and from them to infer the disk-density profile, temperature, kinematics, and mass transfer rates. With *HDST*'s exquisite spatial resolution, which corresponds to ~1 AU for the nearby Galactic star-forming complexes, it should even be possible to perform these measurements on individual disks in a spatially resolved fashion.

Rocky planets and life are made of heavy elements processed by billions of years' worth of galactic accretion, star formation, ejection, and recycling. How do the abundances of these elements influence the planets formed in these systems? The composition of interstellar gas that makes protostellar clouds can be heavily modified by relative depletions onto dust by the time it makes it into the disk and planets, and the exact composition of the disk has large effects on the resulting planets.

Addressing this issue requires relating the metal content of nascent planetary systems to the observed planets and disk. For many nearby young, gas-rich systems, at the point when the central star becomes visible in UV light, there is still sufficient gas contained in the disk, especially if it has a flared structure, that the star can be used as a background source to measure relative chemical abundances in absorption by the foreground disk gas. With its limited UV sensitivity, *Hubble* struggles to reach the brightest nearby systems. The high UV sensitivity of *HDST*, 50–100× that of *Hubble*, will bring within reach potentially hundreds of systems within 300 pc. *HDST* should be able to observe enough systems to examine the composition as a function of height in the disk, and with stars of different ages, giving a 3D time dependent picture of how disks evolve toward planetary systems.

The formation of planets from protoplanetary disks may also depend sensitively on how long the gas and dust remain in the disk during planetary accretion. The formation of rocky worlds, as opposed to gas and ice giants, may depend on the relative abundances of gas and dust in the disk. Protoplanetary disks are complex structures of gas and dust that evolve rapidly, forming planets and then reacting dynamically and chemically to the growth of those planets. In regions where dust has been consumed or dissipated, diffuse molecular gas is the chief observable signature of disk composition and dynamics. The gas that remains appears in far-UV rotational-vibrational emission bands of $H_2$ and CO that arise when the molecules are excited by light from the central star. *Hubble*'s efficient Cosmic Origins Spectrograph has accumulated a sample of young stars, but each must be individually observed for several hours, and so large samples are prohibitive. Moreover, COS cannot fully sample a large range of cluster age, because more distant regions are out of reach. With its wide-field UV MOS capability (3–5 arcmin FOV, R ~5,000), *HDST* will be able to survey hundreds to thousands of young stars in Galactic star-forming regions over a wide age range throughout the Milky Way. Combined with robust UV/optical photometric estimates of protostellar accretion rates (as described in 4.4.1), these gas-disk spectra will reveal the critical late stages of protoplanetary disk evolution, when the gas and dust is dissipating and planets are reaching their final mass. These UV observations also complement observations of colder gas and dust appearing in the infrared, radio, and submillimeter for a complete picture of how planets form. Characterizing the molecular gas content, elemental abundances, kinematics, and mass flows within protoplanetary disks will be a key element of understanding the environments in which planets form, and a major step on the path from Cosmic Birth to Living Earth.

### 4.4.3 Stellar Influences on Planetary Habitability

As discussed in Chapter 3, the driving aim of *HDST* is the detection of statistical samples of Earth-like planets in their habitable zones and identification of biosignature gases in their atmospheres. Recent work indicates that the host star influences the habitability of such worlds beyond merely keeping water in the liquid phase and powering metabolic processes. *HDST* will search for life around main-sequence stars of the F- to M-spectral types. These stars, particularly the later spectral types, continuously irradiate their planets with strong fluxes of UV and X-ray photons, and episodically bombard planets with higher energy photons from stellar flares associated with coronal mass ejections (CMEs) and magnetospheric activity. The incident flux of UV/X-ray radiation and energetic particles affect the atmospheric chemistry of water ($H_2O$) and carbon dioxide ($CO_2$), and in some circumstances can promote *abiotic* production of important biosignature gases, including molecular oxygen ($O_2$), ozone ($O_3$), and nitrous oxide ($N_2O$). The photochemistry of these molecules depends sensitively on the shape and time variability of the UV/X-ray emission. Energetic particles from CMEs associated with flares can substantially reduce the quantity of observable ozone, even if it is biological in origin, for periods of weeks to years. Thus for careful evaluation of potential biosignatures on exoEarths, it will be necessary to characterize the far- and near-UV emission of their host stars with sensitive UV monitoring close in time before and after the planet atmosphere observations. Such observations are difficult and expensive for *Hubble*, which has the UV sensitivity to reach only a small percentage (< 5%) of the stars already identified with planets in the nearby Galaxy, and which cannot reach M stars beyond 15 pc in less than 20 hours. *JWST* will not be able to address this problem at all. Yet assessing the possibility of false positives for biosignature detections is an essential component of any intensive exoplanet characterization effort.

This problem is a strong UV driver for *HDST*. With its exquisitely sensitive UV spectroscopy, enabled by a 25× *Hubble* collecting area and photon-counting low-background detectors, *HDST* will be able to survey stars for their UV emission properties, flare frequency and strength, and the UV radiation flowing through the stars' habitable zones. The reach of *HDST* will encompass all the planet host candidates for its own planet surveys, including M dwarfs to 50 parsec and K stars to 200 parsec. This study will make use of intermediate-resolution UV spectroscopy of point sources (0.1–0.3 micron, R ~30,000), much like *Hubble*'s Cosmic Origins Spectrograph, but with up to 50× more raw sensitivity for the important far-UV and near-UV emission lines that serve as proxies for total stellar

UV and X-ray emission (e.g., C III 977, O VI 1032, H I Lyα, Fe XXI 1354, C IV 1550, and He II 1640). Thus two unique *HDST* capabilities—exoplanet characterization and UV spectroscopy—mutually reinforce one another in the search for life beyond Earth.

## 4.5 Our Solar System

At the very smallest scales probed by *HDST* lies our own planetary system, the best-studied example of the end state of the formation process, and the most powerful laboratory for testing theories of planetesimal-to-planet formation and internal structure, end-state habitable-zone development, migration and tides, atmospheric stability, composition, and stellar-planet interactions.

It is a significant challenge to extrapolate the diversity of objects in the Solar System and the exquisite detail with which they can be observed and used as proxies for objects only hinted at in other systems. Remote sensing has been enormously useful, both for improved understanding of the individual objects and for probing their origins and subsequent evolution. Systemic study is the key to narrowing the parameter space of planetary formation and how their modern states respond to their external environment. From this also comes greater understanding of how evolutionary processes operate, and how they play out for worlds in other star systems. Such investigations were a cornerstone of the most recent decadal survey (Building New Worlds).

Human exploration of the Solar System has been a joint enterprise between targeted *in situ* missions and remote study. The perspective from Earth has benefited from more capable instrumentation, along with the ability to observe over a much longer time period. Moreover, the sheer number of Solar System targets dwarfs our resources for visiting them, and may lie beyond the easy reach of spacecraft. *Hubble* has played a critical role in remote studies of the Solar System by providing wide-field imaging, reduced background, and the elimination of atmospheric absorption in the UV. This has allowed it to "punch above its weight" for its size. Its UV capability has been particularly useful, because reduced solar continuum enables higher contrast

> **Big Questions**
>
> *How did the Solar System's protoplanetary disk evolve?*
>
> *What is the nature of objects beyond the orbit of Neptune?*
>
> *What physical processes determine planetary surface and atmospheric conditions?*
>
> *How do planets and their satellites interact with the Solar wind?*

study of signatures from airglow, auroras, attenuation, exospheres, and plasmas. Hence, the decommissioning of *Hubble* will cripple our ability to remotely monitor small bodies, atmospheric structure, and magnetospheric processes throughout the Solar System. *HDST* will restore and greatly expand our capabilities in these areas, even as it identifies new worlds to compare with ours.

### 4.5.1 Magnetospheric Processes and the Sun–Planet Connection

The movement of mass, momentum, and energy from the Sun and solar wind within the near-space environment of a planet is a major factor defining its detailed energy balance. These same factors govern the transition from atmosphere to exosphere and magnetosphere. Within atmospheres, charged-particle deposition (e.g., aurora) combines with solar UV and X-ray absorption to control the photochemical and thermal environments in the upper atmosphere of the major planets and throughout the thin gas columns of the icy satellites. Layered energy deposition can be tied to altitude-dependent circulation patterns, turbulence, and thermal lapse rates that are strongly dependent on diurnal effects and the solar cycle. In some cases (e.g., Titan) it will affect lower altitudes through the production of aerosols that reduce atmospheric transmission, or precipitate onto the surface.

Observations of near-space magnetospheric plasmas (e.g., the Io plasma torus) are useful probes for the sources and sinks of material, the strength and structure of local magnetic fields, and the current structures connecting upper atmospheres to the external space environment. The individual emission features can be intense and energetic, with complex velocity, thermal, and turbulence structure. However, they are typically buried beneath much brighter solar continuum in the visible, making UV observations essential for their characterization.

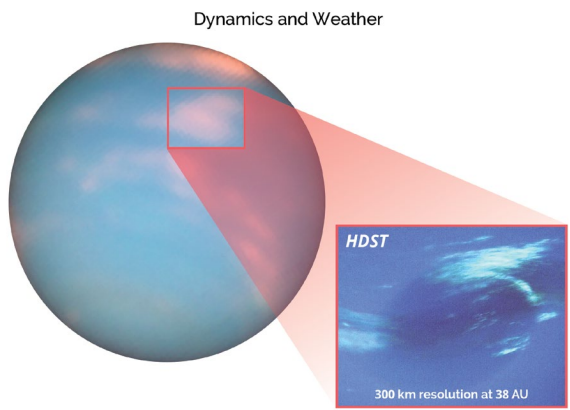

Figure 4-12: A visualization of *HDST*'s imaging performance for weather and atmospherics dynamics in the outer planets. This example using Neptune shows that *HDST* will resolve cloud patterns with 300-km resolution.

Observations of thin atmospheres, exospheres, and plasmas are typically pursued in the continuum-free UV via emission lines from fluorescence, recombination, collisional excitation, and metastable decay (Figure 4-12). *Hubble* is able to spatially resolve the brightest atmospheric features at the level of a scale height only in the Jupiter system, and beyond Saturn even the brightest auroral signatures are at or below detection. In contrast, *HDST* would extend scale height sampling to as far as Uranus, would greatly improve the detailed understanding of auroral signatures, and would provide the first characterization of ice-giant system plasmas and upper atmospheric circulation. At the same time, *HDST* will permit synoptic monitoring of the global magnetospheric response for the two giant planets, including: current satellite structures, plasma structures (e.g., the Io torus), and their connection into the planetary aurorae.

### 4.5.2 Volatility and Volcanism

Sublimation (a change from solid to vapor without melting) of surface ices from planetesimals and satellites has been detected out to the orbit of Pluto, with the extent being strongly dependent on temperature and species. The extent of volatility is strongly dependent on temperature, with activity dropping rapidly at large distances from the Sun. Beyond 3 AU, CO dominates small-body sublimation, but loss of water takes over as objects lie closer to the Sun. There are also evolved objects with minimal activity due to an intrinsic lack of volatile species and to the formation of a de-volatilized regolith (the layer of rocky debris and dust from meteor impacts). Improved understanding of gas production, composition, com-positional and structural uniformity, and isotopic ratios in small bodies are key to characterizing their formation and migration in the planetary disk, as well as for determining their contribution of volatiles to the terrestrial planets. These factors remain statistically under-constrained for both long and short period comets, especially for weakly active objects far from the Sun. For other populations, such as main belt comets, the Jupiter-Trojans, Centaurs, and the sub-dwarf protoplanet Kuiper belt, direct measurements of volatility are rare. *HDST* UV observations could provide two orders of magnitude improvement in our ability to measure elemental (O, H, C, and S) and molecular (e.g., CO, $S_2$) volatility indicators that would open new populations and distance scales for study.

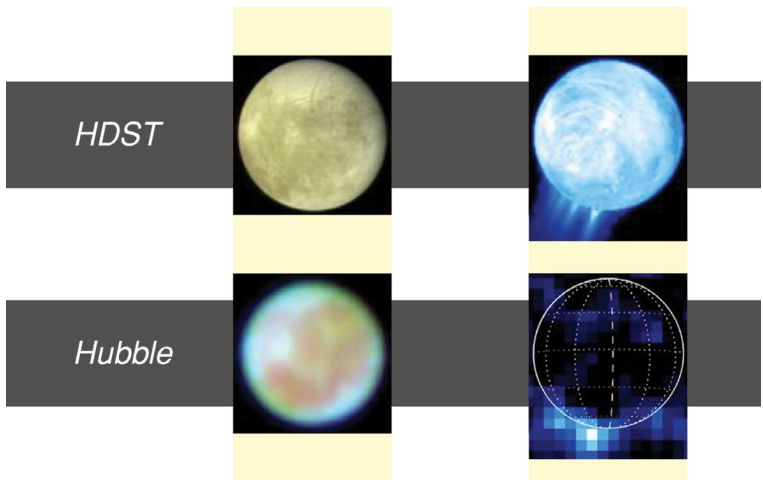

Figure 4-13: Two views of the Galilean satellite Europa, including visible band images at left and UV-band emission-line images of water vapor ejecta with *Hubble*/STIS and re-simulated at *HDST* resolution. At the orbit of Jupiter, *HDST*'s spatial resolution is 35 km.

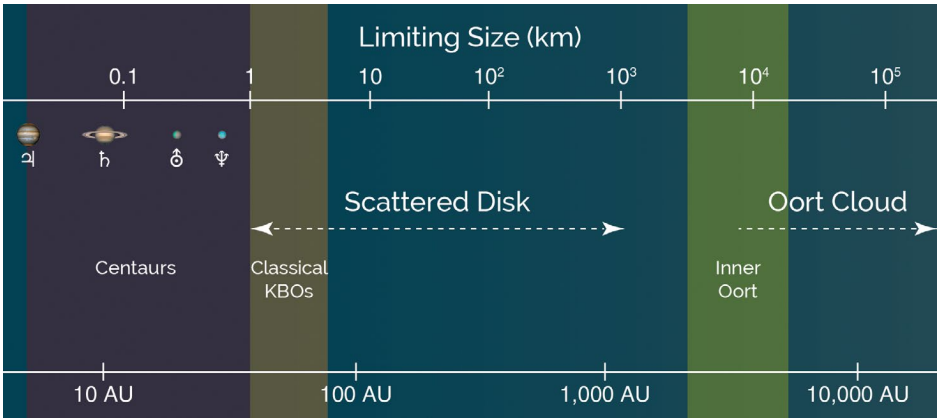

Figure 4-14: The minimum size detection limit (in km) for Kuiper Belt Objects and Trans Neptunian Objects at various heliocentric distances extending into the inner Oort cloud. The limits provided assume a 4% albedo and negligible contribution from thermal emission.

All icy satellites have sublimation-driven atmospheres with significant diurnal and latitudinal variability in their densities. However, neither the regional migration patterns, nor the diurnal/eclipse evolution of the various atmospheric columns are understood. Several objects (e.g., Enceladus, Io) further supplement atmospheric production with geysers and/or volcanism occurring at regions of high surface temperature (see also Figure 4-13). Monitoring these atmospheres requires either eclipse conditions or observation in the continuum-free UV, with the brightest signatures coming from aurora-like processes. Surface ices related to volatility are present in spectra covering the NUV to IR and may display spatial structure of a less than one percent of the surface area.

### 4.5.3 The Organization and Mass Distribution of Small Bodies

The distribution of planetesimals and dwarf planets is a guide to formation of the planets and their dynamical evolution. However, there is only limited information on the number, size, and location of these objects, particularly at heliocentric distances beyond Neptune's orbit (Figure 4-14). Current capabilities have revealed a classical Kuiper Belt that extends between the 3:2 and 2:1 resonances of Neptune (39 to 48 AU), followed by a poorly understood drop ("the Kuiper cliff"). The more remote, scattered population has more inclined and eccentric orbits with aphelia reaching 1000 AU, with the objects detected to date in a possible resonance with an unseen super-Earth(s) beyond 250 AU. The edge of the scattered belt is thought to coincide with the inner edge of the Oort cloud, which may extend to > 200,000 AU. The current size-detection limit at the inner classical belt is ~10 km, while Dwarf planets (1000 km diameter) can be observed out to about ~500 AU, which is less than half the orbital axis of Sedna, the first major dwarf planet discovered in the scattered belt. Only the largest Kuiper Belt objects (KBOs) have been measured spectroscopically.

While the *Wide Field Infrared Survey Explorer* has been able to exclude the presence of planets larger than Saturn out to as far as 82,000 AU on the basis of thermal emission, smaller and cooler objects are detectable primarily via reflectance of the solar visible continuum over much of this range. *Hubble*'s Advanced Camera for Surveys currently delivers the deepest detection limit for reflected light, exceeding even that of even the largest existing ground-based telescopes. Assuming this magnitude-limit advantage projects forward to *HDST*, a 12 m class diameter would project to an even larger gap in sensitivity when compared with ground-based facilities, including both the E-ELT and TMT. Using the *Hubble* approach of targeted searches and statistical extrapolation of small-field surveys, *HDST* would revolutionize our understanding of these populations by both reducing the detectable size and expanding the distance over which Trans-Neptunian objects (TNOs), KBOs, and small-to-medium-sized planets can be observed. It would be capable of detecting KBOs from 1–1.5 km in diameter over the entire classical belt and putting new limits on the population beyond the Kuiper cliff. *HDST* could also enable a spectroscopic compositional survey of objects as small as 10 km throughout the classical belt. Finally, it would expand the range over which 500 km and larger dwarf planets can be detected to include the entire scattered belt, and it will be able to identify the presence of an Earth- to Neptune-sized planet over distances extending into the inner regions of the Oort cloud from 5000–15000 AU (Figures 4-14 and 4-15).

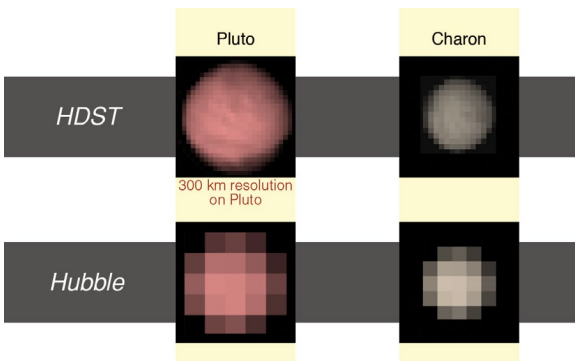

Figure 4-15: Two views of Pluto and Charon. At bottom, *Hubble* images that span the two bodies by only a few pixels. At top, a simulated *HDST* views that resolves surface features as small as 300 km.

## 4.6 Synergies with Other Astronomical Observatories

*HDST* will be the most capable general-purpose UVOIR space observatory ever—by orders of magnitude—in many metrics. In the 2030 era, *HDST* will be one of many powerful observatories making similarly transformative gains in their respective domains. As with *Hubble*'s operation today, *HDST*'s unique capabilities will complement studies done with other facilities, and in other wavelengths. The impressive capabilities of ground-based observatories in the upcoming decade (e.g., 20–40 m optical telescopes, LSST, the fully completed ALMA, and potentially the Square Kilometre Array [SKA]) and anticipated space-based facilities at other wavelengths (e.g., the *Advanced Telescope for High-Energy Astrophysics* [*Athena*], *eLisa*, and possibly other facilities), will lead to the same powerful synergy between ground and space telescopes that exists today, and that has always made their combination more powerful than the sum of the individual parts. There is no doubt that *HDST*'s scientific impact will be greatly enhanced when it acts in concert with these other facilities, in the same way that *Hubble* always has.

### 4.6.1 Multiwavelength Studies at Uniform Spatial Resolution

The laws of physics have distributed unique and informative diagnostics of matter in its diverse physical states across the electromagnetic (EM) spectrum. The expansion of the Universe then moves these diagnostics across the EM spectrum in predictable ways. The goals of astronomy are therefore best advanced with access to these diagnostics over all of cosmic time, at resolution sufficient to untangle the physical scales of interest. Thus reaching the significant 10 mas threshold in the UV, optical, infrared, and sub-mm/radio with future facilities constitutes a strategy for advancing all areas of astronomy on a broad front (Figure 4-2). Reaching 10 mas resolution supports an array of unique UV/optical science aims and complements these other bands with imaging at comparable physical scales. This angular scale corresponds to the physical scale of proto-planetary disks, planetary systems, and star clusters in the nearby Universe, and the building blocks of galaxies in the distant Universe (see Figure 4-4).

High spatial resolution UVOIR observations with *HDST* are a natural complement to ALMA observations of high-redshift galaxies and protoplanetary disks. At higher energies, the interaction between *Hubble* and X-ray observatories like *Chandra*, and the *X-ray Multi-Mirror Mission* (*XMM*) have shown the value of studying AGN and X-ray point sources in these two complementary wavelength regimes; this synergy will continue during the era when *HDST* can operate in tandem with *Athena* and potentially other facilities. Likewise, in the same way that *Hubble* has been of great value for current synoptic studies—by providing high-resolution views of the progenitor environments for supernovae, GRBs, and other unusual transients—so *HDST* will be invaluable to interpreting data from LSST, from higher-cadence observing programs, and from gravitational wave-detectors in space (*eLISA*) and on the ground (LIGO).

### 4.6.2 *HDST* and the Next Generation of Ground-Based Telescopes

In addition to being an essential ingredient in multiwavelength astrophysics studies, *HDST* will operate in tandem with even larger optical and near-infrared telescopes on the ground (see the box on Pages 48–49). There is a clear history from which to draw when considering how *HDST* might work in tandem with the next generation of 20–40 m class telescopes. For more than two decades, *Hubble* has operated simultaneously with ground-based telescopes with $> 10\times$ its collecting area, much like *HDST* would be relative to the ELTs of the 2020s and 2030s. In the cooperative efforts between *Hubble* and large ground-based telescopes like Keck, Gemini, Subaru, Magellan, and the VLT, the larger ground-based facilities have consistently provided the best high-resolution and highly-multiplexed spectroscopy of faint sources. They have also excelled at wide-field imaging with natural seeing and diffraction-limited imaging and spectroscopy in the near-IR over small fields (< 1 arcmin) using AO. Meanwhile, space-based facilities have offered diffraction-limited spatial resolution over large fields (3–4 arcmin), PSF stability and precision, low sky backgrounds, and UV access.

This natural scientific division of labor manifests across many fields of astrophysics, but one clear example can be found in the search for high-redshift galaxies. As of 2015, the most distant galaxies yet detected ($z$ ~10) were all found by *Hubble* operating at its near-IR photometric limits. The ground-based telescopes were then essential for confirming these galaxies with spectroscopy, which required large collecting areas but not high spatial resolution. In this same fashion, and in other fields of astronomy, the space and ground observatories have naturally found problems that they can address with complementary techniques, with each facility contributing where it has the greatest competitive advantage.

To leverage these relative strengths, joint investigations are often pursued by single teams of investigators who have optimized their observational strategy across multiple facilities in pursuit of unified scientific goals. Given that the jump in aperture from *Hubble*'s 2.4 m mirror to *HDST*'s 12 m class mirror is similar to the jump from Keck and VLT to the ELTs, and the strong precedent set by the present generation of facilities, this natural division of labor will likely continue into the next generations of facilities in the 2020s and 2030s.

*HDST* provides tremendous advantages in the time required to reach a given SNR for broadband imaging (Figure 4-17), in the absolute limiting sensitivity (Figures 4-2 and 4-16), and for low-resolution spectroscopy (Figure 4-17), compared to even the largest proposed ground-based facilities. This advantage holds not just in the UV and optical, but in the near-IR as well. The exposure time advantage of *HDST* relative to E-ELT (which we take as the largest, limiting case) diminishes at higher spectral resolution, but still offers very significant gains at $R = 100$ at all optical and near-IR wavelengths, and at $R = 2000$ in the visible. These trends clearly identify where the observational synergies between *HDST* and planned very large ground-based telescopes will be.

At longer wavelengths, the utility of *HDST* for mid-IR observations (3–10 microns) should be studied further. The scientific value of mid-IR wavelengths has been repeatedly proven by *Spitzer*, and will only become more important in the *JWST* era. However, depending on its exact operating temperature (expected to be somewhere in the 250–295 K) range, *HDST*'s sensitivity at these wavelengths may not be comparable to that of future large ground-based telescopes. Figure 4-16 shows the 10-σ point-source sensitivity limits in the 1 to 5 microns wavelength range for a 10,000-second broadband imaging integration for a selection of observatories. *JWST* will provide a large stride in mid-IR capability relative to its predecessors; any detailed science case for mid-IR coverage by *HDST* will need to take *JWST*'s discoveries into account at the appropriate time.

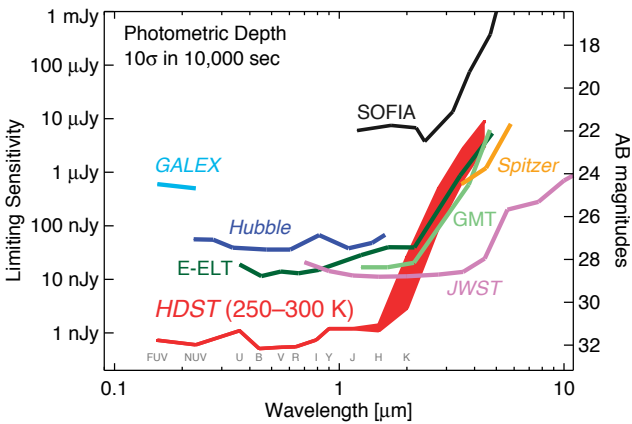

Figure 4-16: The 10-sigma limiting sensitivities as a function of wavelength are shown for various facilities in the near-IR and mid-IR for a 10 ksec broadband imaging integration. *HDST* sensitivities are shown for two different operating temperatures. Even without operating at cryogenic temperatures, *HDST* is competitive with *JWST* and the ELTs at wavelengths of less than 2 microns, and with *Spitzer* at longer wavelengths.

One significant development that might change the relationship between future space and ground-based observatories is the likely maturation of Multi-Conjugate Adaptive Optics (MCAO) and Ground-Layer Adaptive Optics (GLAO). For large-aperture ground-based telescopes, these are expected to enable intermediate-to-high Strehl ratio (~40–80%) performance over fields of view up to 1–2 arcmin for wavelengths > 1 micron. This technology will deliver high-resolution images at or near their diffraction limits in the near-IR. The science cases for these facilities emphasize high-resolution images and spectra of galaxies, proto-stellar disks, stellar populations, stellar dynamics, and Solar System objects as compelling scientific applications of this promising technology. With the maturation of multi-fiber and image-slicing techniques, these facilities also offer the exciting prospect of dissecting galaxies, stellar populations, and proto-stellar disks in three dimensions—high-2D spatial resolution (~10–20 mas) and a spectrum at every point on the sky with integral-field spectroscopy in the optical and near-IR.

Even with their superior resolution in the near-infrared, ground-based telescopes by their very nature must contend with the limitations imposed by poor weather, varying atmospheric conditions (even when the sky is clear), time-variations in bright sky backgrounds, image-quality variations caused by thermal fluctuations in the telescope and instruments, and numerous other unpredictable factors associated with operation on mountaintops. As such, space will remain the optimal environment for low-background, high-stability near-infrared observations that require any combination of the following:

- Stable PSF performance over wide fields of view and across the whole sky;
- Diffraction-limited angular resolution and precise wavefront control over fields of view larger than 2 arcminutes, or over any field of view at wavelengths shorter than 1 micron;
- Sensitivity at sub-nanoJansky levels;
- Ultra-high contrast imaging (> $10^8$ intensity ratio between adjacent sources);
- Extreme photometric precision (< 0.0001 mag) and accuracy in crowded fields;

- Efficient near-IR grism spectroscopy over wide fields, exploiting low sky backgrounds;
- Continuous coverage of optical and near-IR wavelengths in bands obscured by atmospheric absorption.
- High stability of one or more of these performance parameters over tens to hundreds of hours of exposure time spread over years of repeated observations.

In conclusion, human ingenuity will continue to drive the pursuit of many astrophysical investigations from whatever current state-of-the-art facilities are available, regardless of whether they are operated in space or on the surface of the Earth. However, access to the ~4 octaves of wavelength from 0.1 to 2 microns with the unparalleled sensitivity and resolution provided by *HDST* will be an essential part of future astrophysical investigations.

## 4.7 A Transformative Program

As a general purpose astrophysical observatory, *HDST* will carry out a broad array of science investigations. With 4000 to 5000 hours of on-sky

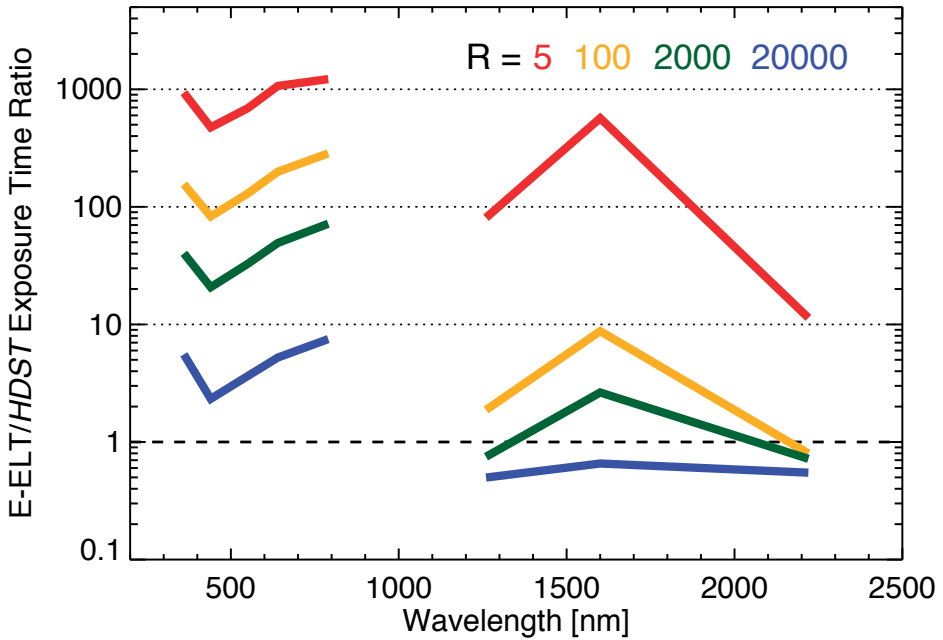

Figure 4-17: The ratio of integration time needed to reach an SNR = 10 on a 38 m E-ELT to that needed on *HDST* as a function of wavelength for point sources. Four different spectral resolutions (R = 5, 100, 2000, and 20,000) are shown for sources with AB mag of 30, 28, 26.5, and 24, respectively.

### HDST 100-hour Highlights

| Epoch | Name | Observations | Science Goal | Time [hr] |
|---|---|---|---|---|
| z = 1-4 | Deep/Wide Galaxy Survey | 1 hour/band in VRIJHK to AB = 32 | Statistically significant galaxy samples down to $M_V \sim -12$ | 120 |
| | Ultra-Deep Field | 20 hours/band in VRIJHK to AB = 34 | Detect faintest dwarf satellites in Milky-Way-like galaxies | 120 |
| z = 0.5-1 | Map the CGM (Emission) | R ~1000 emission in NUV over 30 square arcmin | Map H I, C IV, O VI in 300 galaxies at 1000–2000 Å | 100 |
| | Map the CGM and AGN (Absorption) | R = 20,000 spectra in FOV of emission field | Map diffuse CGM metals using background galaxies | 150 |
| <100 Mpc | Resolved Flows | 100 clusters in 10 galaxies | Probe stellar cluster content and outflow dynamics | 120 |
| | The IMF at Low Metallicity (dwarf Spheroidals) | 1–7 hrs in V and I bands | Measure IMF to 0.1 $M_\odot$ in six classical dSphs at [Fe/H] < −1.5 | 70 |
| <100 kpc | How Stars Form at Low Metallicity | UVOIR images of Magellanic Clouds | Determine protostellar accretion rates at low metallicity | 80 |

**Table 4-1:** Conceptual Astrophysics Treasury Programs with *HDST*

time available in every year, and approximately half of this devoted to peer-reviewed science programs on every possible topic, every observing cycle will bring new discoveries and advances on older questions. To illustrate how even the amounts of time typically available as individual large programs (100-hour scale), of which *Hubble* awards 6–10 annually, Table 4-1 shows a short list of "100-hour highlights" in Table 4-1. All these observations—in terms of depth, areal coverage, the signal they are trying to reach, or the nature of the target—are essentially impossible or prohibitively expensive with today's instrumentation. Yet they are all programs that could be developed by a single team in a single observing year with *HDST*, and could take their place among its most significant advances. That *HDST* can easily do what no current observatory can touch is a testament to its transformative gains in collecting area and spatial resolution, novel instrument modes, and to the flexibility of a general-purpose observatory. Throughout this section, it has been our goal to show how *HDST* can radically advance our understanding of how galaxies, the stars within them, the chemical elements from those stars, and all the building blocks of life are assembled over cosmic time. Many fields of astronomy will be transformed by these capabilities; none will remain entirely untouched.

# Chapter 5 The *HDST* Concept

The vision for a *High-Definition Space Telescope* (*HDST*) blends scientific aspirations—described in previous sections—with technological experience. This chapter maps the natural pathway from scientific goals—which set how the telescope must perform—to the technological requirements that can make these goals a reality. The following chapter uses these requirements to identify areas where focused, early development allows advances beyond initial concepts, in the lead-up to critical mission-design decisions. This task is a community-wide effort, building on the latest achievements and insights of astronomers, scientists, and engineers from across the globe.

## 5.1 Scientific Drivers for Key Capabilities

Every observatory is a technological as well as a scientific achievement. The role of technology is to enable and serve the science objectives that have been described in Chapters 3 (exoplanets) and 4 (the cosmos). This section briefly revisits these broad science cases and connects them directly to the required high-level observatory capabilities.

***ExoEarth Discovery as a Driver for Aperture:*** To find and identify exoEarths, to enable comparative planetology amongst different categories of small planets, and to open up the chance for finding biosignature gases, requires sufficient numbers of planets. As such, characterizing a sample of dozens of exoEarth candidates is a minimum scientific goal. All these candidates must first be identified via imaging, and then studied spectroscopically to characterize their atmospheres.

A telescope with a 12 m class primary mirror is required for detecting and characterizing a statistically meaningful sample of exoEarths, as described in Chapter 3. The collecting area is essential; the planets are so dim that less than one visible-light photon from a nearby exoEarth will strike the telescope's primary mirror each second. In the same second, many billions of photons from the exoEarth's star will flood the telescope, requiring precise separation, suppression, and/or shadowing of the star if the exoEarth is to be detected and studied. Achieving this starlight suppression requires extraordinary optical performance, with imaging resolution at the physical diffraction limit, which is set by the ratio of the wavelength to the aperture diameter, $\lambda/D$. The larger the

aperture—with its correspondingly greater collecting area—the sharper the resolution, and the easier it becomes to isolate and detect exoEarths, while simultaneously enabling many other background-limited science observations that benefit from the reduced noise in each resolution element. The starlight suppression also requires an extremely stable telescope system. Most of the next subsection is devoted to a detailed description of the technologies required to achieve the necessary aperture, stability, and contrast that will allow exoEarths to be found.

*High-resolution imaging across the Universe:* There are many examples across astronomy of important phenomena and objects that are hidden in plain view, obscured by the blurring of atmosphere or telescope optics. As detailed in Chapter 4, great advances in astrophysics require instruments able to isolate and identify stars in crowded fields, the building blocks of galaxies, star-forming regions, and planetary systems at small angular scales. As a byproduct of the large aperture and precisely controlled optics that are required by the search for exoEarths, *HDST* will produce stable images with exquisite angular resolution (~0.01 arcsecond or 10 mas at 500 nm), 4–5× better than *Hubble* and far better than today's best ground-based telescopes.

The science drivers identified in Chapter 4 show that when *HDST*'s ultra-high resolution is coupled to large format, gigapixel detector arrays, the observatory will have survey power unprecedented for any telescope on the ground or in space. Large-format detector arrays have already flown in space (e.g., *Kepler*, *Gaia*), and the next chapters highlight several enhancements that would further increase the detector capabilities. The high science return from wide-field imaging is well established, with examples ranging from the Palomar Sky Survey (photographic; ground) to the Sloan Digital Sky Survey (digital; ground) to the *Hubble* Deep Field and its progeny (digital; space). Indeed, nearly all of *Hubble*'s highest-cited publications, and a share of a Nobel Prize, come from data collected as part of its wide-field surveys, and provided to the community in its online archive. To maximize *HDST*'s survey capability, its focal plane can be populated with multiple wide-field instruments that can be operated in parallel. For the study of extended objects—nebulae, clusters, nearby galaxies, galaxy clusters—or for an unbiased census of distant galaxies, the images, spectra, and catalogs from *HDST*'s panoramic and panchromatic observations will serve the broad science needs of the astrophysical community. And, as noted in Chapter 4, many of these surveys can be executed in parallel with long-duration exoplanet observations, providing immense survey volumes while already pursuing another high-priority objective. *HDST* will be an efficient observatory, nearly always operat-ing in a mode where it is simultaneously executing its dual mission of focused discovery and broad-based surveys.

*Broadband far-UV to near-IR Science:* Unraveling the complex origins of galaxies, stars, and planets, and searching those planets for biosignature gases, requires access to a broad range of physically diagnostic emission and absorption features across the electromagnetic spectrum. The key astrophysical drivers are (1) atomic transitions originating in the ground state, allowing one to probe gas at a wide range temperatures (10 K–2 million K), down to very diffuse densities, with sensitivity to most atomic species; (2) molecular transitions that probe gas at high densities and temperatures of 10–1000 K; and (3) the black-body peak from either cool nearby objects or redshifted galaxies with hotter characteristic spectra. The first of these requires UV sensitivity between 100–300 nm, as most ground state transitions are at > 5 eV. The latter two requirements are strong drivers for near-IR sensitivity at > 1 micron, which also increases sensitivity to objects obscured by dust. *HDST* will therefore not only achieve the resolution and photometric precision required for exoplanet searches, but will also reach regions of the spectrum that are inaccessible from the ground.

Access to the UV is a high scientific priority, as there will be no large UV-capable space telescope once *Hubble* is retired (circa 2020). Ten of the 12 instruments on *Hubble* have included space-UV observing modes. Three of these instruments include high-resolution UV spectrographs, optimized for the study of filamentary gas in the "Cosmic Web" and interstellar medium, hot stars, star-forming regions, and protoplanetary disks. While *Hubble* has served as a pathfinder, *HDST*'s large collecting area and high UV throughput will provide a 100–1000× increase in the number of spectroscopically accessible targets, revolutionizing the study of the origins of cosmic structures. In the UV, atomic spectral lines provide the most compelling science, but molecular spectral features are also scientifically important, particularly for cooler and/or denser gas, and even as possible biosignature gases in the atmospheres of exoplanets. Coverage down to 100 nm is required by *HDST*'s UV science drivers. Coverage down to 90 nm would add scientific value, but this extension should be considered in balance with additional costs and optical-band performance

Molecular bands are dominant in the near-IR, and thus there is a strong motivation to design *HDST* so its sensitivity spans four to five octaves in wavelength—similar to *Hubble*, from from 100 nm to approximately 2 microns. Near-IR wavelengths provide access to many molecular transitions, rest-frame UV from high-redshift galaxies, and objects obscured by dust, with a background sky level that is significantly darker than is observed from the ground. However, at wavelengths beyond 1.7 microns the internal background generated by a warm observatory will limit its sensitivity. Operating *HDST* at a cooler temperature could allow observations that push further into the IR, but building a cold observatory is likely to significantly increase costs. Some of the IR observing modes may also be better pursued with the next generation of AO-enabled ground-based 20–30 meter telescopes (Chapter 4). As such, careful consideration should be given to the optimal wavelength cut-off and the temperature requirements for the mission. For the reasons stated above, the nominal *HDST* design assumes a room-temperature telescope.

| Epoch | Aperture | Wavelength | |
|---|---|---|---|
| | | UV (100–300 nm) | Near-IR (to 2 µm) |
| z = 1–4 | Resolve ALL galaxies to 100 parsec or better, to individual massive star-forming regions (Section 4.1.1) | Detect massive star formation in the smallest pre-galactic building blocks (Section 4.1.1) | Observe building blocks of galaxies in rest-fame optical (Section 4.1.1) |
| z = 0.5–1 | Identify stellar progenitors and hosts for diverse transients (Section 4.2.2) | Detect emission and absorption from gas accreting and recycling into galaxies (Section 4.2.1) | Map galactic star-formation and gas dynamics with rest-frame optical diagnostics (Section 4.2.1) |
| | Reach 100s of background QSOs for outflow and IGM/CGM studies (Section 4.2.1) | Detect hot plasma ejected by SMBHs acting as feedback on their galaxies (Section 4.2.1) | |
| <100 Mpc | Resolve stars down to 1 M⊙ out to the nearest giant ellipticals and clusters out to 30 Mpc (Sections 4.3.1 and 4.3.3) | UV mass functions of young stellar clusters (Section 4.3.1) | Low-extinction and reddening-free stellar population diagnostics |
| | Watch motion of virtually any MW star, LG satellites, and ellipticals in Virgo cluster (Section 4.3.4) | Use UV MOS/IFU to dissect multiphase gas feedback flows in nearby galaxies (Section 4.3.2) | |
| <100 kpc | Resolve individual stars in young clusters everywhere in the MW and Magellanic Clouds (Section 4.4.1) | Measure protostellar accretion from UV to Magellanic Clouds (Section 4.4.1) | Peer into protostellar disks to look for planets (Section 4.4.2) |
| | Examine protoplanetary disks at ~1–3 AU resolution to >100 parsec (Section 4.4.2) | Obtain disk abundances of C, N, O, Si, Fe from UV lines (Section 4.4.2) | |
| <50 AU | Resolve surface and cloud features down to 50 km at outer planets and 200 km at Kuiper belt (Section 4.5) | Planetary magnetospheres and the Sun-Planet connection (Section 4.5.1) | Mapping the spatial structure of surface ices on moons (Section 4.5.2) |
| | Census of outer Solar System (TNOs, KBOs) to small mass and large distance (Section 4.5.3) | Detect emission from planetary coronae and aurorae, volcanism, and geysers (Section 4.5.2) | Detection of smallest and furthest objects (Section 4.5.2) |

**Table 5-1:** Examples of general astrophysics drivers across cosmic epoch (Chapter 4) for three key *HDST* capability requirements: aperture diameter, efficient ultraviolet imaging and spectroscopy, and near-IR imaging and spectroscopy.

***Flagship-class science with a core set of instruments:***  The vision for *HDST* includes two distinct categories of instruments, themselves complementary, but able to work together throughout most of the mission. Instruments that require high throughput and/or low wavefront error, but that do not need a large field of view could be placed at the two-mirror or Cassegrain focus of the telescope. This narrow-field instrument suite might include a coronagraphic ExoEarth Discovery Instrument and a high-resolution UV Discovery Spectrograph. A second suite of instruments could take advantage of the wide field available at a three-mirror telescope focus, with a powered folding mirror providing additional correction resulting in diffraction-limited performance throughout. These instruments, such as a Panoramic Imager and Multi-Object Spectrograph, would be used for *HDST*'s wide-field surveys, often in concert with the single-target observations carried out by the central discovery instruments. A brief description of these instruments, and the science drivers for them, is given here:

*ExoEarth Discovery Instrument*—Exoplanet discovery and characterization requires an initial starlight-suppressed detection in visible-light images, followed by a visible/near-IR spectroscopic hunt for molecular signatures. A spectral resolution of $R > 70$ will be required for unambiguous identification of line and band spectral features. Definitive confirmation of habitable worlds may require the broadest possible wavelength coverage.

Table 5-2: This science traceability matrix links *HDST* science investigations to key capability and instrument requirements (Exo-Im, Exo-Sp – Exoplanet discovery instrument imager and spectrograph; UV – High-resolution ultraviolet spectrograph; PI – Panoramic imager; MOS – Multi-object spectrograph). Detector capability requirement includes Photon counting (PC) and High dynamic range (Dyn Rg). The rightmost column provides a "typical" minimum count rate (in counts per second) per telescope or instrument resolution element (angular and spectral; not per detector pixel). Those observations requiring photon counting or high dynamic range are in bold. (*following page*)

| | Science Investigation | Possible Parallel? | Instrumentation | | | | | | Aperture | | | Band | | | Spectral Resolution | Detector Requirement | Photon Min CR |
|---|---|---|---|---|---|---|---|---|---|---|---|---|---|---|---|---|---|
| | | | Exo-Im | Exo-Sp | UV | PI | MOS | Area | PSF | | UV | NIR | | | | |
| **Exoplanets** | Detect Earth-like planets in HZ | | • | | | | | | • | | | | | | 5 | | 0.05 |
| | Obtain orbital parameters | | • | | | | | | • | | | | | | 5 | | 0.05 |
| | Obtain spectra, detect biosignatures | | | • | | | | | • | | | | • | | >70–500 | PC | <0.005 |
| | Characterize planetary systems | | | • | | | | | • | | | | • | | >70–500 | PC | <0.005 |
| | Transit spectroscopy/atmospheres | | | | | | | | • | | | | • | | 100–3000 | Dyn Rg | $10^{+5}$ |
| **IGM+** | Study disk/halo interface | | | | • | | | | • | | | • | | | >6000 | PC | $10^{-5}$ |
| | Probe using brt. QSO, or gal. | | | | • | | • | | • | | | • | | | 10k–100k | | 0.003 |
| | Detect in emission | • | | | • | | • | • | • | | | • | | | >1000 | PC | $10^{-6}$ |
| **Galaxies & SN** | SF and Chemical Enrichment, all z | • | | | • | | • | • | • | | | | • | | 500–5000 | PC | 0.000 |
| | First galaxies/reionization, high z | • | | | | | • | | • | • | | | • | | 5 | | 0.03 |
| | First galaxies SF and metals, high z | • | | | | | • | • | • | • | | | | • | 100–2000 | | 0.01 |
| | Resolved massive SFR regions z=1–4 | • | | | | | • | | • | • | | • | | | 5 | | 0.03 |
| | Transient stellar progenitors and hosts | • | | | | | • | | • | • | | • | | | 5 | | 0.03 |
| | Resolved SFH, to Msun, <25 Mpc | • | | | | | • | | • | • | | | | | 5 | | 0.03 |
| | Proper motions of stars in LG dwarf | | | | | | • | | • | • | | • | | | 5 | | 0.03 |
| | Low-mass dwarfs and features, all z | • | | | | | • | | • | • | | | | | 5 | | 0.03 |
| **AGN** | SMBH mass function vs. z, env. | • | | | • | | • | | • | • | | • | | | 1k–5k | | 0.01 |
| | Accretion disk and ionized flows | | | | • | | • | • | | • | | • | | | 100–5k | | 0.01 |
| **Stars** | Low Metallicity, high mass | | | | | | • | | • | • | | • | | | 5, 100k | PC | 0.0001 |
| | Circumstellar disk, protoplanetary | | | | | | • | • | | • | | • | | | 100–100k | PC | 0.0001 |
| | IMF and remnant MF microlensing | | | | | | | | | • | | • | | | 5 | | 1 |
| | Globular cluster ages, WDs | | | | | | | | | • | | • | | | 5 | | 0.1 |
| | IMF vary with environment? | • | | | | | • | | | • | | • | | | 5 | | 0.01 |
| **Solar System** | Surface/clouds <200 km @ Kuiper Belt | | | | | | • | | • | • | | • | | | 5 | | – |
| | Outer solar system census | • | | | | | • | | | • | | • | | | 5 | | – |
| | Planetary magnetospheres | | | | | • | • | | | • | | • | | | 5 | | – |
| | Coronae, aurora, volcanism | | | | | • | • | | | • | | • | | | 5 | | – |

*UV Discovery Spectrograph*—With a light-collecting power > 25–100×, depending on wavelength, of the Cosmic Origins Spectrograph on *Hubble* and spectral resolution $R$ ~20,000–150,000 at wavelengths above 100 nm, this instrument will revolutionize our understanding of the largest gaseous structures in the universe, the atmospheres of the hottest stars, and perhaps even the atmospheres of transiting exoplanets.

*Panoramic Imager*—Gigapixel-class cameras will be optimized for the UV, visible and the near-IR, allowing *HDST* to capture vistas of the universe in deep and wide surveys of unprecedented resolution and sensitivity. Anticipated fields of view are ~6 arcminutes across, with exceptionally stable image quality thanks to *HDST*'s exquisite wavefront stability and pointing control. This new era of ultra-stable wide-field imaging will also enable high-precision photometry and astrometry, and a broad suite of new science investigations exploring the transient and dynamic universe, including transiting planets.

*Multi-Object Spectrograph*—As a partner to the panoramic imager, these multi-object spectrographs will patrol nearly the same field of view (~6') and wavelength range (UV to near-IR), providing large numbers of low-to-moderate resolution spectra ($R$ ~100–5000) of selected targets across the field. Recent developments in remotely controllable and reconfigurable shutters—such as the microshutter array built for *JWST*—allow spectroscopic surveys of galaxies, black holes, and stars; and absorption-line and emission-line mapping of the intergalactic medium. The development of low-noise, photon-counting detectors will further enhance the low surface-brightness sensitivity of this instrument.

*HDST*'s combination of aperture, resolution, UVOIR wavelength coverage and a core set of powerful instruments will enable a diverse range of science investigations. Table 5-2 captures the traceability of instrument and other key requirements to example science investigations, as discussed in Chapters 3 and 4. Also indicated are observations that could be carried out in parallel with exoplanet or UV point source observations or exoplanet observations. Past design concepts (e.g., Mag30cam; Brown, et al. 2005) have shown that exoplanet observations of bright host stars can remain sky-limited at arcminute distances, given anticipated mirror surface properties.

## 5.2 *HDST* Capabilities and Technologies

### 5.2.1 The Telescope

***Aperture:*** What sort of optical telescope can accomplish the scientific goals described in previous chapters? The first consideration is aperture size, driven by key science requirements for resolution and sensitivity and goals for exoEarth yield. Of these, discovering and characterizing

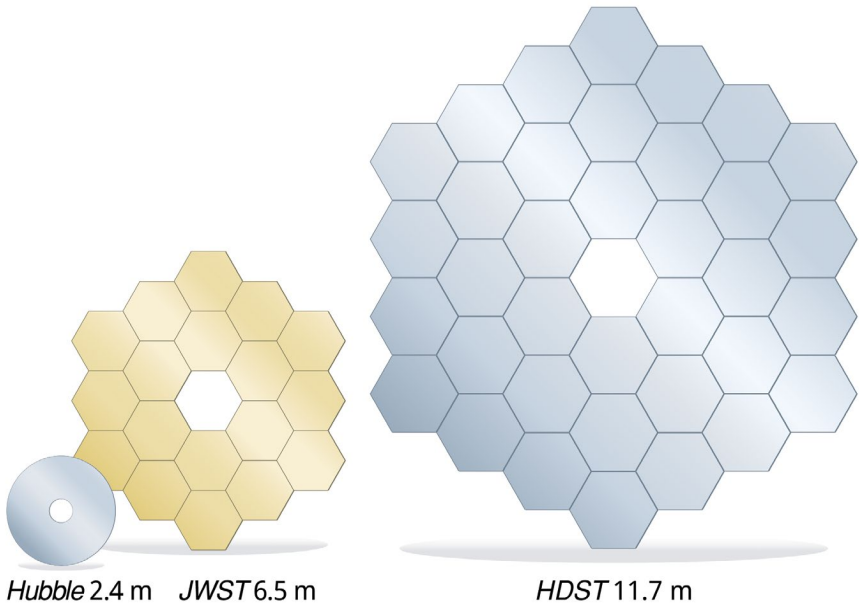

Hubble 2.4 m    JWST 6.5 m                HDST 11.7 m

Figure 5-1: A direct, to-scale, comparison between the primary mirrors of *Hubble*, *JWST*, and *HDST*. In this concept, the *HDST* primary is composed of 36 1.7 m segments. Smaller segments could also be used. An 11 m class aperture could be made from 54 1.3 m segments.

dozens of exoEarths sets the stringent requirement for an aperture of 10 m to 12 m depending on the performance of the exoplanet coronagraph. The astrophysics goals for dissecting galaxies at 100 pc resolution and mapping the cosmic web also require apertures of 10 m or greater. Figure 5-1 sets the scale of a 12 m class aperture in comparison to the primary mirrors for *Hubble* and *JWST*.

A second consideration is whether the primary should be a single large monolithic mirror, or a segmented mirror divided into multiple smaller mirrors, launched in a compact form, and then deployed and phased together after launch. The answer to this question is important, because the launch vehicles available in the coming decades will impose a limit on the aperture size of a monolith.

Space telescopes are highly constrained by launch-vehicle capacities (i.e., fairing payload volume and mass). *Hubble*'s aperture diameter and total mass were limited by what could be carried by the Space Shuttle—as were *Chandra* and *Compton*. Similarly, the aperture diameter and mass of any future mission like *HDST* will also be constrained by the capabilities of the available launch vehicles. Currently, only NASA's planned Space Launch System (SLS) Block 2 vehicle—and then the aperture would be limited to the low end of that range, to about 8 m.

Segmented apertures overcome some of the volume constraint. They can be launched in a folded configuration, like *JWST*, and then unfolded after launch to achieve an aperture larger than the shroud's internal diameter, perhaps as much as 2.5× larger. As discussed in Feinberg et al. (2014), and shown in Figure 5-2, launch vehicles with 5-meter class shrouds may be able to launch segmented mirrors totaling an effective area equivalent of 9–12 m size, by using deployment schemes derived from those developed for *JWST*. An SLS Block 2 with an 8 or 10 m shroud should be able to deploy even larger apertures. Longer term, on-orbit assembly would enable arbitrarily large segmented apertures, though the infrastructure for on-orbit assembly could be costly if it were developed only for *HDST*.

A 12 m class segmented aperture can be launched by currently available and planned launch vehicles. A monolithic aperture larger than about 4 m will require use of the projected SLS Block 2 and will be limited to 6–8 m aperture depending on the dimensions of that vehicle's shroud.

Both segmented and monolithic mirrors have significant heritage on the ground and in space, giving confidence that large apertures could be produced with either approach. In a segmented approach, the large number of mirrors offers opportunities to take advantage of scalable technologies, which can help to control cost. This approach is being developed on the ground, where the quest for 10–30+ m telescopes has led to scalable architectures for segmented apertures where each of the primary-mirror segments may be 1–8 m in diameter. A segmented *HDST* is likely to require segments 1–2 m in diameter. Investments in the technologies for the future are to be balanced against current performance realities—Investments in the technologies for the future are to be balanced against current performance realities—specifically, that segmented apertures present challenges for the starlight suppression technologies required for exoplanet imaging and spectroscopy, as discussed further below.

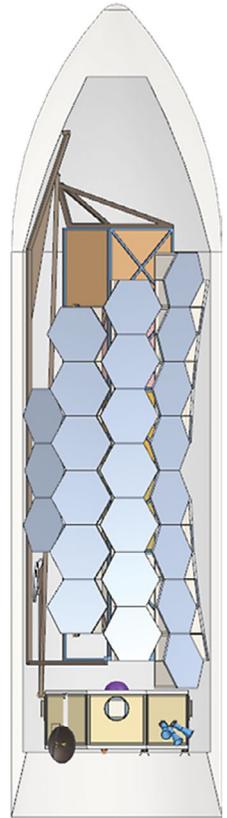

Figure 5-2: A folded 11 m primary mirror, constructed with 54 1.3 m segments, is shown inside a Delta 4-H shroud.

*Mirror Figure:* The image quality of scientific optical telescopes is often measured using "wavefront error," a quantity that captures the deviation from perfect imaging. An Optical Telescope Element (OTE) at the physical limit of angular resolution requires that the wavefront error be considerably smaller than the wavelength of the light being studied. For precision starlight suppression, even smaller wavelength errors may be required. *HDST*'s image-quality requirement is that it be diffraction-limited at 500 nm, corresponding to a total system wavefront error of 36 nm. After including all contributors to image quality, such as observatory jitter and detector resolution, the OTE wavefront error requirement becomes 27 nm, or about $1/20^{th}$ of a wavelength, which will require an active wavefront-error sensing and control system designed to maintain a stable wavefront with minimal error and drift. The OTE wavefront error budget includes all terms that may contribute to an imperfect image, including mirror-shape errors, segment misalignment, thermal and mechanical drift, and the residual from any active wavefront error control loop. An error budget analysis suggests that the required primary-mirror figure-error specification should be $1/30^{th}$ of a wavelength. While further development work is required, both ultra-low expansion glass and Silicon Carbide (SiC) mirror technologies offer lightweight solutions that should be capable of meeting this performance level.

*Mirror Coatings:* *HDST* will have high throughput from the far-UV to the near-IR (100 nm to > 2 microns), enabling science that cannot be done from the ground. *HDST*'s all-reflective 2–3-mirror OTE system will be designed to be highly efficient across the broadest possible wavelength range, while each instrument will be further optimized in narrower wavelength ranges tailored to specific science goals. As with *Hubble*, *HDST*'s usable wavelength range will be defined by its mirror reflectance cutoff at short wavelengths (far-UV), and sensitivity limits due to thermal background noise at long wavelengths (near-IR). UVOIR astronomical telescopes typically use a metallic layer of aluminum as a stable and highly reflective coating material. Aluminum oxidizes easily, however, and thus requires a UV-transparent protective overcoat layer to maintain high reflectance at wavelengths below 250 nm. Beyond the primary and secondary mirrors, subsequent channels and instruments that do not require UV sensitivity may utilize protected silver coatings that have very high reflectance above 350 nm in the visible and near-IR. An important consideration is the impact of OTE coatings on performance of different coronagraph designs, particularly those coatings optimized for the far-UV. This is briefly discussed in Chapter 6 and is the subject of continuing trade-study.

### 5.2.2 Starlight Suppression

*HDST* must block the flood of photons from a star in order to detect and study the faint reflected light from the from the planets in orbit around it. Two methods for starlight suppression—coronagraphs and

starshades—are under development. Both methods have advantages, and both present challenges. This section describes some of the important design considerations for each.

*Coronagraph Requirements:* A classic coronagraph for exoplanet imaging focuses light onto an image-plane mask that obscures the light from the star at the center of the field. The subsequent beam then passes through a shaped aperture stop, which suppresses the immediate diffraction from the edges of the pupil and its obscurations. The effect is to eliminate light from the central star at the center of the field, while passing the light from nearby objects such as planets. As discussed in Chapter 3, coronagraphic starlight suppression depends strongly on aperture diameter (*D*). Additionally, there are a number of other key performance characteristics to consider:

*Contrast.* The detection limit (or "contrast") is the ratio of the intensity of the faintest detectable planet to that of the central star. As an example, light reflected by Earth is 10 billion ($10^{-10}$) times fainter than the Sun. "Raw contrast" is the instrumental suppression without any image processing. "Total contrast" is the contrast after image processing.

*Inner Working Angle (IWA).* The IWA quantifies how close a planet can be to its star, as projected on the sky, and still be detected. When the IWA is small, more planets are sufficiently well separated from their host star that they can be detected and characterized. Because the IWA scales linearly with wavelength (typically expressed in units of $\lambda/D$), a fraction of the exoplanets detected in the optical will fall outside the larger IWA in the near-IR. Nevertheless, orbit characterization and careful observation scheduling should allow many detected planets to be observed in the near-IR, of particular importance for spectroscopic characterization.

*Throughput and bandpass:* The throughput of a coronagraph must be maximized to improve observing efficiency, but will be adversely affected by coronagraphic masks, pupil stops and other optical elements. The bandpass of a coronagraph (i.e., its wavelength range) also affects efficiency, but the ability of coronagraphs to maintain high contrast over a wide spectral band limits it. Bandpass is typically assumed to be approximately 20%. The full spectral range can be observed sequentially by changing filters, or simultaneously with a multi-channel instrument. In a single-channel instrument, the product of bandpass and throughput defines the instrument's efficiency.

ExoEarth yield calculations like those in Chapter 3 show interdependencies between contrast, IWA, and throughput/bandpass, creating a complex tradeoff that establishes no single coronagraph parameter as the dominant performance criterion after the primary driver of telescope aperture. Other astrophysical factors, including the expected intensity and distribution of exozodiacal dust, the size of the stellar disk, and the demographics of the extrasolar planet population, will also play a role in determining the coronagraph architecture and performance requirements that are best suited for *HDST*.

The goals for coronagraphy differ for discovery and follow-up spectroscopy. A coronagraph optimized for exoEarth discovery (survey mode) will need high raw contrast, $10^{-10}$, but at a relaxed IWA of 3.6 $\lambda/D$, for the baseline assumptions discussed in Chapter 3. A coronagraph optimized for characterization will need wavelength coverage up to 1 micron, a spectral resolution of 70, a tighter IWA of 2 $\lambda/D$, but with a relaxed contrast of $10^{-9}$. The *HDST* will need a single coronagraph that can be reconfigured to perform both modes, or two separate survey and characterization coronagraphs. With these capabilities, a coronagraph is predicted to discover and characterize dozens of exoEarths if the *HDST* aperture is 12 m, for a nominal survey time of one year and a parent sample of stars out to 20 parsecs. Table 5-3 provides a summary of starlight-suppression goals for *HDST*, and current technology status for coronagraphs and starshades alike.

*Coronagraph Wavefront Sensing & Control:* Any segmented telescope will need to have its mirrors aligned and shaped through wavefront sensing and control. The stipulations for *HDST* are even more stringent than typical, because the coronagraph places rigorous demands on imaging stability, requiring minimal drift of the wavefront error in a time interval typically associated with the wavefront control loop. If contrast is required at the target $10^{-10}$ level, then the wavefront control loop must keep the contrast stable at the $10^{-11}$ level. The number of photons that can be collected in the dark-hole region sets the relevant time interval, and there are few such photons. Consequently the time interval for control updates is long, about 10 minutes. Within this time interval, the telescope and root mean squared (RMS) coronagraph wavefront error drift must not exceed 10 picometers. Thus the requirement for stability of the OTE is about 10 picometers wavefront error per 10 minutes. This stability requirement is very challenging, and must be a main focus of OTE and mirror technology development efforts going forward. *Critical to achieving this requirement is the performance of the observatory wavefront error sensing and control system, telescope stability and metrology system, and the thermal environment and control.* We discuss a path to achieving this requirement in Section 6.2.

*Starshades:* As described in Chapter 3, starshades are large external spacecraft, tens of meters in diameter, flying between the telescope and the target star and shaped to create a deep star shadow at the telescope. A starshade optimized for observing with a 12 m class *HDST* will be very large, about 80 to 100 m, and must be carefully positioned at about 160,000–200,000 km distance from the spacecraft, tracking the moving tangential line of sight between the star and the telescope with about 1 m precision. It would provide a contrast of $10^{-10}$ and an IWA of less than 50 mas, sufficient to observe a Sun-Earth analog system to distances up to 20 parsecs.

*Starshades vs. Coronagraphs:* Assessing the possible role of starshades for detecting exoplanets requires comparing their performance with that of coronagraphs. Key points of comparison include retargeting time and exoEarth yield, numbers of spacecraft, telescope stability, throughput and bandpass.

Starshades require launching and operating two separate spacecraft—the telescope and the starshade occulter—with significant cost implications. Starshades must also be very precisely shaped to be effective. While a coronagraph can be repointed in a matter of minutes to hours by slewing the telescope, starshades must make large orbital maneuvers covering thousands of km to get into position. This maneuver takes days to weeks in some cases, lowering the total exoplanet yield while using up substantial fuel.

Coronagraphs require extraordinarily stable wavefront and line-of-sight pointing stability to maintain high contrast over long periods of time—a challenge for the telescope designers. Starshade observations can be performed with more conventional telescope stability, making the telescope less of a challenge, and thus less costly to deliver an OTA that meets requirements. Starshades offer the potential for deeper suppression, broader bandpass, and likely a smaller IWA than current coronagraph designs. This makes starshades especially attractive for spectral characterization of exoEarth candidates, particularly at long wavelengths.

Ultimately, the higher exoEarth yield potential makes coronagraphy the preferred method for *HDST*, provided that the needed suppression performance can be achieved. Coronography is therefore the highest priority technology-development item for the immediate future, and recent progress in this area is encouraging. At the same time, starshades have enough attractive features that studies exploring exoEarth yield using one or more starshades—or a combination of coronagraph and starshade—should be performed to ensure that the best system-level solution is understood. In case coronagraph performance falls short—or if other insights occur in the evolving scientific and technical context—continued development of starshade technology is recommended to provide other options. Starshades may also be an attractive second-generation addition to *HDST*, to allow better spectral characterization of potentially habitable worlds initially discovered with *HDST*'s coronagraph.

### 5.2.3 Detectors and other Instrumentation

*Detector technologies:* In the UV and visible, ~5 micron pixels are required to properly sample a point source imaged by the telescope (for *F/*15), critical for crowded-field photometry, precision astrometry and time-series measurements. Given the size of the focal plane, this implies a pixel count of > 1 Gigapixel (32k on a side) for imaging sensor arrays that cover the desired ~6 arcminute field of view. In the near-IR, acceptable pixel sizes will be 10 microns with correspondingly lower pixel counts. Accommodation of gigapixel detectors on *HDST* appears well within the capabilities of current technology (see Figure 5-3). The cameras have pixel counts that are comparable to current equipment on the ground, and only a factor of a few larger than the likely *WFIRST/AFTA* cameras.

For optimal sensitivity, nearly all detector technologies being considered will require active and/or radiative cooling, down to temperatures as low as 100 K. The detector/instrument cooling system must be designed with minimal impact on stability (e.g., from vibration, thermal drift, mechanical coupling) and throughput in the UV from molecular contamination.

Exoplanet coronagraphic imaging and wavefront sensing detectors can benefit from low read noise. The exoplanet coronagraphic characterization detector may also require a dark current lower than 1–10 electrons per hour per pixel (see e.g., Table 5-2). High-efficiency (> 90%) photon-counting detectors with stable photometric performance and high dynamic range would be ideally suited for exoplanet observations, particularly when shorter control-loop response times are required.

The sky background for UV general science observations is typically very low, and to fully exploit these dark skies, detectors with comparably low background noise are required. For this reason, deep UV observations and spectroscopy investigations such as those listed in Table 5-2 will also benefit from the high efficiency, low noise, photon-counting detectors described above. A challenge for the UV is achieving the detector "triple crown"—high quantum efficiency (QE), low read noise/photon counting, and low dark current. The 2000 and 2010 Decadal Surveys recommended continued development of UV detectors. Efforts are underway to optimize large-format UV detector array performance in all three areas using CCD, CMOS and microchannel plate technologies.

*Instrument modules:* The instrumentation for *HDST*, its optics, diffraction gratings, filters, multi-slit modules, and detectors will be designed to capture the broad scientific aspirations of the many investigations envisioned for the observatory. Designs will be modular to allow for a simple interface, for both the management and the hardware. Other flagship observatories have used this modularity to facilitate contributions from a broad set of institutional and international partners, and we envision similar participation for *HDST*. Many of the components of the *HDST* instrument package are conceptually similar to *Hubble* instruments adapted for use in this advanced observatory, and will make use of space-qualified technologies or build on recent development efforts directed towards current and future NASA missions.

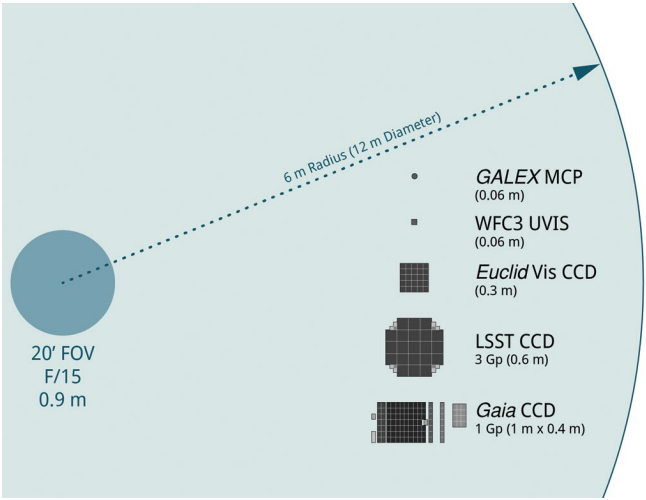

**Figure 5-3:** Comparison of physical sizes of existing and planned detectors with possible "usable" size of the full focal plane of *HDST*. Several channels of wide-field instruments (e.g., panoramic imager with 0.5–1 Gpix detectors and multi-object spectrograph; both with ~6' FOV) will share the focal plane of *HDST*.

### 5.2.4 Spacecraft and Operational Requirements

The *HDST* spacecraft and its operations will likely share many commonalities with the *JWST* architecture and operations. For *HDST*, as for *JWST*, an L2 halo orbit is preferable to a geosynchronous orbit (and most others, including low-earth orbit), because it is thermally stable and has lower fuel requirements for station-keeping. Unlike *JWST*, *HDST* will be a warm observatory, operating nominally above 270 K (vs. 50 K for *JWST*) and will therefore require a less aggressive sunshade than *JWST*, but also a thermal design and the necessary power required to maintain a warm observatory at a stable temperature.

Based on the instrumentation and likely observing scenarios (e.g., exoplanet detection and characterization in Chapter 3, and the 100-hour programs described in Chapter 4) data volume (driven mainly by the large imaging detectors) could be as high as 250 GB per day, compressed. This data rate is about 5–10× the compressed data rate of *Gaia* (also at L2).

The anticipated lifetime of *HDST* is 10 years. The telescope should be serviceable, but a servicing requirement has not been specified. Chapter 6 discusses a possible role that servicing and assembly might have for *HDST* and future large-aperture space telescopes.

### 5.2.5 Capability Requirements for a Single *HDST*

*HDST*'s key capability requirements for exoEarth detection and characterization, and a powerful new general astrophysics program, are summarized below in Table 5-3, along with a comparison to the capabilities of *JWST* and *Hubble* (see also Figure 5-1). These capabilities are also linked to the instrumentation that will largely benefit from them.

In planning this program, this study has carefully considered whether a single observatory is the right platform for addressing both exoplanet and general astrophysics science. For the same reasons, large apertures clearly benefit both major science goals—namely collecting area and angular resolution. Both science programs also share strong drivers for optical and near-infrared wavelength coverage. Given the natural challenges in building large missions, it makes sense to combine both programs into a single long-lived mission, rather than two sequential missions with comparable apertures, both with decade-long development times.

That said, it is worth explicitly noting what potential compromises and challenges may be faced by a combined mission. First, although the large aperture size naturally leads towards segmented concepts, segmentation has an impact on exoplanet coronagraphy. Internal coronagraphs favor telescope designs with minimal obscuration and vignetting, and have no compelling need for wide fields of view. Coronagraphs also place stringent requirements on telescope stability; such requirements would be significantly relaxed with use of a starshade, but with a dramatic impact on search efficiency. Exoplanet missions also have weaker (though not non-existent) science drivers for UV coverage; restricting the telescope to optical and near-IR wavelengths could allow use of simple protected silver coatings to produce a very high-efficiency mirror in the desired wavelength range.

A pure general astrophysics mission would also have somewhat different trade-offs. General astrophysics requires UV-optimized coatings on the telescope mirrors and a telescope design that produces a diffraction-limited PSF over a wide field. The latter would be unlikely to have a major negative impact on exoplanet performance, but the interaction of UV-optimized coatings with internal coronagraphs remains to be fully explored. A general astrophysics mission would also have much weaker requirements on stability, at a level that would potentially be compatible with a starshade.

| | | Capability | HDST (12 m) Increase vs. | | | Instrument | | | |
|---|---|---|---|---|---|---|---|---|---|
| | Parameter | Requirement | Hubble | JWST | exoEarth | UV spect. | Panoramic Imager | Multi-Object Spectrograph | |
| **Large Aperture Space Telescope** | Aperture diameter | 10–12m | x5 | x1.5–2 | X | X | X | X |
| | Wavelength | 0.100 to 2 μm (extended goal: 0.092 to 3–5 μm) | Same w/ high eff. DUV, high-sens. NIR | UV–vis (blue) JWST: 0.6 to >2 8.5 μm | Vis–NIR | UV | UV–Vis–NIR | UV–Vis–NIR |
| | Field of view | 6' | x2–4 | x2–4 | | | X | X |
| | Instrument channel pixel count | 0.5–1 gigapixel | x20–40 WFC3 | x25 NIRCAM | | | X | |
| | Angular resolution | <0.01" (Diff. lim. 500 nm) | x5 @ 500 nm | x1.5–2 @ 1 μm | X | | X | |
| | Survey power (area*FOV) | ~2500 m²–arcmin² | x10–20 | x3–8 | | | X | |
| | Spectral resolution | See instrument | R x1.5–3 w/ high eff. x50–100 multiplex | R x25, similar multiplex | >70–500 | 5,000–150,000 | Grism | 5,000–20,000 + grism surveys |
| **Starlight Suppression** | Raw contrast | $10^{-10}$, $10^{-9}$ | x100–1000 (lower) | x100–1000 (lower) | X | | | |
| | Inner working angle | 3.6 λ/D (det; 0.55 μm), 2 λ/D (sp; 1 μm) | x20 (smaller) | x20 (smaller) | X | | | |
| | Throughput | 0.2 | N/A | N/A | X | | | |
| | Bandpass | 20% | N/A | N/A | | | | |
| | Stability | 10 pm | x1000 @ 1" (lower) | x1000 @ 1" (lower) | X | | Enables some science | |
| | Metrology | <10 pm | N/A | N/A | X | | | |

Table 5-3: *HDST* Summary of Capability Requirements and Comparison with *Hubble* and *JWST*.

In short, until large space telescopes are commonplace, it will be important to maximize the capability of every NASA observatory so that it serves the widest community. At the present time, no requirement for either major science goal is an obvious showstopper for the other. A single and powerful *HDST* will have capabilities that should be able to accommodate the needs of the broadest possible community of astronomers.

# Chapter 6 Taking *HDST* from Concept to Reality

With a strong science case and a clear set of mission requirements in hand, the natural next question is, can it be built? The answer is "yes." This chapter shows how confidence in the *HDST* mission concept follows directly from experience gained from previous and ongoing missions, pathfinder projects, and prototypes. Building on this experience with feasible, smaller investments in key technologies, *HDST* designers and scientists will soon be able to chart the final course towards a specific architecture.

## 6.1 Technology Heritage, Status and Future Synergies

Most technologies needed for *HDST* are either under active development for other missions, or have reached maturity. Although *HDST* may initially appear ambitious, it is actually part of a sequence of scientific and technical precursors, starting with the *Hubble Space Telescope* (1990) and including the *Spitzer Space Telescope* (2003), *James Webb Space Telescope* (2018), and *WFIRST/AFTA* (mid-2020s). These missions have begun the hard work of developing and proving key technologies identified in Section 5.3, including deployment of segmented apertures, thermal and dynamical stabilization, starlight suppression, precision pointing, wavefront control, and other important functions. They provide a firm foundation for the incremental developments remaining to realize the size and performance that *HDST* needs.

*HDST* will also benefit technologically from developments in the commercial world, and from investments made by other countries and other government agencies. Detectors, electronics, structural materials, metrology systems, mechanisms, even large lightweight mirrors, are under development by others, and will provide options for *HDST*.

### 6.1.1 Sources of Heritage for *HDST*

With NASA and the space industry's long experience in designing, building, and launching telescopes, there are many projects that have informed the *HDST* concept. This section briefly touches on relevant mis-sions and development programs, before discussing the specific lessons learned for major *HDST* subsystems.

***The Hubble Space Telescope*** was the first large space telescope for astronomy, and since its 1990 launch and 1993 repair, it has arguably been the world's most productive UVOIR astronomical observatory. With a heavy 2.4 meter diameter monolithic primary mirror made of ultra-low expansion glass, and a massive, stiff, low-expansion composite metering structure to provide stable alignments between the primary and secondary mirrors and the optical instrument bench, *Hubble* is the exemplar of a conservative, passive technical approach to telescope construction. Using mass to provide stability has its costs, though, as it took the full capacity of the *STS Space Shuttle Discovery* to place *Hubble* in its low-Earth orbit. *Hubble* was also the first space telescope designed for on-orbit servicing, which became essential both to compensate for the initial mirror aberration, and to extend the lifetime and performance of the spacecraft and instruments.

***The James Webb Space Telescope (JWST)*** will launch with its 6.5 m segmented primary mirror folded into 3 sections, enabling it to fit into the 4.57 by 16 m shroud of its *Ariane 5* launch vehicle. After launch, the telescope will be unfolded, deploying the primary mirror, the secondary mirror tower, and a 15 × 21 m sunshade, while coasting out to its Earth-Sun L2 Lagrange point halo orbit, 1.5 million km away from the Earth. Hiding in the shadow of its sunshade, its mirror will cool to cryogenic temperatures, about 45 K, to provide a low background for its IR-optical instruments. For all this, *JWST* weighs little more than half of *Hubble*: 6,600 vs. 11,110 kg.

***The Space Interferometry Mission (SIM)*** developed technologies for a space observatory designed to capture the light from stars with multiple small telescopes, combining them coherently to make very precise measurements of the angles between the stars through Michelson interferometry, resulting in astrometry at the microarcsecond level. *SIM*'s primary mission to observe the wobble of the stars induced by motions of surrounding planets required extraordinarily accurate measurement of the geometry of the various collecting apertures. While *SIM* was cancelled before launch, early-phase work led to development of techniques for picometer-accuracy laser metrology: laser distance gauges combined in laser truss networks, used to measure the motions of the critical optical elements in their six degrees of freedom, were shown to be capable of estimating wavefront errors to picometer precision.

***The Terrestrial Planet Finder*** was a mission pre-project that developed two distinct system concepts for exoEarth discovery. *TPF-I*

was an IR-nulling, stellar-interferometer concept, using multiple free-flying space telescopes optically linked to synthesize a larger aperture. Commanded phase differences between the apertures would be used to cancel the light from the observed star, while passing the light from the planets orbiting the star. Multiple apertures, variable baselines and other features were intended to provide deep spectral characterization capabilities in a range of spatial frequencies. The infrared-wavelength range would be especially suitable for certain atmospheric molecules. *TPF-C* was to be a visible space telescope optimized for coronagraphy, making it an important technology precursor for *HDST*. *TPF-C* system studies considered different monolithic aperture configurations, including an elliptical 4 × 8 meter aperture packaged at an angle for launch in a 5 m launch vehicle shroud. Like *HDST*, *TPF-C* also included a large-format imaging camera for general astrophysics.

*The WFIRST/AFTA Project* will use an existing 2.4 m telescope with large secondary-mirror spider obscurations. The addition of a coronagraph to this project therefore places a new emphasis on starlight suppression that can work with non-optimal, obscured apertures. These approaches will also work with segmented apertures, making the *WFIRST/AFTA* coronagraph an important development for an *HDST* telescope.

NASA's Astrophysics-Related Technology Programs have steadily invested in maturing the technology needed for the next generation of missions. The NASA Cosmic Origins Program (COR) has funded development of large, monolithic glass-mirror technology, high-performance detectors, and high-reflectivity UV mirror coatings. The Exoplanet Exploration Program (ExEP) has sustained long-term development of critical starlight-suppression technologies, for coronagraphs and starshades alike. The *Exo-C* and *Exo-S* Probe-class mission studies examined the science and technologies for two small (1.4 m aperture) missions optimized for exoplanet detection, *Exo-C* using a coronagraph with a coronagraph-optimized telescope, and *Exo-S* using a starshade. Also under study is a Rendezvous Starshade Mission proposal, to fly a starshade for operation with *WFIRST/AFTA* after that mission is launched.

Advanced mirror technology projects under non-NASA funding have developed multiple lightweight segmented-mirror technologies that could provide mirrors suitable for an *HDST*, with a relatively small amount of further development.

### 6.1.2 Current and Developing Technologies

While *HDST* will be scientifically transformative, its technological requirements build directly on the progress already achieved by the above past investments in space technology. In subsystem after subsystem, the landscape is favorable for building a facility like *HDST*. These areas are discussed here in turn.

*Mirrors:* Since the launch of *Hubble*, there has been a steady evolution in space telescope mirror technology. *Hubble*'s mirror is a heavy, stiff monolith made of low-expansion glass. In comparison with *Hubble*, *JWST* has more than seven times the aperture area, at about half the mass, by using lightweight mirrors whose position and figure is held by actuators. NASA and others in the optics and space industries have continued to invest in mirror system development, with the goal of making mirrors that are lightweight, stable in the face of temperature changes, and easy to manufacture and build. Thanks to these investments there are a number of options that could provide mirrors for *HDST*.

*Lightweight ultra-low expansion (ULE or Zerodur) glass*, which can use conventional polishing techniques to achieve good optical figure. ULE glass provides a good level of passive thermal stability when operated near the design temperature: 0 to 20 C. With precise thermal control, ULE mirror stability will meet the 10 pm in 10 min target in support of high-contrast coronagraphy. To achieve light weight, such mirrors are made "closed back," with a thin-walled core structure sandwiched between glass facesheets and backsheets.

*Silicon Carbide (SiC)* has a thermal-expansion coefficient (CTE) at room temperature that is 100× higher than glass, but with a thermal conductivity 100× higher as well. This means SiC structures can be thermally controlled to hold their shape just as precisely as ULE, with more precise thermal control, but the SiC structure will equilibrate much faster than ULE. SiC is also a stronger, more robust material than ULE. This means that a SiC mirror will be lighter weight for a given stiffness, even with an open-back structure.

*Open-back low-expansion glass or Zerodur*: Open-back mirrors may be heavier than closed-back mirrors made of the same material to similar stiffness, but thermal control should be more rapid and precise. Similarly, an open-back glass mirror will be heavier than an open-back SiC mirror, but would share some advantages of thermal control.

Analysis indicates that each approach is capable of being thermally stabilized to meet the requirements of a coronagraphic telescope, but this remains to be demonstrated in practice.

When used as segments of a primary mirror, all of these mirror options would use rigid-body positioning actuators to enable precise co-phasing of the segments after launch and deployment. They might also require one or more shape actuators. Shape control can be used to ensure that the mirrors have the required optical figure, by correcting deformations that may occur due to mirror fabrication, mounting and coating stresses, strain relief, and temperature change. Shape control also can make mirror fabrication easier by relaxing hard-to-achieve tolerances, such as segment radius of curvature matching or edge figure error. A potential further benefit of a highly actuated primary mirror is as part of coronagraph wavefront control, with primary shape control replacing one of the deformable mirrors in the coronagraph. These benefits need to be balanced against the potential downsides of shape control. Actuators add new structural elements into the mirrors, which may reduce the passive thermal stability of the mirror, add complexity and cost for the actuators and related control systems, and introduce contaminants for UV-reflective optical surfaces.

All of these approaches offer not just better, lighter mirrors, but potential cost savings as well (although mirrors are usually well under 10% of the cost of a complete space observatory). Other cost-saving improvements are also being made on the manufacturing side. For example, improvements to proprietary polishing methods have occurred at some vendors and could benefit glass and SiC mirrors alike. High-volume production for the glass mirrors needed for large ground-based observatories may lead to cheaper production of *HDST* mirrors. Another way to reduce cost is to use replication methods in manufacturing the mirrors. Active Hybrid Mirrors (AHM) have demonstrated repeatable surface-figure error close to our requirement—under 15 nm after shape control—without any polishing of the mirror itself (Hickey et al. 2010). AHMs are made by bonding a replicated nanolaminate metal-foil reflective surface to a SiC substrate. AHMs require shape control to correct strains from the bonding process, and high-precision thermal control to achieve stability at the levels required for a coronagraph. There is also ongoing innovation in the manufacture of ULE mirrors, notably water-jet-cut lightweight mirror cores, low-temperature fusing of multiple stacked cores to the facesheet and backsheet and low-temperature slumping of the assembled blank on a mandrel (Matthews et al. 2014). These techniques could be used to make deep-core ULE mirrors of 3.5 m size or larger, at mass areal densities a factor of three lower than the approach used for *Hubble* (60 kg/m$^2$ vs. 180 kg/m$^2$).

The best approach for *HDST* should be determined in a future system-level architectural study. There will be design trades between different mirror-system designs, optimizing system mass and thermal performance against risk and cost, to meet the required optical performance. These trades should be grounded by experiment, and we recommend that NASA consider near-term demonstrations to compare thermal performance of multiple mirror approaches. Other trades will consider the degree of primary-mirror shape control. These trades should be done in concert with the coronagraph design and analysis, as there can be a division of labor between the deformable mirrors used in the coronagraph to achieve high contrast and actuators on the primary mirror.

*Wavefront Sensing & Control:* Fully mature wavefront sensing and control systems are an essential part of *HDST*'s development. Modern wavefront-sensing techniques have their origins in the primary-mirror flaw discovered in *Hubble* after its launch. NASA initiated a program to diagnose the problem and design a solution that could be implemented by servicing the instruments. Diagnosing the *Hubble* aberrations led to development of image-based wavefront-sensing techniques. Images of stars taken by *Hubble*'s optical instruments—at multiple focus settings and field points—were processed to generate high-resolution maps of the end-to-end system wavefront errors, and to localize these errors to particular optical elements, to separate the primary-mirror aberrations from other optics in the beam train (Redding et al. 1995). The successful correction of *Hubble* by new instruments installed by astronauts three years later validated image-based wavefront sensing as a method for high-accuracy optical metrology, and so it became the approach adopted for use, and further developed, by *JWST* (Acton et al. 2012).

Wavefront sensing has continued to mature in tandem with the development of segmented telescopes like *JWST* and Keck. *JWST*'s segments will be co-phased to a small fraction of a wavelength after they are deployed, using actuators commanded by wavefront sensing and control techniques like those used during the *Hubble* aberration recovery. The *JWST* primary-mirror segments and secondary mirror, once set, are expected to hold their state for a week or more. This same wavefront sensing and control system can readily be used for *HDST*'s larger apertures, increased numbers of segments, and control degrees of freedom. Further development has come through the *WFIRST/AFTA* coronagraph project, which is developing a Low-Order Wavefront Sensor (LOWFS), designed to operate on rejected light from the target star, to provide near-continuous wavefront measurement. This approach allows correction of wavefront drift effects that otherwise would degrade contrast, without interfering with scientific operation. The *WFIRST/AFTA* coronagraph team is planning a testbed to prove LOWFS performance; if successful, this technology will provide a method for continuous, rather than periodic, wavefront measurement and control that could be used with *HDST*.

*Stability & Metrology:* Wavefront control becomes progressively easier as a telescope becomes increasingly stable. Meeting *HDST*'s requirement on wavefront control therefore requires identifying ways to keep the telescope as stable as possible in the face of inevitable disturbances from changes in the temperature, orientation, and internal configuration of the telescope. Many early lessons were learned through *Hubble*, where it was discovered that its stability was limited not by its exemplary stiffness, but by thermal and dynamical disturbances. For example, *Hubble*'s original solar arrays experienced thermally-driven microdynamical "snaps" during the transition from sunlit to shadowed conditions, causing its pointing to jitter temporarily. Likewise, the *Hubble* system-focus experiences "breathing" of a few nm RMS during each orbit, on top of steady evolution over many years due to slow desorption of water from its epoxy-based composite structures. These subtle effects all have a measurable impact on the science data.

*HDST* will require outstanding dynamical stability for extremely low disturbance of the wavefront and line-of-sight pointing. This will necessitate high levels of isolation from on-board disturbance sources, such as attitude-control reaction wheels, cryocoolers, filter wheels and antenna or solar array drives. Structures will need to be well damped, so that recovery times after attitude maneuvers are short. There are several good isolation options for protecting the observatory from dynamic disturbances. One is a non-contact isolation system that physically disconnects the observatory from the spacecraft and uses voice coils or other methods to retain control in six degrees of freedom. This method isolates the telescope from reaction-wheel dynamics and has been assessed at TRL 5/6 for large observatories. When combined with additional levels of isolation and damping, such a system should be able to achieve picometer stability, making it a credible option for *HDST*. Another option is to use a micro-propulsion system to provide fine pointing, either replacing reaction wheels completely, or using wheels only for large slews, disconnecting the telescope during observations. Active and passive damping methods help control motions of structures like a secondary mirror tower.

Another strategy for increasing stability is to provide a fine line-of-sight pointing stage using a fine-steering mirror. *Hubble* does not have a fine-steering mirror, and the entire observatory must be pointed as a rigid object. *JWST* does have a fine-steering mirror, but *HDST* will need approximately 10× better performance. This requirement can be readily met because *JWST*'s fine-pointing sensor performance is limited primarily by sensor noise from the fine guidance camera; *HDST* is larger and operates at much shorter wavelengths, and thus will provide more photons and a finer point-spread function than *JWST*, minimizing this particular error term.

Existing technology such as non-contact isolation systems, damping and isolation component technologies, even microthrusters and steering mirrors, are all viable options for addressing *HDST*'s stability requirements. The final technical solution for maximizing *HDST*'s dynamical stability will depend on system-level integrated modeling to help identify the best approaches.

The stability of *HDST*, particularly during spacecraft moves, could be significantly enhanced by using picometer metrology to continuously monitor its structure. In laser metrology, laser distance gauges are combined in laser-truss networks that are capable of measuring the motions between the critical optical elements. A significant investment in picometer-accuracy laser metrology was made by *SIM*, which demonstrated sub-10-pm systematic measurement error and 1 picometer per root Hz measurement noise, using large, very stable beam launchers and cornercubes. Metrology was also proposed for use in *TPF-C*, to keep its secondary mirror aligned. Laser-truss technology has continued to develop since the *SIM* project ended, with new, much smaller and lighter beam launchers and cornercubes, suitable for mounting on lightweight optics such as primary-mirror segments, but operating at higher (~1 nm) error levels. These could be used to provide continuous high-bandwidth metrology of all key optical assemblies of an *HDST* telescope, enabling continuous control of line-of-sight or alignment-wavefront errors, by moving the fine-steering mirror, the secondary mirror, or primary-mirror segments. Such a system, if made sufficiently accurate, could simultaneously be very useful for high-contrast coronagraphy, by providing a method to keep the telescope phased during roll calibration maneuvers, or by providing high-bandwidth metrology of the beam train for feed-forward compensation by deformable mirrors in the coronagraph. Further development has the potential to achieve compact, lightweight, low cost, picometer-accuracy metrology, consistent with the needs of an *HDST* equipped with a coronagraph.

*Thermal Environment and Control:* *HDST* can take advantage of many of the more aggressive strategies used by *JWST* for thermal stability and control. Like *JWST*, *HDST* can operate in an "L2 halo orbit." Here L2 refers to the second Lagrange equilibrium point in the Sun-Earth system. The *JWST* spacecraft will slowly orbit this empty point in space, circling the Sun far beyond the Earth and Moon. By hiding behind a large 5-layer sunshade, *JWST* completely blocks the light from the Sun, Earth and Moon, keeping its temperature very cold (~50 K) and very nearly constant; *HDST* can follow a similar strategy, but with fewer layers in its sunshade. This will create a stable thermal environment, but at a higher temperature. *HDST* will use heaters to keep temperatures at the desired operating point, and precision thermal control augmented by wavefront sensing and control to keep thermal deformations at an absolute minimum.

Optimizing *HDST* for UV through near-IR imaging allows it to operate at much warmer temperatures than *JWST*. This difference will allow *HDST* to utilize lower-cost optical materials, incur less thermal stress in the structures, simplify component and system design, manufacturing and qualification, and lower the costs of system integration and testing, much of which has to be done under cryogenic conditions for *JWST*. The downside of a warm *HDST* telescope would be (1) higher background noise for imaging in the mid-IR bands, beyond about 1.7 microns wavelength, and (2) the need to provide higher levels of power for heating parts of the telescope to maintain the warmer temperatures. Note that warmer temperatures for the telescope do not rule out cold temperatures in the instruments. Exactly how far out in the IR to go will be an important decision for *HDST* scientists and system designers.

***Starlight Suppression—Coronagraphs:*** Of the technological areas with substantial heritage that we discuss here, starlight suppression is one of the most rapidly evolving, showing dramatic improvements in likely capabilities of both coronagraphic and starshade approaches, thanks to sustained, long-term investments by *TPF*, *WFIRST/AFTA*, and NASA's Exoplanet Exploration Program.

For coronagraphs, investments have led to the development of multiple designs and technologies, culminating in testing of subscale-coronagraph instruments. Key technologies include occulters, apodizers, deformable mirrors, "speckle-nulling" control algorithms, ultra-low noise detectors, continuous wavefront sensing and control, image calibration, and PSF subtraction image processing. Their performance has been demonstrated using laboratory testbeds, most notably the High-Contrast Imaging Testbed (HCIT), a highly stable, vacuum, high-contrast coronagraph testing facility, and some on-sky results from ground-based observatories.

Most past coronagraph development focused on designs for space telescopes that are optimized for coronagraphy—telescopes with unobscured apertures like *TPF-C*. Development began with the classic Lyot coronagraph design, with an occulting mask to block light at the center of the field, and a Lyot pupil stop to suppress light scattered to the edge of the beam. Use of high-quality deformable mirrors and precisely shaped occulters, along with effective control algorithms, helped to reach *TPF* Milestone 2 in 2009, demonstrating a contrast of $5 \times 10^{-10}$ at an IWA of 4 $\lambda/D$ and a 10% bandpass on HCIT. Since then, multiple new coronagraph designs have been built and tested including: Phase-Induced Amplitude Apodization Complex Mask Coronagraph (PIAACMC, Guyon et al. 2013); Shaped Pupil Coronagraph (SPC); Apodized Pupil Lyot Coronagraph (N'Diaye, Pueyo and Soummer 2015), Vector Vortex Coronagraph; and the Hybrid Lyot Coronagraph (HLC). There are also Visible Nulling Coronagraphs (VNCs), that use pupil-plane interference effects to null on-axis light, following the *TPF-I* model, but with a single aperture; an example is provided in Lyon et al. (2015). The best raw contrast performance yet measured was produced by the HLC, achieving $1.2 \times 10^{-10}$ raw contrast at 2% bandpass, and $1.3 \times 10^{-9}$ at 20% bandpass on the HCIT. The current demonstration status of these approaches, compared to *HDST* goals, is shown on Table 6-1.

Perceptions changed once designers began to study the "unfriendly" *WFIRST/AFTA* obscured pupil, with its lacework of shadows caused by its secondary mirror and support spiders. New research has shown that, counter to earlier thinking, coronagraphs can be designed to provide high contrast even with a complex obscured pupil. Recent work on the HLC design has demonstrated deep polychromatic contrast for *WFIRST/AFTA*, at some cost in throughput and PSF quality, provided the wavefront control system for the telescope's deformable mirrors is co-designed with the coronagraph. Work on the PIAACMC approach has also been promising.

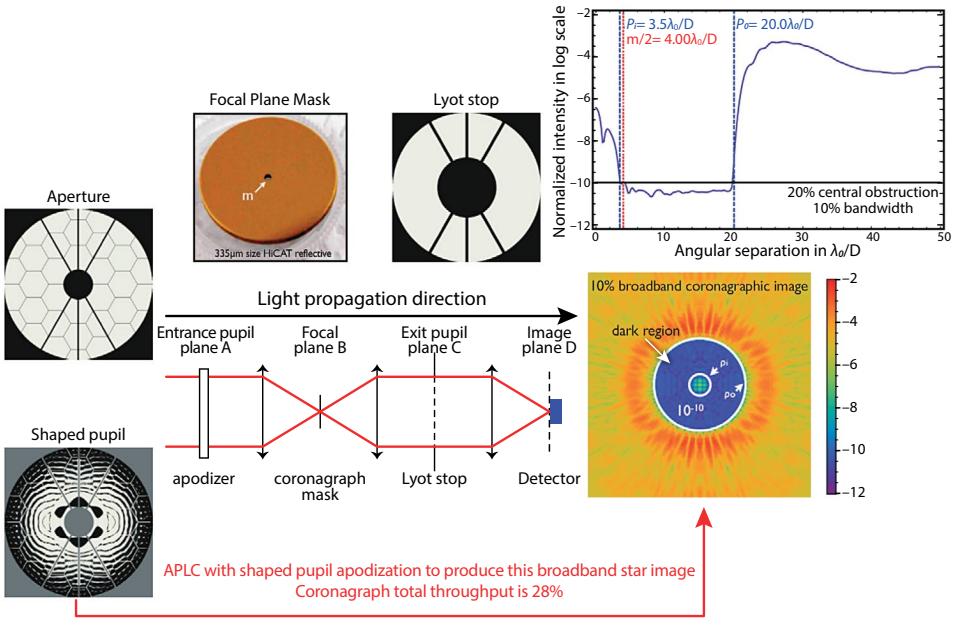

Figure 6-1: A coronagraph designed to yield high contrast with an *HDST*-type segmented, obscured aperture (M. N'Diaye et al., in preparation). This design combines shaped pupil and Lyot coronagraph techniques to obtain $1 \times 10^{-10}$ raw contrast or better, for an IWA of 3.6 $\lambda/D$.

PIAACMC uses beam shaping to compensate the effects of obscurations, and is theoretically able to achieve good performance with any pupil. Current demonstrations for its performance on *WFIRST/AFTA* show a raw contrast of $5 \times 10^{-10}$ at an IWA of 2 $\lambda$/D in monochromatic light.

Total contrast performance of a coronagraph can be significantly improved over its raw contrast by Point Spread Function (PSF) calibration and image post-processing methods. For instance, rolling the telescope about its line of sight to a target star will cause speckles due to telescope aberrations to remain fixed against the changed scene, allowing identification and removal of those artifacts in post-processing. Similarly, observing the same scene at multiple different wavelengths changes the location of diffraction-induced artifacts against a fixed scene, enabling their removal. PSF subtraction has achieved factors of 100–500× contrast improvement (raw contrast/total contrast) for *Hubble* images taken with the NICMOS coronagraph (Soummer et al. 2011). For *WFIRST/AFTA* and for *HDST*, with much higher raw contrast, less but still important amounts of improvement can be expected. The acting assumption for the *WFIRST/AFTA* coronagraph is that total contrast will be 10–30× higher than raw contrast.

How close is the current state of the art to meeting the *HDST* goal of detecting and characterizing dozens of candidate exoEarths? The exoEarth yield of a 10 m aperture would be about two dozen under the assumptions for natural parameters ($\eta_{Earth}$ = 0.1, $\eta_{zodi}$ = 3), as discussed in Stark (2015). Yield for a 12 m under the same assumptions would be about three dozen, to provide the numbers of detections needed for a strong statistical result, as discussed in Chapter 3. The 12 m aperture would also assure strong scientific return in the event the natural situation is less favorable. The coronagraph performance needed to produce these results can be divided into two different collection modes. In survey mode, the coronagraph would be configured for wavelengths shorter than 0.55 micron, and optimized for a deep contrast of $1 \times 10^{-10}$, but at a larger IWA of 3.6 $\lambda$/D. In characterization mode, the coronagraph would provide a spectral resolution of 70 for wavelengths up to 1 micron, but with lower contrast of $1 \times 10^{-9}$ at an IWA of 2 $\lambda$/D, to cover the same observing zone. These and other parameters represent the goals for *HDST* coronagraph technology, as summarized on Table 6.1.

At a very preliminary level, researchers are beginning to explore coronagraphs specifically designed for the (obscured and segmented) *HDST* pupil. The Active Correction of Aperture Discontinuities (ACAD) approach, using two deformable mirrors in the beam train in combination with an Apodized Pupil Coronagraph (APC) to reduce the scattering effects of the pupil obscurations and segment gaps, is predicted to achieve $5 \times 10^{-10}$ raw contrast, with an IWA of 3 $\lambda/D$ and a 10% bandpass (L. Pueyo 2015). Another design combining an APC devised specifically for an *HDST*-like pupil with a Lyot stage is predicted to achieve raw contrast of $1 \times 10^{-10}$ from 4 to 20 $\lambda/D$ with a bandpass of 10% and core throughput of 18%. This is very close to *HDST* goals, even before PSF calibration processing (see Figure 6.1; M. N'Diaye 2015). One challenge for this approach is in the fabrication of the apodizing mask. A PIAACMC design for a 12 m segmented telescope (Guyon 2015), similar to the *WFIRST-PIAACMC*, obtains a high throughput of 70%, a very small IWA of 1.2 $\lambda/D$, a dark hole extending to 30 $\lambda/D$ (limited by DM actuator pitch) and a point-source raw contrast better than $1 \times 10^{-10}$. At small IWA, the stellar angular size increases the raw contrast above $1 \times 10^{-9}$, and future designs will reduce this term by jointly optimizing the focal plane mask design and wavefront correction state (as was successfully done for the *WFIRST/AFTA* coronagraph). This design has the potential to increase the discovery space close to the host star and allow long wavelength characterization out to approximately 2 microns (as shown in Figure 3-18, which includes, for the design at hand, residual photon noise due to stellar angular size).

Table 6-1 shows that, at least for unobscured apertures, current lab demonstrations are also approaching *HDST* science goals. Past and ongoing efforts have pushed the limits in both IWA and contrast to *HDST* goal levels, but not all on the same experiment, and not for the *HDST* aperture. To truly understand the likely performance of an *HDST* coronagraph will require a concerted effort to address the particular *HDST* goals, combining strengths of multiple approaches and implementations. This will require additional, *HDST*-specific study funding, to develop segmented-aperture designs, and to demonstrate them in lab tests.

The ongoing *WFIRST/AFTA* Coronagraph demonstration efforts provide an excellent starting point for *HDST* coronagraph design and demonstration. The nascent efforts to achieve similar performance with segmented apertures promise rapid progress towards *HDST* goals. Additional investment to address *HDST* coronagraphy is recommended, to provide accurate prediction of *HDST* coronagraphic star-suppression performance in advance of the 2020 Decadal Survey.

***Starlight Suppression—Starshades:*** Parallel to the improvements in coronagraphy, starshade technologies are developing equally rapidly. Current starshade designs differ by the size and shape of the petals that surround the fully obscuring central disk. Shade size, petal number and size, and the operating distance from the space telescope are parameters

## Starlight Suppression Goals and Achievements to Date

| | Method | Description | Pupil | Raw Contrast | IWA | OWA | Bandpass | Throughput | Comments |
|---|---|---|---|---|---|---|---|---|---|
| **HDST Goals** | Science goal | 24 exoEarths at D=10 m, 36 exoEarths at D=12 m | | | <50 mas up to λ=1 μm | >1.5 asec | | | 50 mas corresponds to Sun–Earth separation at 20 parsec |
| | Coronagraph goals | Survey mode | Obscured, segmented | 1.00×10⁻¹⁰ | <3.6 λ/D up to λ=0.55 μm | >75 λ/D | 20% | 10–30% | Challenges are (1) telescope stability, (2) throughput, (3) bandpass, (4) contrast |
| | | Characterization mode | | 1.00×10⁻⁹ | <2 λ/D up to λ=1 μm | >90 λ/D | R=70 | | |
| | Starshade goals | Required performance for a 10 m aperture, >80 m Star-shade, at 166 Mm distance, with Fresnel number F=12 | Any | 1.00×10⁻¹⁰ | <50 mas up to λ=1 μm | Inf | λ≤1 μm | 100% | Challenges are (1) large size of Starshade spacecraft, (2) deployment, (3) retargeting time, (4) petal shape precision, (5) formation flying sensing |
| **Coronagraph Demos** | Lyot coronagraph demonstration (2009) | Demonstrated on the subscale High-Contrast Imaging Testbed (HCIT-1 or HCIT-2) | Unobscured, circular | 5.00×10⁻¹⁰ | 4 λ/D | 10 λ/D | 10% | | Demonstration for TPF Milestone 2 |
| | Hybrid Lyot coronagraph demonstration | | | 1.20×10⁻¹⁰ | 3.1 λ/D | 15.6 λ/D | 2% | 56% | Current best demonstrated results; achieved with a linear mask |
| | | | | 3.20×10⁻¹⁰ | | | 10% | | |
| | | | | 1.30×10⁻⁹ | | | 20% | | |
| | Shaped-pupil coronagraph | | | 1.20×10⁻⁹ | 4.5 λ/D | 13.8 λ/D | 2% | 10% | |
| | | | | 2.50×10⁻⁹ | | | 10% | | |
| | PIAA coronagraph | | | 5.70×10⁻¹⁰ | 1.9 λ/D | 4.7 λ/D | 0% | 46% | |
| | | | | 1.80×10⁻⁸ | 2.2 λ/D | 4.6 λ/D | 10% | | |
| | Vector vortex coronagraph (2013) | | | 5.00×10⁻¹⁰ | 2 λ/D | 7 λ/D | 0% | | From Serabyn et al. (2013) |
| | | | | 1.10×10⁻⁸ | 3 λ/D | 8 λ/D | 10% | | |

Table 6-1. Starlight Suppression Goals and Status.

## Starlight Suppression Goals and Achievements to Date (*Continued*)

| | Method | Description | Pupil | Raw Contrast | IWA | OWA | Band-pass | Through-put | Comments |
|---|---|---|---|---|---|---|---|---|---|
| Coronagraph Demos (Continued) | Visible Nuller Coronagraph (2012) | Demonstrated in ambient lab conditions at GSFC | Segmented pupil | $5.30 \times 10^{-9}$ | $1.5 \lambda/D$ | $2.5 \lambda/D$ | 2% | | 6% bandpass to be demonstrated in early CY15 |
| | Shaped-pupil coronagraph | Demonstrated on HCIT | Obscured AFTA pupil | $5.90 \times 10^{-9}$ | $4.5 \lambda/D$ | $10 \lambda/D$ | 2% | 11% | First lab demo of high contrast with AFTA pupil; 10% bandpass milestone scheduled for September 2015 |
| | Hybrid Lyot coronagraph demonstration | Demonstrated on HCIT | | $6.92 \times 10^{-9}$ | $3 \lambda/D$ | $9 \lambda/D$ | 0% | 13% | Work in progress for AFTA, 10% bandpass milestone scheduled for September 2015 |
| | Future PIAACMC coronagraph demonstration | To be demonstrated on HCIT | | $\sim 1 \times 10^{-8}$ $\sim 1 \times 10^{-9}$ | $1.3 \lambda/D$ $2 \lambda/D$ | $9 \lambda/D$ | 10% | 58% | Work in progress for AFTA; predictions based on modeling by Krist (2015) |
| Starshade Demos | Tabletop lab demo | Princeton, F=588 | | $1.00 \times 10^{-10}$ | 400 mas | 800 mas | 0% | 100% | |
| | Desert demo | Demonstrated at km scale using artificial star, F=240 | Any | $\sim 1 \times 10^{-08}$ | 70 arcsec | n/a | 60% | 100% | NGAS TDEM demo; updated numbers from recent demo expected in summer CY15 |
| | Large-scale shape TDEM-1 | Full-scale petal | | n/a | n/a | n/a | n/a | n/a | Petal built to $1^{-10}$ system contrast requirements |
| | Large-scale deployment TDEM-2 | 2/3-scale truss | | n/a | n/a | n/a | n/a | n/a | Truss with petals deployed to 1e-10 system contrast requirements |
| | Future tabletop lab demo | Princeton, F=15 | | $3.00 \times 10^{-10}$ | equiv 82 mas, for D=2 m | TBD | >50% | 100% | Approximately to scale, matching the target Fresnel number; Princeton TDEM demo starts CY15 |
| | On-sky demo | NGAS demo at McMath Solar Observatory, F = 19 | | $1.00 \times 10^{-6}$ | 75 arcsec | N/A | 60% | 100% | First on-sky use of a starshade, using a large flat mirror to direct starlight past a small mask and into a camera. |

that are driven by the telescope aperture size and wavelength, and by the desired IWA and contrast. Some of these quantities are incorporated into a dimensionless optical term, the Fresnel number (denoted by the letter F), which captures the tradeoff between starshade size and distance and suppression performance. A large Fresnel number indicates a mask that is optically far from the camera, hence large. A plausible target value for the Fresnel number for *HDST* is F = 12, for a smaller, less distant mask, but one that is aggressively shaped. Critical starshade technologies include precision petal-edge shaping, formation flying, and deployment from a folded launch configuration. The most serious limitation of starshade operations is inefficiency. There is a large time delay incurred when changing target stars, which significantly reduces the scientific yield, while incurring substantial costs in fuel consumption.

Progress has been made in developing deployment models for starshades. A 30–40 m starshade compatible with *WFIRST/AFTA* can be folded up and packaged in an Enhanced Expendable Launch Vehicle (EELV) 5 m diameter shroud, and then deployed after launch. A ⅔-scale pathfinder starshade was built and tested through many deployment cycles. It utilizes a well-established radial-truss design for the central disk, supporting petals that unroll and deploy with great precision. This approach can provide starshades up to 60 m in diameter. Other designs also under development are projected to be able to deploy up to 80 m in diameter. Yet a 12 m *HDST* would need a starshade of 100 m, operated at a distance up to 200,000 km, for optimal performance. A design to provide a starshade of that size is a priority for *HDST*.

In practice, the size, distance, and formation-keeping required for a starshade to operate make it impossible for ground-based testing near full scale, so much attention is being given to developing and validating the optical models used for design and for predicting performance. Lab testing has demonstrated performance of $1 \times 10^{-10}$ contrast, but at a Fresnel number of 588. Subscale testing over longer distances on the desert floor has provided dramatic demonstrations, achieving contrast of $1 \times 10^{-8}$, but again with a non-representative Fresnel number (240). These demonstrations have been useful in confirming certain predicted effects, such as the impact of light leaks caused by edge imperfections. First on-sky results have been achieved at the McMath-Pierce Solar Telescope, using a flat heliostat mirror to direct starlight past a subscale mask and into a small telescope, operating at a Fresnel number of 19 and a contrast of $1 \times 10^{-6}$ (Warwick et al. 2015).

*Design Modeling:* Another key technology, maturing over time, has been the development of analyses and computer models of telescope and instrument performance. Models that integrate the structures, thermal systems, optics, controls, and scientific performance of an observatory are essential for quantifying design decisions at all levels and stages of development. *JWST* and other more recent projects have applied a systematic approach to integrated modeling, including verification and validation of model predictions. A significant part of testbeds such as HCIT is their role in verification and validation of such models. Comparing model predictions to testbed data has advanced the understanding of coronagraph errors, where even so trivial a thing as a dust particle in the beam train can have a major impact. Modeling and model verification and validation will continue to be a vital activity for *HDST*.

*Detectors:* The progression from *Kepler* to *Gaia* to *Euclid* and *WFIRST/AFTA*, is ushering in the era of gigapixel-class detectors in space. Some detector and sensor systems have seen rapid growth in capability similar to "Moore's Law" for computing technology, and a strategic program should be developed to exploit industry gains as the *HDST* design progresses. Specialized detectors such as low-noise and photon-counting silicon-based Charge-Coupled Devices (CCDs, for the UV and visible), photon-counting microchannel plate detectors (for the UV) and, sensitive Mercury-Cadmium-Telluride arrays (for near-IR) have been customized for astrophysical applications, while also benefiting from other technology drivers and competition in a broader market.

The programmatic division in detector development effort, split between the UV, visible, and near-IR wavebands is based on physical differences in required detector technologies. In the UV, the low efficiency of early silicon-based devices, and the utility of solar-blind detectors with ultra-low noise, has driven much of the effort towards photon-counting photo-emissive devices such as microchannel plate detectors (MCPs). Significant development work on MCPs preceded their use on *EUVE*, *FUSE*, *GALEX*, and *Hubble*'s UV instruments: STIS, ACS-SBC, and the Cosmic Origins Spectrograph (COS). The NASA Astronomy and Physics Research and Analysis (APRA) and Cosmic Origins program has supported recent technology development for MCPs for the UV, which has mainly been focused on increasing quantum efficiency (QE; currently 8–25% across the UV), dynamic range, format, and resolution, as well as improving lifetime and gain performance and stability associated with high-voltage operation, photocathodes and the MCP bulk substrate. This work has also leveraged technology development supported by the particle physics community. Because of their low-noise, photon-counting and solar-blind capabilities, MCPs remain a viable candidate for use in *HDST*'s UV instruments.

UV-sensitive CCDs were developed for *Hubble* instruments using phosphor coatings to boost quantum efficiency. This performance was originally limited by QE, stability and operational issues. More recently, surface processing of thinned, back-side illuminated CCDs (e.g., "delta-doping") has raised internal device QE performance to near-unity, limited by silicon reflectance to ~40% in the UV. Recent COR and APRA-supported work on anti-reflective coatings and surface processing of a new generation of CCDs, electron-multiplying CCDs and CMOS devices has obtained high QE (> 50%) in the UV and near-photon-counting noise performance. When coupled with band-limiting optical coatings to limit red leaks, UV-sensitive silicon-based devices will become competitive with MCPs. In particular, the UV-CCD effort can exploit the associated development of large-format visible CCD-imaging arrays in astronomy and the rapid gains seen in industry for scientific and commercial applications. Operational challenges associated with high voltage for MCPs and contamination control for cooled-silicon detectors will continue to be one of the many system and instrument-level trade-offs driving final selection of UV detectors for *HDST*.

Visible and near-IR arrays can build on significant development work both in ground and space, most notably the development of visible silicon CCD detectors and large format near-IR-sensitive HgCdTe arrays for *Hubble*, *Gaia*, *JWST*, *WFIRST/AFTA*, *Euclid* and LSST. Recent work supported by COR and other agencies has focused in three main areas: 1) development of large format gigapixel arrays for astronomy instruments and missions and lowering cost per pixel, 2) pushing towards low-noise photon-counting devices, 3) addressing second-order performance issues (e.g., charge transfer efficiency, dark current/other noise sources, radiation damage, power). *HDST* will benefit from progress in all of these areas. In particular, low-noise, photon-counting devices are required for exoplanet spectroscopic characterization, and these same devices may also be suitably modified to produce low-noise detectors for the UV.

*Coatings:* The *HDST* primary and secondary mirrors are expected to obtain high UVOIR reflectance using an aluminum metallic layer plus a protective dielectric overcoat. For *Hubble*, aluminum with a magnesium fluoride overcoat provided high reflectance from 115 nm into the near-IR, and similar industry-standard coatings have also been used on *GALEX* and other space astrophysics and smaller missions. To reach shorter wavelengths, *FUSE* employed a LiF overcoat, which is hygroscopic, presenting a challenge during handling on the ground, and which has a lower reflectance above 115 nm. NASA COR is currently supporting efforts to improve reflectance of broadband coatings at even shorter wavelengths (< 92–110 nm). Recent developments show promise for producing high-reflectance UV-Near-IR broadband mirrors, including atomic layer deposition, a new generation of protective materials, and nanotechnologies and surface processing that may allow the aluminum layer to maintain its intrinsic high reflectance. It has been noted that coatings will produce variations across the aperture that may impact imaging precision and downstream starlight suppression; these effects are being studied. In particular, further work is required to understand the impact of polarization by metal and dielectric-metal films on coronagraph performance, and to develop strategies for mitigating any negative effects. As was the case for coronagraphy with segmented apertures (which was originally thought to be unable to deliver high performance) it appears likely that focused study can find solutions that obtain the required starlight suppression performance with these coatings. Until such detailed system-level studies are performed, it is impossible to reach strong conclusions (pro or con) regarding the compatibility of internal coronagraphs with particular mirror coatings, including UV-optimized mirror coatings.

More broadly, community efforts have been directed towards four main goals: 1) achieving end-of-mission UV-Near-IR reflectance performance as close as possible to that of pristine, bare Al; 2) enhancing Al coatings to boost reflectance in specific regions (e.g., UV, visible) while maintaining high-broadband performance; 3) developing and optimizing a combined reflectance and coronagraphic polarization impact figure-of-merit; and 4) obtaining ultra-high reflectance in specific wavelength bands (using other metallic, dielectric or combined coatings). Progress has been made in all areas, with the first being the highest priority goal.

*Servicing:* One of the lessons learned during *Hubble*'s operation was the value of servicing. Five servicing missions were flown during the first 19 years of the *Hubble* mission. These installed new optical instruments and replaced worn-out spacecraft components, such as gyros and solar panels. The new instruments compensated for the primary-mirror aberrations and used improved designs and technology to significantly enhance *Hubble*'s scientific capabilities, thus keeping it at the forefront in UV spectroscopy, wide-field high-resolution imaging, and other areas. Without the repairs, *Hubble* would likely have failed in the early 2000s due to loss of gyros. Its life has been extended by many years, hopefully until 2020 or beyond. This experience speaks powerfully to the advantages of servicing space telescopes.

The NASA Space Shuttle infrastructure that made astronaut servicing of *Hubble* possible has been retired. However, it could be replaced by robotic spacecraft or by future astronaut capabilities. NASA studied robotic servicing as an alternative to the last servicing mission, and has continued active development. Robotic servicing and refueling has already been successfully demonstrated, in 2007, by the NASA and the Defense Advanced Research Projects Agency (DARPA) Orbital Express mission. NASA completed a study (NASA 2010) based on the *Hubble* experience, and maintains the Satellite Servicing Capability Office. Robotic servicing is planned for *Hubble*'s end-of-life de-orbit or boost, and is being considered for the *WFIRST/AFTA*. Indeed, the NASA authorization act in 2008 included Section 502, calling for such servicing capabilities to be investigated. Robotic servicing to enable replacement of instruments and critical spacecraft elements would present a very attractive option for *HDST*, and for other spacecraft as well. When fully developed, robotic servicing will enable the use of relatively small launch vehicles to service equipment—even at remote locations such as the Sun-Earth L2.

### 6.1.3 Heritage Summary

The entire rich heritage described above demonstrates how astrophysics-related technology programs have successfully invested in technologies that have application to a future *HDST*. This is not an accident, since the highest-priority technologies for an *HDST* mission were identified in the Astro2010 decadal survey as targets for development, a recommendation reinforced recently by the report of the NRC, and acknowledged in the NASA 30-year Roadmap for astrophysics. What has changed since the 2010 Decadal Survey is not a need for a new direction; rather, in response to new scientific insight, there is a need for incremental additional development, building off current programs and projects. The following section describes how well-targeted investments will lead naturally to the remaining improvements needed for *HDST*.

## 6.2 Critical Enabling Technologies Prioritized

The baseline *HDST* identified in Section 5.2 has a 12 m class segmented primary mirror which folds to fit into the shroud of an EELV-class launch vehicle, and uses a coronagraph for starlight suppression. With this notional mission, which of the technologies just reviewed are critical for an *HDST* observatory? Which are enhancing? Which need additional development?

Table 6-2 provides a summary of the key technologies, noting the performance needed for *HDST*, the current status, and the heritage—all drawn from the discussion in the previous section. Many of these technologies are close to being ready for *HDST*, but some need further development. A subset of these developing technologies is critical to meeting *HDST* science requirements, and thus must be matured before an *HDST* mission can be confidently proposed.

| Technology Category | Technology Needed for HDST | | | | Current Status | | | |
|---|---|---|---|---|---|---|---|---|
| | Technology | Performance Goal | Details | Heritage | Current Performance | Maturity of Goal Perf. | Priority |
| Coronagraph | Segmented aperture coronagraphy | Raw contrast <1×10⁻⁹ | Image-plane and/or pupil-plane coronagraph designs | WFIRST/AFTA coronagraph, ExEP studies, *TPF* | Unobscured aperture, contrast <1×10⁻⁹ | Developing | Highest |
| | Continuous speckle nulling WF control | WF sensing error <5 pm | | | <5 pm | Developing | Highest |
| | PSF subtraction | 10–30x contrast reduction | PSF matching, roll calibration, etc. | *Hubble* | 100x contrast reduction on noisier images | Developing | Highest |
| Segmented mirror system | Mirror segments | <20nm WFE; <5pm WFE drift/10 min | Improve production to reduce cost and lower mass; UV performance | Non-NASA MMSD; NASA AMSD, COR/AMTD, Industry R&D | ULE and SiC substrates to 1.4 m size, <30 nm WFE, actuated | Substrate: TRL 4+; System: TRL 3 | Highest |
| Ultra stability | Mirror thermal control | pm stability for coronagraph | Combining passive and active methods | Non-NASA; NASA various | nm stability | TRL 4 | Highest |
| | Dynamically stable structures | Picometer stability | Dynamically stable structures and fixtures | *JWST* | nm accuracy | TRL 3 | Highest |
| Starshade | Advanced starshade design | D ≥80 m, F=12 | Deployment; edge precision; long life | NASA ExEP; Industry R&D | D to 40 m | Developing | Highest |
| Sensitivity and throughput | Ultra-low noise and UV-sensitive detectors | Detector noise & QE: Read: <0.1-1 e-; Dark <0.01 e-/s; QE(FUV): >50% | Exoplanet spectroscopic characterization and UV general astrophysics | NASA COR, commercial sources | Low noise, high QE photon-counting Vis-NIR and UV detectors | TRL 4–6 | Highest |
| Wavefront control | Metrology (pm) | Picometer precision | Compact, lightweight laser truss metrology | *SIM*, Non-NASA | nm accuracy | TRL 3 | Highest |

Table 6-2: Critical technologies, prioritized.

NASA is already developing most of these technologies: for *JWST*, under the COR and ExEP Programs; or for the *WFIRST/AFTA* Coronagraph project, as discussed in Section 6.1. Additional development specifically addressing *HDST* constraints (such as the segmented aperture) and *HDST* goals will be required, however. The good news is that the further development needed for *HDST* builds on these existing efforts, exploiting progress already made, enabling future progress at lower cost, while substantially reducing risk associated with the mission.

Some of the key technologies, and their application, will be the subject of trade studies normally carried out as part of the mission design process. For instance, is it better to have active segment figure control, using actuators which might decrease thermal stability; or is it better to polish the mirrors to absolute perfection, taking the risk that uncontrollable figure errors might creep in after launch? Another example is the choice of a vibration suppression methodology, which will depend essentially on observatory configuration, mass properties, attitude control methods, etc. These sorts of system architecture trades require good understanding of the traded technologies and options. For some technologies this will require near-term development. Completion of such trades, however, should be delayed until a full system design study is underway.

The immediate, high-priority technology investment areas are those that are fundamental to mission feasibility, and that need further development prior to the 2020 Decadal Survey. These pertain to the hardest problems, especially starlight suppression, and govern the fundamental *HDST* architectural approach. Will a coronagraph meet *HDST* science requirements, or is a starshade the only way forward? Is there a way to provide sufficient stability for a coronagraphic instrument? There are the beginnings of an approach, utilizing active thermal control of mirror segments and the underlying structures, plus active controls combining technologies such as continuous coronagraphic wavefront sensing, picometer metrology, deformable mirrors and segmented deformable mirrors, but the path forward relies on the performance of these technologies being well established. The *WFIRST/AFTA* coronagraph is exploring elements of this, but *HDST*-specific analysis and development is required as well. By investing now in studies and demonstrations matched to *HDST*, NASA will have a strong basis for a decision in 2020.

The very first priority is the development of *high-contrast coronagraphs* for the *HDST* baseline mission. The challenge is to design an instrument that can meet the contrast, IWA, bandpass and throughput requirements, for a segmented, obscured aperture *HDST*. Elements of this activity will include: coronagraph design, drawing on the various image-plane and/or pupil-plane methods; performance modeling and model validation; contrast control algorithms; microarcsecond line-of-sight pointing control; wavefront picometer stabilization; image post-processing; plus devices such as occulters, apodizers, and beam-shaping mirrors. This work should include small-scale demonstrations.

| Technology Category | Technology Needed for Candidate Architectures ||| Current Status |||| |
|---|---|---|---|---|---|---|---|---|
| | Technology Provided | Performance Needed | Details | Heritage | Current Performance | Maturity | Priority | Scope |
| Ultra stability | Structural Thermal Control | pm stability for coronagraph, nm for starshade | | Non-NASA; NASA various | µm stability | TRL 3 | High | Architecture |
| | Non-contacting vibration isolation | 140 dB Isolation | | Industry R&D | 80 dB isolation | TRL 5 | High | Architecture |
| | Micro-thruster pointing control | Ultra-low vibration; uas pointing control | An option for low-disturbance LOS pointing, in lieu of RWs | Industry | | TRL 3 | High | Architecture |
| Monolithic mirrors | 4 m monolithic mirrors | WFE < 20 nm; to 14" thick and 4 m wide for f >60 Hz | 4 m monolith may provide a reduced-performance option | NASA COR/AMTD | to 14" thick and 30 cm wide | TRL 4 | High | Architecture |
| | 8 m monolithic mirrors | WFE < 20 nm; to ~30" thick and 8 m wide for f >60 Hz | Launch requires development of SLS Block 2 10 m fairing | NASA COR/AMTD | to 14" thick and 30 cm wide | TRL 3 | Low | Architecture |
| Spacecraft | Deployment | 2-fold, 6.5 m aperture | S/C architecture dependent | JWST | 2-fold, 6.5 m aperture | TRL 6 | Medium | Architecture |
| | Sunshade | T to ~90K, gimballed | S/C architecture dependent | JWST | T to 30K, fixed | TRL 6 | Medium | Architecture |
| Wavefront sensing & control | WFS&C for initial alignment | <10 nm | Can use qualified methods | Non-NASA; NASA JWST | <10 nm | TRL 6 | Low | Architecture |
| Coronagraph | Segmented DMs for coronagraph WF control | Segmented to match PM, <10 pm WFC | Needs control concept development | Industry R&D | Segmented facesheet, <100 pm WFC | TRL 3 | Medium | Device |
| | DMs for coronagraph WF control | <10 pm WFC | Demo'ed at 32x32, needs scaling up to 48x48 | NASA ExEP | Continuous facesheet, <100 pm WFC | TRL 4 | Highest | Device |
| Segmented mirror system | RB Actuators | pm accuracy, 1 Hz operation, infinite life | | Non-NASA; NASA JWST; Industry | cm stroke, nm accuracy | TRL 4–6 | Medium | Device |
| Throughput | Coatings | Below 120 nm >50–70%, UV/Vis: >85–90% | | NASA COR | High-reflectance FUV mirror coatings | TRL 4–6 | High | Device |
| Science data processing | PSF Subtraction | Additional 1×10⁻¹ contrast | Low cost | Hubble | | Operational | Highest | Science |

Table 6-3. Key architecture-specific, device, and science technologies.

The second-highest priority is to address *segmented-mirror system technologies*, including the mirrors, structure and mounts. The key issues are thermal and dynamic stability, but others are also important: optical performance; cost; mass; robustness; efficiency of manufacture; and the potential for control. Mirror substrate technologies are relatively mature, at TRL 4+ or better, but the other elements of the mirror systems—thermal control, stable composite structures, actuation, mounts, etc.—are less so. NASA, through the COR/AMTD program, has developed innovative monolithic mirror technologies. Investment should now also be made to advance UVOIR segmented-mirror technologies, as needed for evaluation of *HDST*. This investment can be made without great expense, by exploiting existing hardware and facilities to perform near-term analysis and testing that will establish high readiness for mirror segment systems, tracing to the coronagraph-level stability requirements.

While all aspects of stability are important, a key challenge is *dynamic stability*, the resilience against disturbances that are too fast to be easily correctable by sensing and control strategies, due to signal-to-noise limitations in the measurements. Experience from *JWST* suggests tip/tilt modes of the mirror, and substrate dynamic deformations driven by mirror rigid-body modes, can be significant factors in system stability. While isolating disturbance inputs is essential, it is also critical that large structure technologies like advanced composites, and mounting structures like hexapods, be made stable into the picometer regime.

The integrated modeling team working on the *ATLAST* design concept recently completed a preliminary set of system wavefront predictions with a model that includes the key elements in the stability of the system. The integrated model they developed leverages *JWST* models and included reaction wheels, isolation system, mirror segments and the backplane. While more fidelity should be added, initial results indicate that *single picometer class stability is feasible*, even with attitude control reaction wheels running over a large range of wheel speeds. Key next steps include adding more of the support structure, and demonstrating that the model itself is predictive down to the picometer level. Technology demonstrations that physically measure representative subscale structures using the model predicted post-isolator disturbances from these models would validate the approach taken.

| Technology Needed for Starshades | | | |
|---|---|---|---|
| Technology Category | Technology Provided | Performance Needed | Details |
| Starshade | Starshade modeling and model validation | Validation at F=15 or less | Full-scale starshades are not possible on the ground—need other methods to prove out |
| Starshade | Starshade operations | Optimized for exoEarth yield and characterization | Operating strategies that minimize effects of retargeting |
| Starshade | Starshade formation flying | Shadow control < 1 m over >100,000 km sight lines | Includes metrology, propulsion, etc. |

| Current Status of Starshade Technology | | | | |
|---|---|---|---|---|
| Technology Provided | Heritage | Current Performance | Maturity | Priority |
| Starshade modeling and model validation | NASA ExEP | Models not yet validated at traceable Fresnel number | Developing | High |
| Starshade operations | NASA ExEP | Days to weeks retargeting time | Developing | High |
| Starshade formation flying | NASA ExEP | | Developing | High |

Table 6-4. Key Starshade Technologies.

High observatory *throughput from the UV into the Near-IR* is also a top priority. This goal requires specialized broadband coatings, handling and contamination control for the primary and secondary mirrors, and the instrument optics. As part of future study, lifetime issues such as environmental exposure during development and testing, monolayer build-up on-orbit and ice buildup on cold parts, must continue to be assessed, as must the impact of observatory operational and handling considerations such as mirror temperature, thermal cycling, purge, contamination, etc.

*Instrument and detector technologies* also require investment to overcome similar challenges in realizing high throughput and QE, particularly in the UV. Effort has been directed at achieving high QE in the UV for photo-electric (e.g., micro-channel plate) and solid-state detectors (e.g., CCD, CMOS) using new photocathodes and/or detection layer/bandgap engineering. A generic goal of QE > 50% across the UV band would represent a significant improvement in capability, and NASA has supported efforts that are achieving significant progress towards this benchmark. These capabilities would be widely useful in many possible future instruments, benefitting low-cost as well as flagship missions.

Developing a more comprehensive set of capability requirements from UV to Near-IR is more complex. For a given instrument concept and sensitivity goal, trade-offs between detector QE, noise, wavelength range, stability and dynamic range will define the required performance envelope. Recommend is establishing detector performance goals that are aimed at broadly defined mission-instrument designs with support for targeted prototype detector development in these areas. The task of defining detector capability goals also benefits from more detailed design of starlight suppression systems. Exoplanet spectroscopy (R = 70–500) requires moderately low dark current (< 1 electron/pixel/hour) and very low read noise (< 1–2 electrons/pixel/read) or photon-counting detectors. Electron Multiplying Charge-Coupled Device (EMCCD) detectors, and low noise CMOS active-pixel sensors already show considerable promise, although future work is required to also demonstrate dynamic range and low spurious noise in a realistic, end-to-end environment for both exoplanet imaging and spectroscopy.

Stellar-image based wavefront sensing and control is an essential capability for a coronagraph-equipped *HDST*. This capability is well on its way to maturity under *JWST* and other projects. Maintaining wavefront quality without an external (star) reference is also vital for *HDST*. *Nanometer to picometer accuracy metrology* capable of continuously monitoring the optical configuration without an external reference will permit continuous correction of the line of sight, secondary-mirror alignment, and primary-mirror figure, using the fine-steering mirror, deformable mirrors (in a coronagraph), or secondary- or primary-mirror actuators. Among other things, picometer metrology will enable use of bright calibration stars for coronagraph speckle-nulling control. Once calibrated, the telescope would slew to a much dimmer science target, with the calibrated state maintained through use of the metrology. Without metrology, the slew represents a significant disturbance that would upset the calibrated state, necessitating a recalibration on the much dimmer science star.

The critical technologies identified in Table 6-2 are fundamental to any 12 m class *HDST* with a segmented aperture and a coronagraph. There are other key *HDST* capabilities that may require technology development, but that will be specific to a particular system architecture. Table 6-3 lists important technologies that are either architecture-specific or that are already under development for other projects, or that are already at TRL 5 or better. These are lower priority for *HDST* at this time—but at least some of these will need to be matured in due course.

*Technology investments for alternate versions of HDST:* Although envisioning a particular architecture for *HDST*, it is possible that alternate versions of the telescope could meet the *HDST* science requirements. The 12 m class aperture baseline was developed given our current understanding—but that understanding will evolve in the coming years, possibly changing the areas where technology investment is needed. For example, it is possible that exoEarths are less common than currently projected, or that "normal" exozodi levels are much higher, or that characterization of exoplanet atmospheres is impossible without longer wavelength spectroscopy. It is also possible that the IWA or contrast achievable with the baseline *HDST* coronagraph does not meet its goals. Each of these contingencies might demand larger segmented apertures, 16–20 m say, to meet *HDST* science goals. This larger form of *HDST* could be launched and deployed from a larger launch vehicle, such as the SLS Block 2, with an 8.4–10 m shroud diameter. Alternatively, it could be launched in pieces and assembled on orbit.

The opposite situation—where either natural conditions are more favorable than assumed, or where ultimate IWA performance is better than projected—cannot be ruled out, either. With relaxed performance requirements, a smaller aperture size might be able to accomplish some *HDST* science goals. In this case, a monolithic primary mirror could be considered, possibly for launch with a SLS Block 2 launch vehicle. Depending on the diameter of the shroud that is ultimately adopted, such a mirror could be as large as 8 m—a size well beyond the current light-weight space mirror state of the art and requiring further technology investment.

Another key architectural factor is the choice of starlight-suppression method. Neither coronagraphs nor starshades have yet been proven to work at *HDST* scale or with the required performance. Progress is being made, as summarized in the previous section, but until a definitive demonstration is achieved, it is important to keep options open. In particular, although starshades are not part of the *HDST* baseline, they provide a vital alternative architecture, should coronagraphy fall short of meeting *HDST* requirements. Hence, study of *HDST*-specific starshades is a high priority. This work also builds nicely on work ongoing for other projects, namely the NASA *Exo-S Probe* study, and other investments by both the ExEP Program and industry. The chief challenge going beyond existing efforts is the much larger size (to 100 m or more) and distance (100,000 km or more) that come with the larger *HDST* apertures, along with the resulting reduction in mission efficiency, if only one starshade is in use. Starshade studies should include methods for deploying or erecting 100 meter-class starshades on orbit, and operational strategies that permit access to the maximum number of exoEarth candidates.

If a pure starshade approach is selected, the job of building the telescope becomes much easier, because the ultra-stability needed for coronagraphy would no longer be required. Rather, an *HDST* for use with a starshade could meet its contrast requirements with the same wavefront error as needed for UV and visible imaging science: diffraction limited at 500 nm wavelength. That requirement can be met with telescope wavefront stability of 10 nm, rather than the ultra-stable 10 pm per 10 minutes needed for coronagraphy at $1 \times 10^{-10}$ raw contrast. Recommendations for starshade technology development are summarized in Table 6-4.

| | Key Far-Term Technologies | | |
|---|---|---|---|
| Technology Category | Technology Provided | Performance Needed | Details |
| Far term | Servicing | Robotic infrastructure required | High cost; high value; scope is multi-project |
| Far term | On-orbit assembly of starshade or telescope | Robotic infrastructure required | Needed for largest structures; cost impact not known |

| Current Status of Key Far-Term Technologies | | | | |
|---|---|---|---|---|
| Technology Provided | Heritage | Current Performance | Maturity | Priority |
| Servicing | Hubble | Astronauts, Using STS | Mature, abandoned | Low |
| On-orbit assembly of starshade or telescope | NASA OPTIIX, DARPA, others | | Early phase | Low |

Table 6-5: Key Far-Term Technologies.

Finally, on-orbit servicing and on-orbit assembly are far-term technologies that might prove to be of great value to *HDST*, although they really represent separate programs. By repairing or replacing failing components, upgrading instruments, and replenishing expendables, servicing *HDST* could be a cost-effective life extender and utility enhancer—and possibly a mission saver, as it was for *Hubble*. Although the space shuttle infrastructure that served *Hubble* so well no longer exists, robotic servicing vehicles could be built to provide much useful capability. A venture of this magnitude would be beyond the *HDST* project—it would be a NASA or possibly a commercial venture with broader scope, including servicing of other satellites.

Similarly, on-orbit assembly is a capability that could enhance *HDST* mission capabilities and possibly lower costs. It is the only way to produce an even larger telescope—of 20 m aperture diameter or larger—on orbit. On-orbit assembly could also provide a useful option for a 100 m starshade. The technologies required would parallel those for servicing, and would draw heavily on the robotic devices and remote manipulators used for *ISS* construction and *ISS* and STS operations. This path is worthy of serious attention, but is beyond the scope of this *HDST* study. Recommendations for far-term technologies are summarized in Table 6-5.

## 6.3 Investments to Make Now for the Future of *HDST*

The key technical feasibility issues identified can be resolved, starting now, within the scope of current NASA programs, and building on the progress made on projects such as *Hubble*, *JWST*, the *WFIRST/AFTA* Coronagraph, and even ground-based observatories. Examples of small research projects that would directly address the highest-priority *HDST* technologies, and which could be funded now through the ROSES call, include:

- Generate and compare multiple segmented aperture coronagraph designs for *HDST* coronagraphy, building on *WFIRST/AFTA* but not restricting options to those *WFIRST/AFTA* has selected, given *WFIRST/AFTA*'s different aperture and other constraints. Noting that progress in this field often comes from combinations of techniques, do studies on a non-competitive basis that does not discourage hybrid approaches.

- Address the ultra-stability challenges, starting with mirrors to demonstrate picometer-level wavefront stability under active thermal control using multiple mirror types and materials. Use ground models for design, to improve NASA testing infrastructure. In the process, evaluate other mirror factors such as cost, mass, manufacturability, and technical readiness.

- Continue and expand exoEarth yield studies, to inform design trades that will emerge as system performance and cost are better understood, to find the best compromise between system affordability and science performance. Include starshade options, as a complement to the coronagraph or as a stand-alone.

- Continue development of ultra-low noise detectors and promote their use in large format arrays. A goal should be low read noise or photon-counting with low dark current in the visible, with an added extended goal of maintaining high sensitivity in the UV and Near-IR. Demonstration of the required performance parameters for an Exoplanet discovery detector, and prototype ~0.25 Gpix arrays for general astrophysics should be achievable within this decade.

These projects, or others to the same effect, could be proposed this year and begun next year, in time to produce useful answers well before the 2020 Decadal Survey.

Longer term projects with higher levels of funding could provide hardware demonstrations of key technologies needed for *HDST*. Ground testbeds such as the HCIT could be used to demonstrate coronagraph performance, for instance. This work could be done under the COR and ExEP program offices. Currently planned space-flight demonstrations from *JWST*, the *WFIRST/AFTA* Coronagraph, and/or Probe missions such as *Exo-C* and *Exo-S* will also serve to mature *HDST*'s needed technology. If these proceed on schedule, all key *HDST* technologies could be demonstrated at TRL 5 or higher by the mid-2020s, in time to support project start.

Other issues, especially cost, but also including some feasibility and performance issues, can only be addressed at a system level. *HDST* system architecture and design studies could begin, starting soon. NASA has announced that it will fund Science and Technology Definition Team (STDT) studies to prepare white papers for submission to the 2020 Decadal Survey starting as early as next year. It is recommended that the community propose, and NASA select, a joint exoplanet and general astrophysics *HDST* study, to prepare such a white paper.

System architecture studies will provide the context for some key technical performance and feasibility issues, such as dynamic stability. This critical area—vibration suppression—will benefit from existing TRL 5 options, such as non-contacting isolation, but should also consider more innovative approaches. One such would be to replace the conventional reaction wheels used for pointing control with ultra-precise, low-thrust reaction-control jets, thereby eliminating the largest on-board disturbance source. Integrated modeling, based on detailed system designs and grounded by experience and experiment, will be the basis for decisions in these areas.

System architecture studies are also needed to resolve optical design trades, such as the right balance between active control vs. passive approaches for ultra-stability, or determining the division of labor between coronagraph optics and telescope optics in providing the needed wavefront error performance. Advances in control-related technologies, such as active figure control of mirrors, picometer metrology, and real-time wavefront sensing and control could enable a more efficient and effective *HDST* architecture.

Finally, on-orbit satellite servicing and on-orbit spacecraft assembly are two key technologies that are beyond the scope of any single project, but that could be of great benefit to *HDST* and many other future missions, were they to be provided. We encourage NASA to consider whether a new on-orbit servicing infrastructure could be developed, perhaps in concert with other agencies or the commercial space sector. Replacement of failed components extended the life of *Hubble*, and replacement of its instruments greatly enhanced its scientific value. We would want the same for *HDST* and other missions as well.

# Chapter 7 The Path Forward

The path to a scientifically revolutionary flagship observatory like *HDST* is well laid out, but challenges must be overcome. NASA is already well on its way to meeting the technological milestones needed for *HDST*. Key technologies have been proven by *Hubble*, are being implemented for *JWST* and *WFIRST/AFTA*, and are under active development by industry and in labs as part of NASA's Cosmic Origins and Exoplanet Exploration programs (Chapters 5 and 6). Given this state of technical readiness, and the unparalleled potential for revolutionary discoveries in both exoplanets (Chapter 3) and general astrophysics (Chapter 4), the question is not whether we will ever find a telescope like *HDST* to be scientifically compelling, but how soon we will start moving actively towards realizing its potential.

The answer depends on the resolution of many interlocking practical issues. These include immediate issues of technical feasibility and readiness for the 2020 Decadal Survey (see Chapters 5 and 6). Beyond this, however, loom larger, but equally important questions: how to approach building flagships with constrained budgets, how to foster the participation of groups beyond NASA's Astrophysics Community, and how to inspire and maintain the engagement of the broad scientific community and the public over the long path to development. While we do not have definitive conclusions for how to approach all these issues, we offer some considerations that bear on how to choose the path to achieving *HDST*'s ultimate success.

## 7.1 Building Flagships in an Age of Cost Constraints

Guided by past Decadal surveys, NASA's astrophysics program deployed an array of missions that balance cost and scientific return over a wide range of scales. These approaches spanned from sounding rockets and balloon flights, to moderate missions focused on single scientific questions, and finally to large general-purpose observatories (deemed "flagships"). Although there are always questions about exactly how this balance should be struck, the value of a balanced program has been endorsed by all recent Decadal surveys and sustained by NASA's Astrophysics Division though many administrations. Each tier of the balanced program serves to advance science and technology in its own way, while mutually reinforcing other tiers in the service of NASA's ultimate scientific priorities. The smaller programs drive technology development at a modest cost, while cultivating the next generation of instrument builders and engineers. The moderate-sized programs answer fundamental astrophysics questions and/or carry out foundational large-area surveys, with short development times that can respond quickly to changes in the scientific landscape. The largest programs have longer development times, but offer the most revolutionary capabilities and engage the largest numbers of scientists across all fields of astrophysics.

The recent history at NASA and NSF, and the uncertain funding for science in general, have taught astronomers caution about budgets. In an era of flat or declining budgets, it is not unreasonable to ask whether NASA can sustain the smaller elements of its program while also continuing to develop more costly flagships at a once-a-decade pace. Flagships run the risk of being perceived to damage the rest of the program out of proportion to their scientific returns by crowding out the smaller projects that are widely recognized as vital parts of the space science ecosystem.

These concerns, however, must be weighed against the equally important role that flagships have played, and will continue to play, as part of a balanced space science program. Smaller missions primarily push back the frontiers of knowledge in targeted areas, having the largest impact when they are able to open a new region of parameter space (wavelength, cadence, etc. and/or leverage new technology. As time passes, however, these unexplored corners of knowledge inevitably become harder to find. Flagships are different. By offering a suite of instruments backing large apertures, flagship observatories can radically transform understanding of the universe in nearly all fields of astrophysics. Reaching the necessary capabilities invariably requires longer development times, however, which naturally leads flagships to be more expensive missions.

Even for flagships, "expensive" need not be synonymous with "unaffordable." History has shown that developing a flagship mission involves navigating financially challenging environments, but history has also shown that when the scientific case is sufficiently compelling, the budgetary hurdles can be overcome. The search for habitable or inhabited worlds beyond our own, along with explaining life's origins in the history of the cosmos, offers exactly the motivation needed to sustain the development and ultimate launch of *HDST*, even in a cost-constrained age.

The naturally long development time of flagships also means that there is never a time when one can afford *not* to be actively developing new missions, no matter how unseasonable the budgetary climate. Technology development typically requires decades, thus the failure to actively plan and manage investments towards well-defined, scientifically compelling goals essentially guarantees that ambitious missions will never be feasible, or that they will be more expensive than they need to be. In contrast, continuing to keep flagship missions as an active part of NASA's balanced program maintains focus on projects of sufficient vision and scope that they "make the pie bigger." A mission as exciting as *HDST* has the potential to bring in advocates and partners across agencies, nations, industries, and astronomical communities. A vision restricted to only modest, simpler missions with their science aims tailored to existing cost constraints, leaves a vacuum that will be filled by others with greater ambitions than ours.

## 7.2 A Strategy for Cost Control for *HDST*

For carrying out the most ambitious science, the issue is not "should flagships be built" but "how can they be built efficiently, managing both cost and schedule to maintain a balanced program"? Thankfully, NASA and its partner international agencies have decades of experience developing and building large space telescope missions. These provide compelling guidelines for how one can build advanced mission architectures, while also incorporating realistic, effective, cost-saving strategies. With this experience—both positive and negative—have also come better methods to estimate cost and manage risk. In the decades since *JWST* was first proposed, NASA has improved parametric cost estimation that includes risk metrics for the development of key technologies, as well as realistic tools for the identification and management of mission cost drivers and potential areas of cost growth. In addition, the cost models for building a large, segmented space telescope will be well grounded, due to NASA's experience launching *JWST*.

Putting these lessons to use, key components for the cost-control strategy for *HDST* include:

- Develop the technology and mature the architecture through formulation, followed by a rapid, intensive, and appropriately funded eight-year development effort that minimizes the impact of increasing staffing levels before sufficient risk has been retired.
- Use a "room-temperature" (250–295 K) telescope architecture to avoid the expense and complexities of cryogenic designs and testing.
- Create an overall system architecture with sufficient design margins such that standard processes can be used in all but the most critical and clearly identified areas.
- Identify technologies that address the most challenging requirements early, and embark on a comprehensive well-funded and

| Area | Program | Relevant Heritage Leveraged |
|---|---|---|
| Mirrors | JWST, TMT/E-ELT, non-NASA projects | Mirror Fabrication, Primary Mirror Testing Methods, Secondary Mirror Testing, Hexapod Designs, Gravity Modeling. Facilities and Infrastructure. |
| Wavefront Sensing & Control | JWST, non-NASA projects | All aspects of sensing and control need significantly better performance of the final phasing algorithm. |
| Metrology | SIM, WFIRST/AFTA, non-NASA projects | Picometer class metrology systems. Zernike Sensors. |
| Deployment Systems | Chandra, JWST | Mechanisms, Latches, Hinges, Motor designs—all leverage prior development systems. |
| Sunshield Systems | JWST | Depending on Architecture: membranes, membrane management, materials testing, deployment methods. |
| Composite Structures | Hubble, Chandra, JWST | Basic backplane design similar to, but improved over, JWST design. Basic laminates will leverage technology used in Chandra. |
| Instruments | Hubble, WFIRST/AFTA, JWST | UV, visible detectors leverage Hubble investments; visible, NIR leverage WFIRST/AFTA and JWST investments. |
| Coronagraphs | WFIRST/AFTA | Greatly leverages WFIRST/AFTA coronagraph development to deal with obscured aperture. |
| Integration & Testing | JWST | Mirror assembly methods, alignment methods, testing approaches, and facilities. |

Table 7-1: Key Technology Heritages

sustained development program that achieves TRL 6 by the end of the formulation phase.
- Develop key technology items in parallel as separable items that cannot serially impact the critical path.
- Use extensive heritage from NASA programs and proven technology efforts to minimize the cost risk and cost uncertainties.
- Employ economies of scale through automation and parallel fabrications of mirrors, actuators, structural elements, detectors, and electronics.
- Use competition wherever possible. Competition between multiple vendors is the fastest, most economical method to advance TRL.

A key element of the cost-savings strategy is to utilize lessons learned and heritage from previous NASA programs and technology efforts, as described in Chapters 5 and 6. Several of the key subsystems from *JWST*—including the deployment system and the wavefront sensing and control system—have direct application to *HDST* and will be at TRL 7 or above at the end *JWST*'s formulation phase. In some cases, it may even be feasible to reach TRL 9 through the use of "build-to-print" designs, or direct reuse of software or algorithms from previous programs. When feasible, use of this approach will greatly reduce the cost of the program. Table 7-1 shows several of the technology areas where heritage designs may be directly used or readily adapted, which will reduce both the development cost and the overall cost/schedule risk to the program.

Because it is likely to be a significant cost driver, the optical system for *HDST* has been given specific attention in this study. Using segmented mirrors leverages the technology knowledge base developed for *JWST*, for which the design, development, and test program has clearly established feasibility, while also providing a firm reference point for evaluating cost. For example, *JWST* and several ground-based segmented telescopes have demonstrated that there is a significant learning curve during production of the first few mirrors. This experience shows that there is an opportunity to capitalize on economies of scale to reduce the overall cost. Therefore, while the mirrors and associated structures do have a tight performance specification, an appropriate early investment to achieve automated, rapid mirror fabrication and installation may minimize the schedule and reduce development risk, and thus keep the mirror development off the critical path. Ground-based telescopes such as TMT and the E-ELT will be fabricating hundreds of glass mirrors in the coming decade at a rate of almost one mirror per day, and these facilities, infrastructure, and experience could offset risk in building *HDST*'s segmented mirrors.

For *JWST*, the cryogenic nature of the optical structure and key telescope systems drove a significant fraction of the cost. Designing, building, and testing an end-to-end cryogenic system consisting of mirrors, structures, and instruments—all with tight performance requirements—had a huge impact on the design and the duration of testing. Delays in the cryogenic system led directly to substantial cost growth. Project test schedules and costs were further compromised by cryogenic testing at each level of assembly. The beryllium mirrors needed to meet *JWST*'s cryogenic requirements had to go through a drawn-out cryogenic polishing cycle. *HDST*'s room-temperature design avoids these difficulties, which should significantly reduce both costs and schedule risk compared to *JWST*. Some of this cost avoidance will be offset by *HDST*'s more stringent performance and stability requirements, but the degree to which this is an issue cannot be known without more detailed study.

The fundamental approach for reducing the total mission cost is designing, building and testing the observatory efficiently, by having all of the key system elements built in parallel and integrated quickly to minimize the time required for maximum staffing. Central to this approach are: (1) sufficient sustained early technology maturation; (2) a low risk and robust system architecture enabling a fast (approximately eight-year) development effort; and (3) accurate system-level engineering with excellent interface control between subsystem modules. Such an architectural approach, including room-temperature operation, allows for early testing that can minimize development uncertainties and associated risk, while also allowing for parallel development relatively deep into Phase D. This parallel development can then be followed by an efficient and fast integration and testing phase.

Also essential to controlling costs is entering into development with a well-understood model of where those costs are likely to be incurred. Accurate, early cost models allow missions to identify items that are likely to drive the yearly and the lifecycle budget. In response, missions can search for strategies for bringing those costs down, either by improving technology, design, and/or manufacturing processes, or by searching for alternatives. These same models also allow realistic budgeting by NASA. This ensures that the project has access to reserves at appropriate times (which is important given that improperly phased reserves was identified by the Casani report as a major issue for *JWST*, as was *JWST*'s lack of mature technology at the time it began). Realistic budgeting also permits NASA to appropriately phase the project budget so that it does not crowd out the smaller missions that are essential parts of a balanced space science program.

While early costing is essential, it is invariably challenging for any complex flagship mission like *HDST*. Traditional approaches using parametric cost models and analogy-based models can be broadly useful in a relative sense, especially when comparing similar architectures, but evaluating the absolute cost of a unique observatory with this level of complexity will be much more complicated. While there is a rough $D^{1.6}$ parametric cost-scaling for the telescope itself, the scaling approach has only been established for small apertures and does not include the cost of the entire observatory, which would include instruments, testing, etc. For *JWST*, the telescope itself was only ~15% of the total mission cost, leaving ~85% of the other observatory costs—the spacecraft and sunshield, instruments, integration and cryogenic testing, ground system and operations, program management, and system engineering—outside the realm where simple scaling models were valid.

For *HDST*, then, one cannot simply scale *JWST*'s total cost by the $D^{1.6}$ scaling parameter. Instead, telescope data needs to be combined with other mission factors like operating temperature, performance, complexity, economies of scale and heritage. In several of these areas, *HDST* is well positioned. It can effectively reuse much of *JWST*'s architecture by using segmented hexagonal mirrors (like *JWST* but with an extra "ring" or two of segments), its sunshade (but with fewer layers), and its deployment strategy (allowing reuse of latches, folding backplane engineering, etc.). With its similar (and potentially identical) mirror segment sizes, *JWST*'s testing and transport infrastructure can also be reused. *HDST*'s stability requirements are potentially an added load on performance and complexity, but these too have substantial heritage from *SIM*. History has shown that the technology development needed to enable a capability also tends to reduce the cost of that capability moving forward. As such, the choice to leverage as much heritage as possible should be considered as a significant cost control measure.

Ultimately, a formal cost estimate that is properly phased with appropriate reserves can *only* come from careful system design and analysis, coupled with detailed requirements developed in pre-formulation. During formulation, these steps must be followed by an equally careful and thorough "work breakdown structure" analysis of the cost. Such an approach must account for key parameters including heritage, complexity, maturity, temperature, and performance of each subsystem, rather than relying primarily on scaling laws based on mass or aperture.

In short, while *HDST* is clearly a flagship-class observatory, there is a clear route to making sure it is a flagship that we can afford. Not only must the mission be properly designed and managed, but it also must enter the mission-development period only after the architecture is finalized, all technologies are at TRL 6, and the testing plan is clear. With the existing technological heritage and the well-defined plan for technological investment outlined in Chapter 6, these goals can be achieved, and the scientific vision of *HDST* realized.

## 7.3 Scientific Impact of Smaller Apertures

It is common to assume that when attempting to control costs, missions should adopt the smallest possible telescope diameter that is compatible with their science. Larger apertures tend to increase the launch weight of the telescope, and historically, cost has been assumed to scale with both aperture and mass (although with significant scatter). Instruments for smaller telescopes also tend to be less expensive. A reduction in mirror size is therefore frequently used as a "descope" option that could be invoked when a mission suffers from cost growth or appears to be prohibitively expensive. However, reductions in aperture do come with quantifiable impacts to the core science of the mission that must be carefully weighed against both budgetary and operational constraints.

The baseline *HDST* assumes a 12 m class aperture, motivated by maximizing the number of exoEarths that can be detected and characterized spectroscopically, while also fitting into likely launch vehicles. There are a few natural descope aperture sizes to evaluate for their impact on core science. The first is a 9.2 m aperture, which corresponds to using the exact same *JWST* mirror-segment sizes, but with an additional "ring" of hexagonal segments. The second is a 6.5 m aperture, which could be built as a non-cryogenic *JWST* clone. A third possibility is switching to monolithic mirrors, which could potentially be launched with up to an ~8 m diameter, comparable in collecting area to the 9.2 m. Adopting the 9.2 m and 6.5 m sizes to bracket the likely range of a scaled-down mission, these two descope options would have ~60% and 30% of the collecting area of the baseline *HDST*, respectively. Their resolution at the diffraction limit would be 30% and 85% lower. We now trace how these reductions in capability would affect a number of high priority science goals.

*Exoplanet Yields:* The search for exoEarths is one of the leading scientific goals for *HDST*. A high yield of exoEarths is needed for comparative planetology, for finding exoEarths that can be characterized at longer wavelengths, and, most importantly, for maximizing the chances of finding exoEarths with biosignature gases in their atmospheres. The calculations presented in Chapter 3 show that the exoEarth yield increases dramatically with telescope aperture, such that larger telescopes will always outperform smaller ones when equipped with starlight suppression systems with comparable performance (IWA, throughput, contrast).

The biggest concern in the search for exoEarths is that if the telescope becomes too small, only the most favorable combination of astrophysical parameters would yield a sample of any appreciable size. Figure 7-1 simulates this effect by assuming reasonable distributions for the major astrophysical unknowns—the exoEarth occurrence rate, the exozodiacal background, and the planetary albedo—and then calculating the corresponding probability distribution of yields that would result, including the effects of Poisson sampling. While these astrophysical parameters have some degree of uncertainty, experiments are underway that should narrow distributions over the coming decade (Chapter 3). For each combination of these parameters, *HDST* could study more or fewer exoEarths than the baseline assumptions adopted in Chapter 3. The largest concern is that when all the parameters are unfavorable (i.e., fewer exoEarths, higher backgrounds, and dimmer planets), the yield of exoEarths may become unacceptably small. The exact definition of "unacceptable" is a matter of debate, but for a flagship mission that aims to transform our understanding of exoEarths and search for biosignature gases in their atmospheres, a sample of 20 exoEarths is a reasonable lower bound that could still give an appreciable chance of finding evidence for life, assuming (generously) the rate at which exoEarths show atmospheric biosignature gases is 5%. For a 12 m, there is a much lower risk of not hitting this threshold. In > 75% of the cases in Figure 7-1, a 12 m *HDST* would expect to find at least 20 exoEarths. For a smaller 9.2 m telescope, however, only ~50% of the likely astrophysical parameter combinations reach this minimum threshold. The impact on yield is more severe for a 6.5 m aperture, for which fewer than 10% of the cases would yield 20 or more exoEarths. These metrics should not be overinterpreted, however, because different choices for the desired sample size and for the distributions of astrophysical unknowns can change the apparent relative performance of different size apertures.

A second complication is that changing to a monolithic mirror would allow different, possibly higher performance starlight suppression systems to be used. This change could potentially recover some of the yield lost to smaller apertures and segment diffraction, but is unlikely to recover all of what would be lost in the descope. If changing to a 6.5 m monolith allowed improved starlight suppression with a narrower inner working angle of 2 $\lambda/D$, the yield would improve, but there would still be only a 30% chance of finding more than 20 exoEarths. An even more radical descope to a 4 m would perhaps open up even higher performance suppression options with inner working angles as small as $\lambda/D$, but would still only have a 15% chance of finding more than 20 exoEarths. The higher performance options that may be available with smaller, monolithic mirrors could also improve coronagraphic throughput. The models in Figure 7-1 assume 20% system throughput, but because the dependence of yield on throughput is not particularly strong (see Chapter 3), even tripling the throughput only increases the yield by 45%; at the maximum of 100% throughput, the gain in yield is 75%. Thus, while an extremely high performance starlight suppression system operating on a 6.5 m may approach or perhaps even surpass the exoEarth yield of a 9.2 m, the yield of a 12 m with less ambitious starlight suppression performance is still significantly larger. In addition, the performance of internal coronagraphs for segmented mirrors is likely to continue to improve. Progress in this field is rapid (as discussed in Chapters 5 and 6) and it is not clear that starlight suppression for monolithic mirrors will always be significantly better. Finally, while increased starlight suppression performance could improve the yield, it cannot compensate for the impact of smaller apertures on the general astrophysics program.

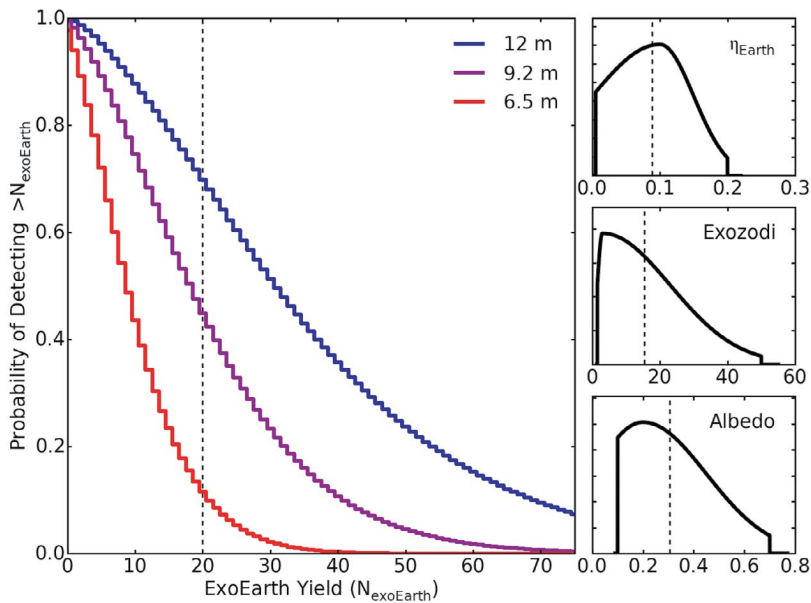

Figure 7-1: The possible range of exoEarth yields (left), given the uncertainties in astrophysical parameters (the earth occurrence rate, the level of exozodiacal light, and the planet albedo; right panels, top to bottom), for the nominal *HDST* 12 m, and the possible 9.2 m and 6.5 m descope options (assuming an IWA of 3 $\lambda$/D, and 1 year of on-sky integration including both detection and spectroscopic characterization). Histograms indicate the probability that a mission would find more than a given number of exoEarths. For a fixed telescope aperture and starlight suppression performance, there is a large range in possible yields that reflect how favorable the true astrophysical parameters are. Larger telescopes offer better protection against the possibility that astrophysical parameters are skewed to values that make exoEarth detection more difficult. Vertical dashed lines in the right panels indicate the median value of each astrophysical parameter.

6) and it is not clear that starlight suppression for monolithic mirrors will always be significantly better. Finally, while increased starlight suppression performance could improve the yield, it cannot compensate for the impact of smaller apertures on the general astrophysics program.

Given the uncertainties, the final mission concept for *HDST* will need to address these issues in detail, weighing scientific aims against system-level decisions about architecture and cost. As time progresses, our knowledge of coronagraphic performance and astrophysical parameters will both improve, and their values may be discovered to be more favorable, making a reduction in aperture size more feasible. However, until that time arrives, a 12 m is the aperture size that is certain to provide a large sample of exoEarths, and guarantees that the results of this flagship mission would be truly transformative rather than incremental.

Given the uncertainties, the final mission concept for *HDST* will need to address these issues in detail, weighing scientific aims against system-level decisions about architecture and cost. As time progresses, our knowledge of coronagraphic performance and astrophysical parameters will both improve, and their values may be discovered to be more favorable, making a reduction in aperture size more feasible. However, until that time arrives, a 12 m is the aperture size that is certain to provide a large sample of exoEarths, and guarantees that the results of this flagship mission would be truly transformative rather than incremental.

**Depth:** *HDST* will be a cutting-edge facility pursuing the deepest, faintest possible observations. Reducing the size of telescope's aperture increases the exposure time needed to reach a desired magnitude target. Descoping from a 12 m will then either require longer exposure times, or relaxing science goals that require faint limiting magnitudes.

Figure 7-2 shows the increase in the exposure time needed for a 10-$\sigma$ detection of a point source when descoping from a 12 m to a 9.2 m (purple lines) or to a 6.5 m (red lines), as a function of the magnitude in *U*, *V*, or *J* (light, medium, and heavy lines). In the background-limited regime where *HDST* is likely to operate, exposure times would be 3× longer for a 9.2 m, and 12× longer for a 6.5 m. These increases would negatively impact the efficiency of the mission. A 9.2 m could do 1/3 the imaging science of a 12 m, or could do the same imaging science as a 12 m but only if the mission were operating for 3× longer. This latter option would increase the total mission cost, and possibly negate some of the savings gained by choosing a smaller aperture. Evaluating the relative costs of building and operating a more reliable, longer-lived, smaller telescope versus a baseline 12 m would require detailed study as part of any descope decision.

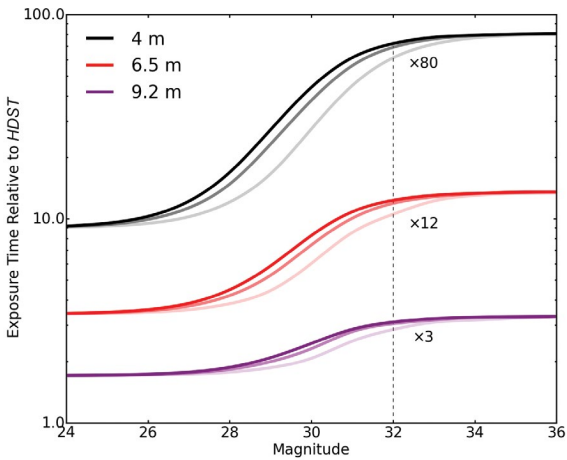

Figure 7-2: The increase in the exposure time for a 10-sigma detection of a point source when descoping from a 12 m to a 9.2 m (purple line) or a 6.5 m (red line), as a function of magnitude. In the background limited regime (>29 mag), exposure times increase by a factor of 3 for a 9.2 m and by a factor of 12 for a 6.5 m. Light, medium, and dark lines are for the *U*, *V*, and *J* bands, respectively. A 4 m aperture (black line) is included for comparison.

Spectroscopic programs would experience similar increases in the exposure time for the near-infrared, since observations at these wavelengths are almost always background limited. In the UV or optical the typical backgrounds are darker, and spectroscopy is more likely to be in the source limited regime, where the impact of a descope would be somewhat less. UV and optical spectroscopic programs would require 1.7× more exposure on a 9.2 m, and 3.2× more exposure on a 6.5 m.

Rather than adopting longer exposure times, most scientific programs would probably chose to scale back their scientific goals. In these cases, the programs would use the same amount of observing time, but would limit themselves to brighter targets. At the same exposure time, observations with a descoped 9.2 m to 6.5 m aperture would have limiting magnitudes brighter by 0.65 to 1.4 magnitudes, respectively, in both the source and the background limited regimes. For general astrophysics, limiting observations to brighter objects leads to a variety of impacts.

First, brighter targets will, on average, be closer. Restricting to nearer targets drops the number of potentially observable sources. For local objects like stars and nearby galaxies, the effective survey volume scales like distance cubed, and thus the 9.2 m would have 6× fewer possible targets, and the 6.5 m would have nearly 50× fewer at the same exposure time. Descoping from a 12 m would therefore significantly reduce the number (and variety) of galaxies for which the ancient main-sequence turnoff would be detectable, the number of white dwarfs that are candidates for UV spectroscopy, the number of stellar clusters whose low-mass IMF could be directly measured, the number of faint optical transients that could be identified, and negatively impact countless other programs. For some scientific programs, this change may still leave large numbers of objects within the smaller survey volume, but when studying rare examples of a class, these restrictions can mean the difference between having a single example, or a larger statistically significant sample. Over cosmological distances the gains in the comoving volume are not as large, due to the rapid increase in the luminosity distance with redshift, but they are still significant. As a result, limiting observations to brighter galaxies and AGN will compromise some studies of evolution over cosmic time, by reducing the redshift range over which a given measurement can be made.

Second, brighter targets will be intrinsically more luminous. When studying objects at a fixed distance (say, galaxies at a particular redshift, or young stars in a particular H II region), a brighter magnitude limit makes low-luminosity objects undetectable. The sources that could be seen by a 12 m *HDST* would be nearly 2× less luminous than those seen with a 9.2 m, and more than 3.5× less luminous than those seen with a 6.5 m in the same exposure time. Observations would then lose the lowest luminosity stars—which impacts studies of the IMF and the white dwarf cooling sequence—and the lowest luminosity galaxies—which are potentially important contributors to reionization at high redshift and which locally trace the extreme limits of galaxy formation and the transition to starless dark matter halos. The added dynamic range is also the key to identifying physically significant breaks, such as the characteristic mass scale of the IMF and its faint-end slope, cutoffs in the luminosity function of star forming regions, and the shape of galaxy luminosity functions, which typically have an exponential cutoff at bright magnitudes and a power-law slope at faint magnitudes. As an example, a steep power-law slope for the $z = 3$–$10$ rest-frame UV luminosity function would imply that increasingly fainter galaxies dominate the luminosity density, the production of ionizing background photons, and the cosmic star-formation rate. *HDST* will reach sufficiently deep magnitude limits that it can detect the expected turnover in the faint end slope, while 9.2 m and 6.5 m apertures are likely to detect only the brighter population above this turnover.

Finally, brighter targets are rarer and have a smaller density on the sky. Efficient multi-object spectroscopy relies on having enough bright sources to use all available slits, and when the surface density of sources goes down, spectroscopic observing campaigns can become unfeasible—although the larger fields of view associated with smaller apertures can potentially compensate in some science cases. The impact is particularly severe for UV absorption-line studies, where the density of background sources is low, and limits the physical scale at which the intervening gas can be mapped. The sky density of UV bright sources increases by a factor of ~5 for every magnitude of additional depth, such that the 12 m *HDST* reaches 7× more background sources than the 6.5 m descope, for fixed SNR and exposure time. This density can be translated into physical resolution when studying nearby galaxies (< 10 Mpc; Section 4.3.2), which have virialized halos that are ~0.3 Mpc in radius, covering 5–10 square degrees on the sky. Within this area, *HDST* could carry out UV spectroscopy of 50–100 QSO sightlines and 1000–2000 background galaxies, which would constrain the composition and kinematics of the circumgalactic medium at physical scales of 10–15 kpc. For a 9.2 m, the accessible source density drops by about a factor of 4, and reduces the physical resolution to 20–30 kpc. A 6.5 m can only resolve scales down to 30–50 kpc. These larger physical resolutions compromise the ability of the measurements to discriminate among theories of gas acquisition and feedback, particularly close to the galaxies' luminous inner regions (< 15 kpc).

At higher redshifts, Figure 7-3 illustrates the impact of possible aperture reductions on high-redshift IGM and CGM science using spectroscopy of UV bright QSOs. At z ~1, half of the currently cataloged sources would be accessible with a 6.5 m telescope, but at z ~1.5–2, spectroscopy of the vast majority of QSOs requires a larger aperture; these redshifts are essential for studying gas flows during the epoch where the star-formation rate density peaks. At all redshifts, the density of background sources is a critical factor for studying the gaseous environment of rare populations of foreground galaxies. Measuring gas supply and feedback in uncommon or short-lived phases of galaxy evolution (such as massive red galaxies, starbursts, mergers, and galaxies hosting AGN) requires 5–10× larger samples of background sources than can be accessed with a 6.5 m; in contrast, smaller 6.5 m telescopes can only probe the properties of typical "mainstream" galaxies. Finally, many astrophysically important tests rely on having pairs of objects along a line of sight (such as "QSO sightlines passing near foreground galaxies," or "pairs of QSOs probing angular correlations in the structure of the intergalactic medium"). The number of systems that can be studied in such cases depends on the square of the source density, which compounds the effects of descoping to smaller apertures.

*Resolution:* HDST will deliver unprecedented resolution at optical and UV wavelengths. The resolution scales inversely with telescope diameter, such that the physical scales that can be resolved will increase by a factor of 1.3 for a 9.2 m descope option, and by a factor of 1.8 for a 6.5 m. For studies that require resolving a particular physical scale (such as resolving structures in debris disks, or identifying star forming regions in distant galaxies), the drop in resolution limits the volume of space within which appropriate targets can be found. The number of systems that can potentially be resolved drops by a factor of 2.2 for a 9.2 m descope option, and a factor of 6.3 for a 6.5 m for sources at non-cosmological distances.

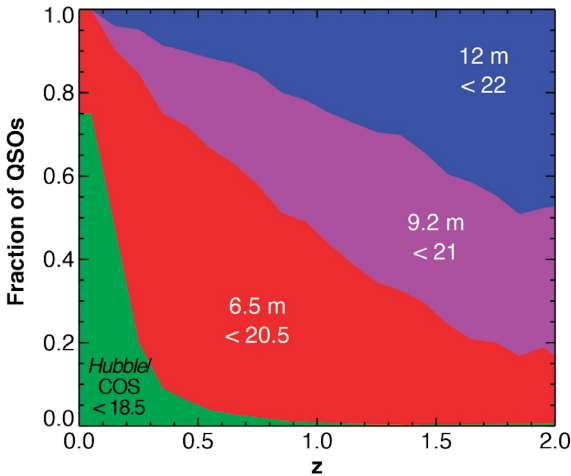

Figure 7-3: The fraction of currently cataloged UV-bright QSOs in bins FUV magnitude (AB < 18.5 for *Hubble*, < 20.5 for a 6.5 m, 21 for a 9.2 m, and < 22 for a 12 m *HDST*; left panel) as a function of redshift. Larger telescopes access significantly more sources and provide the majority of the most valuable high-redshift sources that support the evolutionary studies of galaxies, the CGM and IGM, and AGN (Chapter 4).

The largest impact of resolution will be felt in "crowding-limited" observations. In dense stellar fields, like those found in the centers of stellar clusters or in nearby galaxies, the ability to measure photometry of individual stars is not limited by their photon flux, but by the ability to separate adjacent stars from one another. The brightest stars are rare enough that they can easily be distinguished from each other, but fainter stars can become too densely packed to be detected individually. The limiting magnitude of the stars is then set by the magnitude at which the stellar density becomes higher than a given telescope+instrument can resolve. This limiting magnitude will be brighter when the apparent stellar surface density is higher, as is found for intrinsically high stellar densities (such as in galaxy bulges or the centers of globular clusters) or for more distant systems (which have smaller angular separations between stars).

For a desired limiting magnitude (as would be needed to reach a particular feature on a stellar color-magnitude diagram), a 12 m *HDST* will be able to make measurements in regions whose angular stellar surface density is a factor of 1.7× higher than a 9.2 m, or a factor of 3.4× higher than with a 6.5 m. A 12 m can therefore study the stellar populations in higher mass galaxies, in their denser central regions, or in galaxies that are farther away. The number of galaxies that can be potentially targeted with a 12 m *HDST* increases by a factor of 5 compared to a 9.2 m, and by a factor of 40 compared to a 6.5 m, greatly increasing the range of galaxy types and environments that can be studied. Similar scalings apply to globular and stellar clusters.

For a particular system (say, a specific galaxy or globular cluster in the local volume), larger apertures will be able to measure fainter stars. The exact change in the limiting magnitude with decreasing telescope aperture will depend on the details of the stellar luminosity function, which sets how the stellar density changes with apparent magnitude. Two examples in Figure 7-4 show the impact on observations of an old stellar population (which typically dominates the number density of stars in galaxies—assumed here to be 10 Gyr and 1/10$^{th}$ solar metallicity), and on a young stellar population (typical of an active star forming region—assumed to be 5 Myr and solar metallicity). The large blue dot marks the assumed depth of a 12 m *HDST* observation. When reducing the telescope size to 9.2 m (purple arrow) or 6.5 m (red arrow), the maximum stellar density that can be observed decreases. Comparing to the luminosity function on the left panel shows that these changes in the limiting density brighten the limiting magnitude by 1 magnitude and 2.6 magnitudes for the old stellar population, and by 2.4 and 4.6 magnitudes for the young population, for a 9.2 m and 6.5 m descope option, respectively.

The brighter limiting magnitudes limit the information that can be gleaned from the resulting observations. The right hand panels of Figure 7-4 show how the limiting magnitude affects the features that would be observed in the color-magnitude diagram of a galaxy at ~4 Mpc. For the old stellar population, a 12 m *HDST* observation would nearly reach the turnoff of ancient 10 Gyr stars, giving direct constraints on the early history of star formation at high redshift. In contrast, a 9.2 m could only detect younger main-sequence turnoffs, and a 6.5 m would be limited only to very recent turnoffs and the red end of the horizontal branch. For the young stellar population, a 12 m *HDST* observation could detect ~1.6 solar mass stars, but reducing to a 9.2 m would nearly double the limiting mass, and reducing to a 6.5 m would bring the limiting mass higher still to close to 5 solar masses, leaving the low-mass stars unresolved. Not only does the reduction in aperture significantly decrease the possible leverage on the high-mass IMF, it restricts the ages of the young stellar populations that can be studied in a given cluster. Clusters as old as 2 Gyr host 1.6 solar mass stars, but 3 solar mass stars can only be found in 500 Myr-old clusters, and 5 solar masses in 100 Myr-old clusters. Smaller apertures will therefore only be able to resolve main-sequence stars in younger clusters.

Changes to resolution also compromise proper-motion studies of nearby stars. The smallest angular proper motion that can be measured scales in inverse proportion to telescope aperture diameter, for a fixed exposure time and time baseline (Section 4.3.4). A smaller telescope can achieve the same precision as a larger one, but must either observe over a correspondingly longer time baseline or observe objects that are proportionally closer. For example, the nominal 12 m *HDST* will observe proper motions at the level of ~1 km s$^{-1}$ out to ~150 kpc, encompassing many of the known ultra-faint dwarfs and stellar streams, as well as the many new ones expected to be discovered by LSST. By constrast, a 6.5 m telescope will only achieve this level of precision at 42 kpc, a volume 6× smaller. This region still encompasses some of the known ultra-faint dwarfs, but puts most of the expected new discoveries beyond reach.

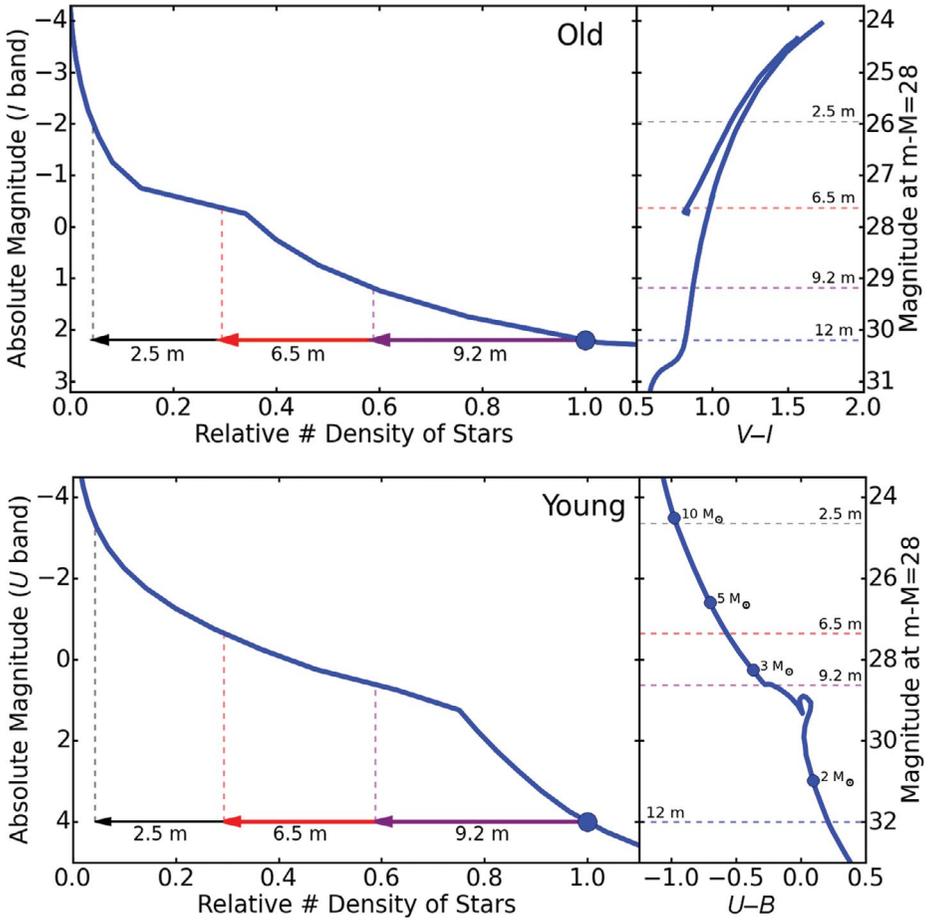

Figure 7-4: The change in limiting magnitude for crowding-limited observations of stellar populations when descoping from a 12 m to a 9.2 m or 6.5 m aperture. As shown from the stellar luminosity functions on the left, descoping to a smaller aperture restricts the observations to lower stellar surface densities, and thus to rarer, more luminous stars, leading to brighter limiting magnitudes. The top panel shows the impact on observations of an old stellar population that could resolve ~9 Gyr old main-sequence turnoffs with a 12 m *HDST*. The CMD on the right panel shows that a 9.2 m could no longer detect the old turnoff, and a 6.5 m could only detect much younger turnoffs and the red horizontal branch. The bottom panel shows the impact of smaller apertures on observations of young stellar populations, for which *HDST* could detect ~1.6 solar mass stars with ~2 Gyr lifetimes, while a 9.2 m and 6.5 m could only detect ~3 solar mass stars with 500 Myr lifetimes, and 5 solar mass stars with 100 Myr lifetimes, respectively. The rightmost axis gives apparent magnitudes at a distance of ~4 Mpc; there are many hundreds of galaxies within this volume.

In summary, the baseline 12 m *HDST* mission offers systematically higher science returns than smaller apertures, in both the exoplanet and general astrophysics programs. This level of performance offers transformative gains that should not be prematurely traded away prior to determining a reliable life cycle budget for a 12 m *HDST*.

## 7.4 *HDST*: Forging Connections

A mission as potentially revolutionary as *HDST* will naturally be a mission that draws together nations, industries, divisions, technologists, scientists, and people.

*Nations:* As the true successor to *Hubble* and *JWST*, *HDST* is well-suited for leadership by the United States. NASA and its partners have spent decades developing the expertise to design, build, and operate a large UVOIR segmented space telescope. This project is sufficiently compelling that we expect other nations to welcome the chance to participate, continuing the long history of cooperation between NASA and other nations' space agencies. With ESA's large projects now allocated to X-ray and gravitational-wave missions, the door remains open for US leadership in a large UVOIR mission like *HDST*.

*Industry:* NASA's industry partners have long been vital collaborators in building flagship missions. They have participated in every stage of major NASA missions, from design, through assembly, to launch. With their strong record of technological innovation, industry leaders—large and small—would undoubtedly contribute substantially to *HDST*. A key aspect of the NASA-industry relationship is that there is both collaboration and iteration of the requirements and designs based on what industry demonstrates is feasible and what NASA assesses to be needed. Having several industry partners studying design concepts and participating actively in technology development assures that the best ideas get considered and pursued. As demonstrated on numerous programs, competition between industry teams can lead to rapid progress. Meanwhile, NASA's role is to be an informed consumer that can iterate the requirements and plans based on industry progress and feedback.

*Divisions & Agencies:* While development of flagship missions has historically been the purview of NASA's Science Mission Directorate, *HDST* offers a number of opportunities to draw in expertise and support from other divisions and agencies. For example, on-orbit satellite servicing and on-orbit spacecraft assembly are two key technologies that are beyond the scope of any single project, but that could be of great benefit to both *HDST* and other future missions, were they to be provided. We encourage NASA to consider whether a new on-orbit servicing infrastructure could be developed, perhaps in concert with other agencies or the commercial space sector. Servicing could capitalize on the upfront investment in *HDST* by both extending its life and revitalizing its suite of instruments, in the same manner as was so successfully employed for *Hubble*, while also providing the public with moments of drama that reveal the excitement of space exploration. Likewise, launching *HDST* would be an obvious use for a large, advanced launch vehicle like the SLS Block 2, giving NASA's Human Exploration and Operations Directorate an important milestone on the agency's path to Mars.

*Scientists:* As a true "observatory," *HDST* is in the proud tradition of other NASA flagships. The beauty of general-purpose facilities is that the entire astronomical community decides how they are best used, and that they let individuals pursue ambitious "small science" on the biggest, most capable telescopes. The resulting science is nimble, responding to the rapid shifts in understanding and discovery. It is also at its most creative, by drawing on the minds of thousands of scientists, each bringing their own insights and vision. A suite of capabilities, when coupled to the genius of many, has always led to science that extends the reach of a single telescope far beyond our initial vision.

*HDST* also merges together the dreams of both those who seek worlds like our own, and those who quest to understand how such worlds eventually came to be. Both dreams depend on *HDST*'s large aperture and resolution, and can easily work together in a single mission. A rich program of parallel observations turns every exoplanet observation into a powerful general astrophysics survey, and every general astrophysics result helps stoke the excitement of the public and the astrophysics community, keeping the interest and relevance of *HDST* high for decades.

*The Public:* As a mission that aims to answer one of the most basic questions of human existence—are we alone?—*HDST* will capture the imagination of all. But, during the needed years of careful searches for Earth-like systems, *HDST* will provide the world a steady diet of breathtaking images, with a depth and clarity that is hard to comprehend, making images from *Hubble* look blurry. Knowing the depth of the public's affection for *Hubble*, we are certain that the level of the public's engagement with *HDST* will repay them for their investment in it.

# Chapter 8 Summary and Closing Thoughts

The vision outlined in this report is for a 12 m class space telescope with exquisite sensitivity from the UV through the Near-IR and superb image and wavefront quality. This observatory would allow direct detection of Earth-like planets and characterization of their atmospheres, along with a rich program of astrophysics covering every stage of the pathway from cosmic birth to living earths. We have demonstrated the ample technical heritage that can make this ambitious mission a reality in less than two decades, and have outlined an approach for technology development and program execution that should allow this project to fit within a profile consistent with past NASA flagship missions. In short, building this transformative mission is within our abilities, if only we move forward with the needed development in a timely manner.

One could conceive of a more focused mission than we described here—one whose only goal would be to find Earth-like planets and assess their habitability. However, the concept we present here is much more than an experiment. It is an observatory, with the richness, nimbleness, and versatility that entails. Rather than being a narrow project for a subset of the community, this mission would be a platform for thousands of scientists to explore their ideas, capitalizing on the genius of many to keep the observatory relevant and exciting for decades. As a general purpose facility with revolutionary capabilities, we cannot begin to predict what the landscape for discovery will look like 10 years after its launch, but with a world-wide community of scientists using *HDST* as a laboratory for exploration, we can be certain that its future discoveries will be those that transform how we all—scientists and public alike—see our place in the universe.

# Afterword:
# To Seek Life, to Explore the Cosmos

The most powerful and productive space telescope ever built—the *Hubble Space Telescope*—will be in its waning years by 2019. *Hubble*'s successor, the *James Webb Space Telescope* (*JWST*), will be launched in late 2018. *JWST* will take up the mantle of *Hubble*, and will glimpse the earliest galaxies, young exoplanets, and, possibly, unpredictable cosmic phenomena. *JWST* will deliver stunning new science and exciting breakthroughs that have been *Hubble*'s hallmark, and could operate until the late 2020s.

However, great space observatories require 15 to 20 years to design, build, and launch. Midway between the decadal surveys of 2010 and 2020, this report boldly proposes an ultraviolet-optical-infrared (UVOIR) telescope with capabilities that are over a thousand times beyond *Hubble*'s, and even beyond *JWST*'s.

Life is almost certainly out there somewhere beyond the surface of the Earth. To find life elsewhere is perhaps the most important endeavor that astronomers could attempt to tackle. We believe that people everywhere share the desire to understand themselves and this wonderfully wild world on which we live: a solitary sphere orbiting a relatively ordinary star. Discovery of life elsewhere will unite humanity in a deep intellectual manner. The proposed telescope will be able to detect signatures of life on planets outside our own Solar System, and begin to address the issue of whether life is ubiquitous.

This report shows that we know how to do this now. The techniques exist. And this same technical capability will give this observatory a view of the entire history of stars in the universe. It will lay bare the origins of the atoms of carbon, nitrogen, oxygen, and iron that make life possible in a universe that once consisted of nothing but hydrogen, helium and lithium, and elucidates the twin mysteries of dark matter and dark energy.

Throughout history, every thousand-fold increase in telescope technical capability has led to the discovery of new astrophysical phenomena or classes of objects. *Hubble* was no exception. *Hubble*'s science has been seminal: from distant galaxies, super massive black holes, proto-stars, Pluto's small moons, exoplanets, and cometary impacts on Jupiter (to name just a small sample of *Hubble* results), to its contribution to the Nobel-prize-winning discovery of dark energy, *Hubble* has revolutionized cosmology and physics in the past two decades. *JWST* will revolutionize astronomy in the coming decade.

The *raison d'être* of the proposed telescope is to seek signatures of life elsewhere, and it will also explore the history of the universe from cosmic birth to today's cosmos. In its lifetime, though, the telescope proposed in this report will almost certainly find "something"—perhaps several "somethings"—that are as unexpected and different from everything we know, as dark energy is from the rest of the Universe.

This engaging and imaginative report makes a strong case for a magnificent telescope that would launch into space in the 2030s. It richly deserves to be read and debated in the astronomical community, and society at large. The path forward is clear, construction of this facility will require partnerships with other nations around the globe, and its results will resonate with all of humanity.

Mordecai-Mark Mac Low and Michael Shara
*Curators of Astrophysics*
American Museum of Natural History

# Acknowledgements

This report represents not just the work of the committee, but also of our many colleagues and collaborators who have provided support, knowledge, and insight throughout this process. We owe a tremendous debt to many, without them this report would not have been possible.

We begin with a warm thanks to the former AURA president, William Smith, for initiating the call for this report and to the AURA vice president, Heidi Hammel, for her critical support and many helpful discussions. On the logistics side, we owe special thanks to Nanci Reich from AURA, who tirelessly organized telecons and meetings for a group of people with nearly incompatible schedules. Technical editor Sharon Toolan and layout specialists Ann Feild and Pam Jeffries did a superb job bringing the report to its final polished state. The cover page artwork and several of the key figures were prepared by graphic artist extraordinaire, Christine Godfrey.

Crafting a plan for a flagship mission naturally involves discussion and participation at high levels. We are appreciative to Mike Garcia and Arvind Parmar from NASA and ESA, respectively, for being *ex officio* members of the committee. We also had many fruitful interactions with NASA's COPAG and EXOPAG groups, chaired by Ken Sembach and Scott Gaudi. We owe significant thanks to Alan Dressler, the lead of the beautifully written and highly influential "*HST* and Beyond" report that was the inspiration for this effort. Alan was generous with his time in the early planning, providing us with many insights from his past experience. Matt Mountain and Antonella Nota provided useful insights in the formative stages of this process. We are especially grateful to our "red team" reviewers: Alan Dressler, Jonathan Lunine, Jennifer Lotz, Karl Stapelfeldt, Ken Sembach, Makenzie Lystrup, Steve Murray, Chris Carilli, and Martin Barstow. They provided invaluable insights and a critical reading of a preliminary draft of this report. The report was significantly improved due to their careful review. Additional thanks go to Garth Illingworth, Bruce Margon, and Neill Reid for valuable discussions.

The broad scientific aims of this mission concept were refined after discussions with many, many people. These include Eric Agol, David Alexander, Martin Barstow, Tom Brown, Mark Clampin, Andy Connolly, Roger Davies, John Debes, Shawn Domagal-Goldman, Mike Fall, Kevin France, Matt Greenhouse, Darryl Haggard, Suzanne Hawley, Sally Heap,

Jeremy Heyl, John Jameson, Jason Kalirai, Bill Keel, Gerard Kriss, Jennifer Lotz, Thomas Maccarone, Vikki Meadows, Adam Meyers, Knut Olsen, Marshall Perrin, Klaus Pontoppidan, Hans-Walter Rix, Massimo Robberto, Aki Roberge, Steve Rodney, Donald Schneider, Ohad Shemmer, Greg Snyder, Harley Thronson, Ben Weiner, John Wisniewski, Pieter van Dokkum, Roeland van der Marel, the attendees of the "Near-Field Deep-Field Connection" meeting (February 2014, Irvine CA) and the "Science with the *Hubble Space Telescope* IV" conference (March 2014, Rome Italy).

In the fast-moving field of exoplanets and starlight suppression technology, we have drawn on advice from many people. First among these is Chris Stark, who provided invaluable help with calculating exoplanet yields and mapping out the trade space of aperture and coronagraph performance. Stuart Shaklan, Jeremy Kasdin, Rémi Soummer, Laurent Pueyo, Bob Brown, Mamadou N'Diaye, Avi Mandel, and John Trauger were also very generous in sharing their in-depth knowledge of both the current state and future expected performance of coronagraphs and starshade concepts.

Fleshing out the details of the *HDST* mission concept relied heavily on the technical expertise of a wide range of people. We are extremely grateful for the many individuals who talked with us about aspects of telescope design, engineering concepts, and detector performance. These include Bala Balasubraminian, Jim Beletic, Matthew Bolcar, Julie Crooke, Don Figer, Jim Janesick, Andrew Jones, Michael Kienlen, Alice Liu, Ben Mazin, Shouleh Nikzad, Manuel Quijada, Norman Rioux, Oswald Siegmund, John Vallerga.

Many people contributed either figures or data for figures during the preparation of the report. These include Scott Gaudi, Galen Henderson, Andrew Jones, Mansi Kasliwal, Mark Kuchner, Mamadou N'Diaye, Laurent Pueyo, Greg Snyder, Rémi Soummer, and Chris Stark. We thank John O'Meara for his perspective on aperture trade-offs.

We must thank the breakfast group at the "Science with the *Hubble Space Telescope* IV" conference—Matt Mountain, Giovanni Fazio, and Priya Natarajan—for helping us converge on the report's current title while riffing on an earlier suggestion from Garth Illingworth.

Finally, the members of the committee wish to thank our families and friends for tolerating our being distracted by phone cons, travel, and writing during the past year.

## Chapter 5:

## Chapter 6:

## Suggested Reading

# Acronym Definitions

ACAD: Active Correction of Aperture Discontinuities

ACS: Advanced Camera for Surveys

ACS-SBC: Advanced Camera for Surveys Solar Blind Channel

AGN: Active Galactic Nuclei

AHM: Active Hybrid Mirrors

ALMA: Atacama Large Millimeter Array

AO: Adaptive Optics

APC: Apodized Pupil Coronagraph

APRA: Astronomy and Physics Research and Analysis

AU: Astronomical Unit

AURA: Association of Universities for Research in Astronomy, Inc.

CCD: Charge-Coupled Device

CGM: CircumGalactic Medium

*CHEOPS*: *CHaracterizing ExOPlanet Satellite*

CMOS: Complementary Metal Oxide Semiconductor

CO: Carbon monoxide

COR: Cosmic ORigins

COS: Cosmic Origins Spectrograph

D: Diameter

DARPA: Defense Advanced Research Projects Agency

DMS: dimethyl sulfide

E-ELT: European Extremely Large Telescope

EELV: Enhanced Expendable Launch Vehicle

ELT: Extremely Large Telescope

EM: ElectroMagnetic

EPICS: ExoPlanet Imaging Camera and Spectrograph

*EUVE*: *Extreme UltraViolet Explorer*

ExEP: EXoplanet Exploration Program

exoEarth: Earth-like exoplanets

FOV: Field Of View

FWHM: Full Width at Half Maximum

*FUSE*: *Far Ultraviolet Spectroscopic Explorer*

*GALEX*: *GALaxy Evolution Explorer*

GLAO: Ground-Layer Adaptive Optics

GMT: Giant Magellan Telescope

GPI: Gemini Planet Imager

GRB: Gamma-Ray Burst

HARPS-N: High Accuracy Radial velocity Planet Searcher for the Northern hemisphere

HCIT: High-Contrast Imaging Testbed

*HDST*: *High-Definition Space Telescope*

HgCdTe: Mercury-Cadmium-Telluride

HLC: Hybrid Lyot Coronagraph

IFU: Integral Field Unit

IGM: InterGalactic Medium

IMF: Initial Mass Function

IR: InfraRed

*ISS: International Space Station*

IWA: Inner Working Angle

*JWST*: *James Webb Space Telescope*

KBO: Kuiper Belt Object

LBTI HOSTS: Large Binocular Telescope Interferometer Hunt for Observable Signatures of Terrestrial Systems

LiF: Lithium Fluoride

LIGO: Laser Interferometer Gravitational-wave Observatory

LOWFS: Low-Order WaveFront Sensor

LSST: Large Synoptic Survey Telescope

mas: milliarcsecond

$M_\odot$: solar mass

MCAO: Multi-Conjugate Adaptive Optics

MCP: MicroChannel Plate

MIRI: Mid-InfraRed Instrument

MOS: Multi-Object Spectrograph

Mpc: MegaParseC

NASA: National Aeronautics and Space Administration

NIR: Near InfraRed

NIRCam: Near-InfraRed CAMera

NIRISS: Near-infrared Imager and Slitless Spectrograph

NIRSPec: Near-InfraRed SPECtrograph

nm: nanometer

NSF: National Science Foundation

OPTIIX: Optical Testbed and Integration on ISS eXperiment

OTE: Optical Telescope Element

OWA: Outer Working Angle

PC: Photon Counting

PFI: Planet Formation Instrument

PIAACMC: Phase-Induced Amplitude Apodization Complex Mask Coronagraph

*PLATO*: *PLAnetary Transits and Oscillations of stars*

PSF: Point-Spread Function

QE: Quantum Efficiency

QSO: Quasi-Stellar Object

R: Spectral Resolution value, equal to central wavelength divided by spectral bandwidth

RMS: Root Mean Squared

RV: Radial Velocity

SAFIR: Single Aperture Far-Infrared Observatory

SDSS: Sloan Digital Sky Survey

SiC: Silicon Carbide

*SIM*: *Space Interferometry Mission*

SKA: Square Kilometer Array

SLS: Space Launch System

SMBH: SuperMassive Black Hole

SNe: Supernovae

SNR: Signal-to-Noise Ratio

SPC: Shaped Pupil Coronagraph

SPECS: Submillimeter Probe of the Evolution of Cosmic Structure

SPHERE: Spectro-Polarimetric High-contrast Exoplanet REsearch instrument

STDT: Science and Technology Definition Team

STIS: Space Telescope Imaging Spectrograph

STS: Shuttle Transport System

TEC: Thermo-Electric Cooler

*TESS*: *Transiting Exoplanet Survey Satellite*

TMT: Thirty-Meter Telescope

TNO: Trans-Neptunian Object

TOO: Target Of Opportunity

*TPF-C*: *Terrestrial Planet Finder Coronograph*

*TPF-I*: *Terrestrial Planet Finder Interferometer*

TRL: Technology Readiness Level

UDF: Ultra-Deep Field

UFD: Ultra-Faint Dwarf

ULE: Ultra-Low Expansion

UV: UltraViolet

UVOIR: UltraViolet–Optical–InfraRed

VLTI: Very Large Telescope Interferometer

VLT-NaCo: Very Large Telescope-NAOS CONICA (Nasmyth Adaptive Optics System (NAOS) Near-Infrared Imager and Spectrograph (CONICA)

VNC: Visible Nulling Coronagraph

WBS: Work Breakdown Structure

*WFIRST/AFTA: Wide-Field Infrared Survey Telescope/Astrophysics Focused Telescope Assets*

*WISE: Wide-field Infrared Survey Explorer*

*XMM*: *X-ray Multi-mirror Mission*